\documentclass[a4paper,11pt]{article}
\usepackage{jheppub} 
\usepackage{lineno}

\usepackage[colorlinks=true,linkcolor=blue,citecolor=blue,urlcolor=blue]{hyperref}

\usepackage[T1]{fontenc}
\usepackage{blindtext}
\usepackage{amsmath}
\usepackage{amsfonts}
\usepackage{amssymb}
\usepackage{multirow}
\usepackage{graphics,graphicx}
\usepackage{bm}
\usepackage{epsfig}
\usepackage[normalem]{ulem}
\usepackage[usenames]{color}
\usepackage{epstopdf}
\usepackage{float}
\usepackage[utf8]{inputenc} 

\usepackage{makecell}
\usepackage{footmisc}

\usepackage[most]{tcolorbox}
\usepackage{rotating}

\usepackage{tikz}
\usetikzlibrary{decorations.pathmorphing}
\usetikzlibrary{decorations.markings}
\usetikzlibrary{calc}
\usetikzlibrary{math}

\allowdisplaybreaks
\newcommand{\Ds}{\displaystyle}

\newcommand{\nn}{\nonumber}

\renewcommand{\(}{\left(}
\renewcommand{\)}{\right)}
\renewcommand{\[}{\left[}
\renewcommand{\]}{\right]}

\renewcommand{\vec}[1]{\bm{#1}}

\newcommand{\red}[1]{{\color[rgb]{1,0,0} #1}}

\title{Phenomenology of genuine twist-three distributions from a global QCD analysis}
\author[a,b]{Guillermo Portela}
\author[a]{and Alexey Vladimirov}
 
\affiliation[a]{Departamento de Física Teórica \& IPARCOS, Universidad Complutense de Madrid, \\ Plaza de las Ciencias 1, 28040 Madrid, Spain}
\affiliation[b]{Departamento de Física Fundamental and IUFFyM, Universidad de Salamanca. \\
Plaza de la Merced S/N, E-37008 Salamanca, Spain}

\emailAdd{gportela@usal.es}
\emailAdd{alexeyvl@ucm.es}

\abstract{We present an extraction and phenomenological study of genuine twist-three parton distribution functions (PDFs) through a global QCD analysis of the structure function $g_2$ in deep-inelastic scattering (DIS), its moment $d_2$, and single-/double-spin asymmetries in semi-inclusive DIS (SIDIS). The unified description of these processes clearly confirms the universality of genuine twist-three PDFs and the validity of QCD factorization theorems at this subleading level. The extraction is accompanied by several validation tests, tests of sum rules, comparisons with earlier literature, computation forces and average transverse momentum of partons. As a by-product, we also obtain Sivers and worm-gear-T transverse momentum dependent (TMD) distributions, which allows us to construct a tomographic picture of the proton based on substantially broader data.}

\begin{document}
\maketitle

\section{Introduction}

The exploration and understanding of hadron structure is one of the definitive frontiers of modern particle physics. The dynamics of partons—quarks and gluons—is well understood at infinitesimal distances, but remains a puzzle at the scale of a hadron \cite{Gross:2022hyw}. The main source of knowledge about internal hadron structure is parton distribution functions (PDFs), which are, in a broad sense, hadron matrix elements of QCD operators. During the last few decades, huge efforts have been put into the determination of various PDFs, especially collinear PDFs \cite{Gao:2017yyd, Nocera:2026est}, transverse momentum dependent (TMD) PDFs \cite{Angeles-Martinez:2015sea, Boussarie:2023izj}, and generalized parton distributions (GPDs) \cite{Diehl:2003ny, Ji:2004gf, Belitsky:2005qn, Lorce:2025aqp}. These elements provide us with intuition about inter-parton interactions, and also form a foundation for practical QCD applications.

The effort to determine parton distributions has been largely concentrated on those that provide the leading contributions to factorization theorems, such as unpolarized and helicity PDFs, unpolarized TMDPDFs and GPDs. Undeniably, these distributions are the most practically important since they represent the main contribution and normalization for any process. Yet, at their core, they are number densities of partons, or closely related to them, and thus carry limited information about the dynamics of parton interactions. The details of this dynamics, such as the distribution of forces and transverse momentum correlations, are instead encoded in more involved parton distributions \cite{Politzer:1980me}, which include polarized TMDs and PDFs of higher twist, and other multi-parton correlators. However, these involved observables receive much less attention due to their broad variety, and thus the reduced importance attributed to any single one of them.

Nevertheless, theory tells us that many of these involved objects are not independent, but are indirectly related to each other and can be expressed via a few fundamental distributions -- the so-called genuine higher-twist PDFs, which are functions of many variables with intricate properties. Recently, in a proof-of-concept study \cite{Vladimirov:2025qrh}, we demonstrated how genuine PDFs of twist-three can be determined from data via a unified fit of multiple processes. In this work, we elaborate on this approach, present a comprehensive exposition of this determination, and analyze the immediate outcomes of this new knowledge.

Factorization theorems are the foundation of parton physics, allowing us to describe high-energy processes. In essence, they provide a systematic way to express the cross-section as a series in the hard-probe virtuality $Q$ \cite{Politzer:1980me, Shuryak:1981kj, Ellis:1982cd}. Each subsequent term in the power series is characterized by a set of parton operators with increased dimension and corresponding PDFs. Often, such PDFs are referred to as dynamical twist PDFs \cite{Jaffe:1996zw}. Although there are various operators of higher dynamical twist, eventually they can be reduced to a smaller basis of genuine twist PDFs. These operators are characterized by the geometrical twist ("dimension-minus-spin") \cite{Gross:1971wn, Brandt:1970kg} and possess an important defining property: the geometrical twist is preserved by the evolution equation. This property guarantees the non-perturbative independence of genuine twist PDFs, and consequently, their universality.

Naturally, the number of PDFs and their complexity grow with the twist. Specifically, there are three twist-two PDFs (unpolarized, helicity, and transversity PDFs), which are functions of a single variable that satisfy a corresponding DGLAP-type evolution equation. There are, in total, four collinear twist-three PDFs (for each quark flavor, plus two additional gluon PDFs), which are functions of two variables and satisfy an evolution equation with a two-dimensional integral convolution \cite{Bukhvostov:1985rn, Balitsky:1987bk, Braun:2000av, Braun:2009mi}. Beyond this, the complexity of the approach grows rapidly; there are about a dozen twist-four distributions with three or four variables, which have a very intricate system of evolution equations \cite{Bukhvostov:1987pr, Braun:2009vc, Ji:2014eta}. Note that in this discussion we concentrate solely on collinear PDFs, omitting the discussion about higher-twist TMD distributions and GPDs, which have also gained interest in recent years; for reviews, see \cite{Rodini:2022wki, Braun:2022gzl, Belitsky:2000vx, Aslan:2018zzk}.

Genuine twist-three PDFs appear at the next-to-leading power correction. They do not have standard names, and even their notation differs from one publication to another. In this work, we utilize the notation inherited from a line of publications \cite{Braun:2001qx, Braun:2009mi, Scimemi:2018mmi, Rodini:2024usc}, which defines them as $\{T,\Delta T, H, E\}$. All of them are given by three-point light-cone operators, and all dynamical twist-three distributions can be derived from them and genuine twist-two PDFs. The distributions $\{T,\Delta T\}$ have an analogy with unpolarized and helicity PDFs, whereas $\{H, E\}$ have an analogy to the transversity PDF. In this work, we consider only the pair $\{T,\Delta T\}$ (and the related gluon distributions), which are defined and described in detail in sec.~\ref{sec:tw3-theory}.

Interest in twist-three PDFs has persisted since the early '80s, which is evident from a significant number of theoretical works. Meanwhile, there are practically no phenomenological works; those that do exist are all dedicated to dynamical higher-twist functions and are thus methodologically incomplete. There are several reasons for this situation.

The primary one is the absence of good-quality data. Sub-leading power effects and polarization effects are suppressed at high energies, and thus the corresponding observables are often statistically insignificant. Furthermore, the corresponding experiments have large systematic and statistical uncertainties.

However, even with a very precise measurement of a single twist-three observable, one would not be able to determine a genuine twist-three PDF. Genuine twist-three PDFs are functions of two variables that are projected onto a single kinematic  point by an integral convolution. Due to this, the consideration of a single observable is inconclusive, since it only gives access to a section of the function. Here, it is important to mention that, nonetheless, the scale evolution operates on the whole domain of the PDFs, and thus integrates information about the rest of the function into its scale dependence.

This discussion immediately leads to a practical conclusion: the only hope to determine twist-three functions is to perform a joint analysis of various processes. This will allow one to gain access to various sections of a two-dimensional function, and thus pin it down from different perspectives. Such a study must be supplemented with proper scale evolution, which will further correlate various observables. Moreover, this global analysis should incorporate data points from many measurements, presumably helping to overcome the poor quality of individual measurements through their sheer quantity.

There are various twist-three effects that could be included into the analysis. An important role here is played by indirect twist-three measurements, in particular, by TMDPDFs. There are eight TMD twist-two functions (not to be confused with the geometrical and collinear twists discussed above) \cite{Mulders:1995dh, Angeles-Martinez:2015sea, Boussarie:2023izj}, four of which match genuine twist-three distributions in the limit of small $b$ \cite{Scimemi:2018mmi}. This limit provides the dominant contribution to the cross-section, making these TMDPDFs an important source of information about twist-three distributions. As a matter of fact, the most precise extractions of parts of twist-three distributions come from the determination of the Qiu-Sterman matrix element from fits of the Sivers function \cite{Cammarota:2020qcw, Bacchetta:2020gko, Bury:2021sue} or from the large-$p_\perp$ part of the single-spin asymmetry \cite{Kouvaris:2006zy, Cammarota:2020qcw, Anselmino:2013rya}.

In this work, we consider four data sources for twist-three effects: the structure function $g_2$ in deep-inelastic scattering (DIS), its moment $d_2$ (which is in many cases measured independently), and the Sivers and worm-gear-T functions (which are TMDPDFs). This combination provides access to various parts of the multi-dimensional function and, as demonstrated here, allows us to reconstruct genuine twist-three PDFs with sufficient confidence. This list of observables is not exhaustive and is motivated by our previous investigations. In the future, we plan to extend it.

Clearly, this kind of analysis is theoretically and technically challenging. The consideration of several distinct processes requires the construction of a framework able to process them simultaneously. This is not purely a technical problem, because such an analysis requires knowledge of the accompanying theoretical (coefficient functions, evolution kernels, etc.) and phenomenological (twist-two PDFs) elements. In this respect, this work represents the final step of a long-term theoretical and numerical effort toward the systematization and numerical implementation of twist-three effects \cite{Scimemi:2018mmi, Scimemi:2019gge, Moos:2020wvd, Bury:2020vhj, Vladimirov:2021hdn, Horstmann:2022xkk, Rein:2022odl, Rodini:2024usc} in collinear and TMD factorizations. Many practical aspects, especially those related to the analysis of TMD data, were developed and tested in our earlier works \cite{Vladimirov:2019bfa, Bury:2020vhj, Bury:2021sue, Bury:2022czx, Horstmann:2022xkk, Moos:2023yfa, Moos:2025sal}, providing us with a solid foundation to start with. Simultaneously, there are many aspects that have never been implemented before, such as a complete twist-three evolution, or the joint analysis of TMD and collinear data, to name a few.

Let us remark that, as previously stated, the proof-of-concept study was published in ref.~\cite{Vladimirov:2025qrh}. The present work represents an update of this earlier study in practically every aspect. This is mainly reflected by the inclusion of more data for the fit, some of which are more precise, and the implementation of a more flexible parametrization for PDFs, with a larger number of parameters. Still, we found that the newer determination is in very good agreement with the results of ref.~\cite{Vladimirov:2025qrh}, which demonstrates the consistency of the approach.

It should be kept in mind that this is the first work dedicated to the problem of determining genuine twist-three PDFs from data. Therefore, in many cases, we make decisions (such as the construction of the fitting ansatz, interpretation of the data, etc.) without the benefit of prior experience. Although we put special effort into checking various possibilities, our results could still be biased. In some sense, the situation here is similar to the first extractions of twist-two PDFs, such as \cite{Gluck:1977ah, Martin:1987vw}. Over time, many of those early choices were replaced and updated, but they formed the base for subsequent exploration.

The results of our determination are very promising: the twist-three distributions $T$ and $\Delta T$ are extracted from the data with reliable statistical accuracy. This acts as a stress test for the universality of parton distributions, since four very distinct observables must be described simultaneously by a small set of functions. As a consequence, for the first time, one can analyze various twist-three effects directly using the extracted distributions. To motivate and initiate these studies, we dedicate a significant part of the paper to discussing some immediate conclusions about hadron structure that follow from the extracted PDFs.

Since this work unifies several historically isolated theoretical and phenomenological branches of QCD, we provide a detailed, self-contained exposition of each component. The remainder of this paper is structured as follows.

In sec.~\ref{sec:tw3-theory}, we review the definitions, formal properties, and evolution equations governing genuine twist-three distributions, and establish the formal notation used throughout the text. Section \ref{sec:DIS} collects the relevant theoretical expressions and available experimental world data (sec.~\ref{sec:DIS-data}) for the DIS observables $g_2$ and $d_2$. Analogously, section \ref{sec:TMD} outlines the SIDIS and TMD framework, with its corresponding data review located in sec.~\ref{sec:SIDIS-data}. The practical implementation of our global analysis, code, parametric ansatz, and statistical treatment is detailed in sec.~\ref{sec:fit}, summarized schematically in the block diagram of fig.~\ref{fig:scheme}. Our results are evaluated across two final sections: sec.~\ref{sec:validation} provides a thorough validation of the fit, error propagation, and stability checks against data sub-selection, while sec.~\ref{sec:results} explores the physical consequences of the extraction, including evaluations of QCD sum rules, quark transverse momentum moments, and effective color forces. To keep the main text concise, additional diagnostic plots are compiled in the appendix. The practical elements, such as the extracted grids, the complete analysis code, and the framework, are publicly available in the repository \cite{REP}.

\section{Definition and properties of genuine twist-three distributions} 
\label{sec:tw3-theory}

In this section we present a summary of the definitions, symmetry properties, evolution equations, and interpretation of the genuine twist-three distributions, which, for brevity, we may call twist-three PDFs. There are several other naming conventions used in the literature. Given the explicit definitions of the matrix elements, a comparison between different conventions is straightforward.

In this work, we follow the notation pattern used in the works \cite{Braun:2009mi, Rein:2022odl, Rodini:2024usc} since it has proven to be practically convenient and general. Specifically, we use two sets of functions
$$\{T_f,\Delta T_f,T_{3F}^+,T_{3F}^-\},\qquad\text{and}\qquad \{\mathfrak{S}^+_f,\mathfrak{S}^-_f,\mathfrak{F}^+,\mathfrak{F}^-\},$$
where the first two functions in each set are quark-gluon-quark twist-three PDFs, and the last two are three-gluon twist-three PDFs. Both sets are related to each other by an algebraic transformation. The need to consider both sets lies in what each one represents: the first set is more physically motivated, and thus all results are presented in it, while the second one significantly simplifies the evolution equations, and thus it is used for the technical implementation. 

\subsection{Quark-gluon-quark distributions}

Quark-gluon-quark twist-three distributions, also referred to as quark twist-three distributions, are the main focus of this work. The complete set of distributions contains four functions -- two chiral-even distributions ($T$, $\Delta T$) and two chiral-odd ($E, H$). These subsets are independent and do not mix in physical observables in massless QCD (similarly to unpolarized and transversity twist-two distributions).

In this work, we consider only chiral-even distributions (for corresponding definitions for the chiral-odd distributions see \cite{Rodini:2024usc}). Their explicit definitions are
\begin{eqnarray}&&\label{def:T}
\langle p,S |g\,\overline{q}(z_{1}n)F^{\mu+}(z_{2}n)\gamma^{+}q(z_{3}n)|p,S\rangle
\\\nn &&\qquad\qquad\qquad
=2\epsilon^{\mu\nu}_{T}s_{\nu}(p^{+})^{2}M\int[dx]\:T_f(x_1,x_2,x_3)\:e^{-ip^{+}(x_1z_1+x_2z_2+x_3z_3)}~,
\\&&\label{def:DeltaT}
\langle p,S |ig\,\overline{q}(z_{1}n)F^{\mu+}(z_{2}n)\gamma^{+}\gamma^{5}q(z_{3}n)|p,S\rangle
\\\nn &&\qquad\qquad\qquad
=-2s^{\mu}_{T}(p^{+})^{2}M\int[dx]\:\Delta T_f(x_1,x_2,x_3)\:e^{-ip^{+}(x_1z_1+x_2z_2+x_3z_3)}
~,
\end{eqnarray}
where $g$ is the QCD coupling constant, $M$ is the proton's mass, $q$ is a quark field and $F^{\mu\nu}$ is the gluon-field strength. The index $f$ denotes flavor of the quark field. In expression (\ref{def:T}, \ref{def:DeltaT}) we have omitted gauge links along $n$, joining the fields at different points in a gauge-invariant manner. So, the operator should be read as
\begin{equation}
\overline{q}(z_{1}n)F^{\mu+}(z_{2}n)\gamma^{+}q(z_{3}n)\xrightarrow{\text{read as}}\overline{q}(z_{1}n)[z_{1}n,z_{2}n]F^{\mu+}(z_{2}n)\gamma^{+}[z_{2}n,z_{3}n]q(z_{3}n),
\end{equation}
where 
\begin{equation}\label{def:WilsonLine}
[z_{1}n,z_{2}n]=\mathcal{P}\exp\(-ig\int_{z_{1}}^{z_{2}}ds\:A^{+}(s n)\),
\end{equation}
and $F_{\mu\nu}=t^A F^A_{\mu\nu}$ with $t^A$ being the infinitesimal generators of the SU(3) group.

The vector $n^\mu$ is the light-cone vector ($n^2=0$) aligned along the large component of the proton's momentum\footnote{
We use the standard convention for components of a vector in light-cone coordinates. Namely, a vector $a$ has decomposition
\begin{equation*}
a^{\mu}=a^{+}\bar{n}^{\mu}+a^{-}n^{\mu}+a_{T}^{\mu}~,
\end{equation*}
where $n^2=\bar n^2=0$, $(n\bar n)=1$, $a^+=(n\cdot a)$, and $a^-=(\bar n\cdot a)$. The transverse component $a_T^\mu$ satisfies $(n\cdot a_T)=(\bar n\cdot a_T)=0$. 
}.
$$
p^\mu=\bar n^\mu p^+ +n^\mu \frac{M^2}{2p^+}.
$$
The transverse plane is specified by symmetric and anti-symmetric tensors
\begin{equation}\label{def:g,eps}
g^{\mu\nu}_{T}=g^{\mu\nu}-n^{\mu}\overline{n}^{\nu}-\overline{n}^{\mu}n^{\nu}, \quad \epsilon^{\mu\nu}_{T}=\epsilon^{\mu\nu-+},\quad \epsilon^{0123}=1.
\end{equation} 
The vector $s^{\mu}$ is the spin-vector of the proton normalized as $s^2=-1$, and $s_T^\mu=g_T^{\mu\nu}s_\nu$ is its transverse component.

\begin{figure}[t]
\centering
\includegraphics[width=0.6\linewidth]{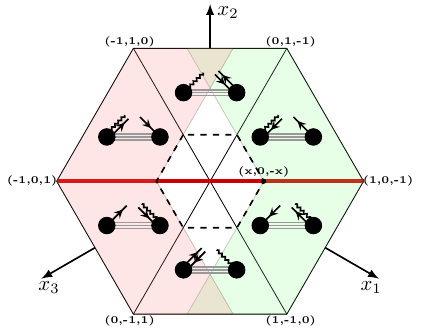}
\caption{Twist-three quark PDFs configuration space. Each sector corresponds to an interference between particular partonic configurations, shown in diagrams (with wavy lines being gluons, and straight lines being quarks). Each little diagram represents the corresponding partonic process, where the fields attached to each left circle are emitted, and the fields attached to each right circle are absorbed.}
\label{fig:Hex_Interp}
\end{figure}

The variables $x_{1,2,3}$ are momentum fractions of the parton fields with respect to the proton's momentum. The integration measure $[dx]$ incorporates restrictions imposed by the conservation of momentum and causality conditions defining the support of twist-three PDFs \cite{Jaffe:1983hp}. Namely
\begin{equation}\label{def:domain}
    -1\leq x_{i}\leq 1\qquad\text{and}\qquad x_{1}+x_{2}+x_{3}=0\:.
\end{equation}
Hence, the measure $[dx]$ is defined as
\begin{equation}
\int[dx]=\int_{-1}^{1}dx_{1}\int_{-1}^{1}dx_{2}\int_{-1}^{1}dx_{3}\:\delta(x_{1}+x_{2}+x_{3})\:.
\end{equation}
The conservation of total momentum indicates that twist-three PDFs are functions of two variables rather than three. Nonetheless, it is convenient to preserve the three-variable notation, as it proves to be the natural setting for understanding the physical interpretation and properties of twist-three PDFs. In particular, such a notation suggests a convenient way to present twist-three distributions using barycentric coordinates \cite{Braun:2009mi}, in which the domain (\ref{def:domain}) takes the form of a proper hexagon (see e.g. fig.\ref{fig:Hex_Interp}).

In what follows, we often use the shorthand notation to indicate the three argument sequence of $x$'s as
$$
(x_{ijk})\longleftrightarrow (x_i, x_j,x_k).
$$
Besides their dependence on the momentum fractions, all PDFs depend on the renormalization scale, denoted by $\mu$. In the majority of expressions we omit the argument $\mu$ for simplicity, but it is always implied.

There are several alternative variants of basis functions that are used in literature. As a classical example, there are the functions $(S^{+},S^{-})$, which naturally appear in the description of DIS structure functions and are used in the works \cite{Braun:1998id, Braun:2001qx, Braun:2021aon} and related series. These are related to $(T, \Delta T)$ by the following transformation:
\begin{equation}
S^{\pm}(x_{123})=\frac{-T(x_{123})\pm \Delta T(x_{123})}{2}.
\end{equation}

However, these functions do not have definite properties with respect to C-parity transformation, and thus mix under the QCD evolution equations. Therefore, it is often convenient to consider combinations with definite C-parity $(\mathfrak{S}^{+},\mathfrak{S}^{-})$, introduced in ref.\cite{Braun:2009mi}. They are defined as
\begin{eqnarray}
\mathfrak{S}^{\pm}(x_{123})
&=&
T(x_{123})-\Delta T(x_{123})\pm T(x_{321})\pm \Delta T(x_{321})
\\\nn &=&-2\(S^{+}(x_{123})\pm S^{-}(x_{321})\),
\end{eqnarray}
and the inverse relations are
\begin{eqnarray}\label{T->Sigma}
T(x_{123})&=&\frac{1}{4}\(
\mathfrak{S}^+(x_{123})+\mathfrak{S}^+(x_{321})+
\mathfrak{S}^-(x_{123})-\mathfrak{S}^-(x_{321})\),
\\
\Delta T(x_{123})&=&-\frac{1}{4}\(
\mathfrak{S}^+(x_{123})-\mathfrak{S}^+(x_{321})+
\mathfrak{S}^-(x_{123})+\mathfrak{S}^-(x_{321})\),
\\
S^+(x_{123})&=&\frac{-T(x_{123})+\Delta T(x_{123})}{2}=-\frac{\mathfrak{S}^+(x_{123})+\mathfrak{S}^-(x_{123})}{4},
\\
S^-(x_{123})&=&\frac{-T(x_{123})-\Delta T(x_{123})}{2}=-\frac{\mathfrak{S}^+(x_{321})-\mathfrak{S}^-(x_{321})}{4}.
\end{eqnarray}

There are several alternative conventions for the naming of twist-three distributions, which are usually alternatives to our $T$ and $\Delta T$ functions with different prefactors and orders of arguments. The translation from one convention to another can be found by a direct comparison of the operator definitions. For example, the twist-three PDFs employed in works \cite{Kanazawa:2015ajw, Rein:2025pwu} and related series are related to ours as \footnote{The relative -1 between functions appears due to the definition of operators in refs.\cite{Kanazawa:2015ajw, Rein:2025pwu} with $F^{+\mu}$.}
\begin{eqnarray}
F_{(qg)}^{\mathbf{1}}(x,x')=-T(-x,x-x',x'),\qquad F_{(qg)}^{\mathbf{5}}(x,x')=-\Delta T(-x,x-x',x').
\end{eqnarray}
Some extra relations between conventions can be found in refs.~\cite{Rodini:2024usc, Braun:2009mi}.

\subsection{Three-gluon distributions}

The three-gluon genuine twist-three distributions are defined as
\begin{eqnarray}\label{def:Tg+}
&&\langle p,S|igf^{ABC}F^{A,\mu +}(z_1n)F^{B,\nu +}(z_{2}n)F^{C,\rho +}(z_{3}n)|p,S\rangle
\\\nn
&&\qquad \qquad\qquad
=(p^{+})^{3}M\int [dx]e^{-ip^{+}(x_1z_1+x_2z_2+x_3z_3)}\sum_{i=2,4,6}t^{\mu\nu\rho}_{i}F^{+}_{i}(x_1,x_2,x_3),
\\\label{def:Tg-}
&&
\langle p,S|gd^{ABC}F^{A,\mu +}(z_1n)F^{B,\nu +}(z_{2}n)F^{C,\rho +}(z_{3}n)|p,S\rangle
\\
&&\nn \qquad\qquad\qquad
=(p^{+})^{3}M\int [dx]e^{-ip^{+}(x_1z_1+x_2z_2+x_3z_3)}\sum_{i=2,4,6}t^{\mu\nu\rho}_{i}F^{-}_{i}(x_1,x_2,x_3),
\end{eqnarray}
where $f^{ABC}$ and $d^{ABC}$ are the anti-symmetric and symmetric structure constants of $SU(3)$. The tensors $t^{\mu\nu\rho}_{i}$ are the set of independent tensors \cite{Scimemi:2018mmi}:
\begin{eqnarray}
t^{\mu\nu\rho}_{2}&=&
s^{\alpha}_{T}\epsilon^{\mu\alpha}_{T}g^{\nu\rho}_{T}+s^{\alpha}_{T}\epsilon^{\nu\alpha}_{T}g^{\rho\mu}_{T}+s^{\alpha}_{T}\epsilon^{\rho\alpha}_{T}g^{\mu\nu}_{T},
\\
t^{\mu\nu\rho}_{4}&=&
-s^{\alpha}_{T}\epsilon^{\mu\alpha}_{T}g^{\nu\rho}_{T}+2s^{\alpha}_{T}\epsilon^{\nu\alpha}_{T}g^{\rho\mu}_{T}-s^{\alpha}_{T}\epsilon^{\rho\alpha}_{T}g^{\mu\nu}_{T},
\\
t^{\mu\nu\rho}_{6}&=&s^{\alpha}_{T}\epsilon^{\mu\alpha}_{T}g^{\nu\rho}_{T}-s^{\alpha}_{T}\epsilon^{\rho\alpha}_{T}g^{\mu\nu}_{T}.
\end{eqnarray}
Other elements of the definition are the same as in the case of quark distributions.

To make expressions (\ref{def:Tg+}, \ref{def:Tg-}) gauge-invariant, they should be decorated by Wilson lines, as follows
\begin{eqnarray}
&&f^{ABC}F^{A,\mu +}(z_1n)F^{B,\nu +}(z_{2}n)F^{C,\rho +}(z_{3}n)\xrightarrow{\text{read as}}
\\\nn &&\quad
F^{A,\mu +}(z_{1}n)[z_{1}n,z_{0}n]^{AA'}F^{B,\nu +}(z_{2}n)[z_{2}n,z_{0}n]^{BB'}F^{C,\rho +}[z_{3}n,z_{0}n]^{CC'}f^{A'B'C'},
\end{eqnarray}
where $[z_1,z_0]^{AB}$ is a straight Wilson line (\ref{def:WilsonLine}) in the adjoint representation. The expression is independent of the position $z_0$.

The gluon distributions $F^{\pm}_{2,4,6}$ possess a large number of symmetries inherited from the symmetries of the operator. As a consequence, the number of independent functions is smaller than the number of irreducible tensors $t_i^{\mu\nu\rho}$. In practice, it is more efficient to work with the basis $(T^{\pm}_{3F},\Delta T^{\pm}_{3F})$ introduced in \cite{Braun:2001qx, Braun:2009mi}. The relation is
\begin{eqnarray}
T^{\pm}_{3F}(x_{123})&=&4F^{\pm}_{2}(x_{123})+2F^{\pm}_{4}(x_{123}),
\\
\Delta T^{\pm}_{3F}(x_{123})&=&-2F^{\pm}_{6}(x_{123}).
\end{eqnarray}
However, these four distributions are also not independent. They satisfy
\begin{eqnarray}
\Delta T_{3F}^\pm(x_{123})=\pm \(T_{3F}^\pm(x_{132})-T_{3F}^\pm(x_{213})\).
\end{eqnarray}
Thus, there are only two independent gluon distributions $(T_{3F}^+,T_{3F}^-)$. In the following, we use the basis $(T_{3F}^+,T_{3F}^-)$ to illustrate the results.

As in the case of the quark distributions $(T,\Delta T)$, the gluon distributions $(T_{3F}^+,T_{3F}^-)$ do not have definite C-parity. The definite-C-parity distributions are \cite{Braun:2009mi}
\begin{equation}
\mathfrak{F}^{\pm}(x_{123})=T^{\pm}_{3F}(x_{123}) \mp T^{\pm}_{3F}(x_{132})\pm T^{\pm}_{3F}(x_{213}),
\end{equation}
with the inverse relation being
\begin{equation}
T^{\pm}_{3F}(x_{123})=\frac{1}{2}\[\mathfrak{F}^{\pm}(x_{123})\mp \mathfrak{F}^{\pm}(x_{321})\].
\end{equation}

This notation can be compared with notation of other groups. For example, the comparison with \cite{Beppu:2010qn} provides
\begin{eqnarray}
T^{+}_{3F}(-x_2,x_2-x_1,x_1)&=&2\(2N_1(x_1,x_2)- N_1(x_1,x_1-x_2)- N_1(x_2,x_2-x_1)\),
\\
T^{-}_{3F}(-x_2,x_2-x_1,x_1)&=&2\(2O_1(x_1,x_2)+ O_1(x_1,x_1-x_2)+ O_1(x_2,x_2-x_1)\),
\end{eqnarray}
and 
\begin{eqnarray}
T^{\pm}_{3F}(-x_2,x_2-x_1,x_1)&=&-x T_G^{(\pm)}(x_1,x_2).
\end{eqnarray}

\subsection{Interpretation} 
\label{sec:interpret}

Twist-three parton distributions are new functions for phenomenological studies, and thus it is not clear which intuition one should utilize in their case. Obviously, their structure is different from the familiar twist-two distributions, which, at first approximation, can be thought of as parton number densities within the proton \cite{Drell:1969wd}. In this simplified picture, twist-three distributions correspond to the interference of parton fields within the proton wave-function \cite{Politzer:1980me, Jaffe:1983hp}.

To obtain a naive parton model interpretation, it is sufficient to consider a distribution in the infinite momentum frame and the light-cone gauge \cite{Diehl:2002he, Burkardt:2002hr}. In this limit only "good" components of parton fields are dynamical, and their expansion in terms of canonical operators reads
\begin{eqnarray}
q(x^-)&=&\frac{1}{2\pi}\int_0^\infty\frac{dp^{+}}{2p^{+}}
\sum_{s}\(\hat{a}_s(p)u(p,s)e^{-ip^{+}x^{-}}+\hat{c}^{\dagger}_s(p)\:v(p,s)\:e^{i\:p^{+}x^{-}}\),
\\
A^{A}_\mu(x^-)&=&\frac{1}{2\pi}\int_0^\infty \frac{dp^{+}}{2p^{+}}
\sum_{\lambda}\(\hat{b}^{A}_\lambda(p)\varepsilon_{\mu}(p,\lambda)e^{-ip^{+}x^{-}}+\hat{b}^{A\dagger}_\lambda(p)\varepsilon^{*}_\mu(p,\lambda)e^{ip^{+} x^{-}}\),
\end{eqnarray}
where $a_s$, $b_\lambda$ and $c_s$ are creation operators for quark, gluon and anti-quark respectively, the $u$, $v$ are quark spinors, and $\epsilon_\mu$ is the gluon polarization vector. Substituting these expressions into the definition of distributions (\ref{def:T}, \ref{def:DeltaT}), one obtains that the configuration space is naturally divided in six sub-sectors delimited by lines $x_i=0$. Each one of them corresponds to a given $qgq$ state interaction within the proton.

We have found the following expressions (for definiteness we orient the spin of the proton along the direction $1$):
\begin{itemize}
\item Sector $(x_1>0,~x_2>0,~x_3<0)$:
\begin{eqnarray}\nn
T_f(x_{123})&\sim & \text{Im}\Big(
\langle p,s|\hat c_R(x_1)~ \hat b_2(x_2)\hat c^\dagger_R(-x_3)
+ \hat c_L(x_1)~ \hat b_2(x_2)\hat c^\dagger_L(-x_3)|p,s\rangle\Big),
\\\nn
\Delta T_f(x_{123})&\sim & \text{Re}\Big(
\langle p,s|\hat c_R(x_1)~ \hat b_1(x_2)\hat c^\dagger_R(-x_3)
- \hat c_L(x_1)~ \hat b_1(x_2)\hat c^\dagger_L(-x_3)|p,s\rangle\Big).
\end{eqnarray}
That is, it represents the interference between an anti-quark-gluon pair and an anti-quark.
\item Sector $(x_1<0,~x_2>0,~x_3<0)$:
\begin{eqnarray}\nn
T_f(x_{123})&\sim & \text{Im}\Big(
\langle p,s| \hat b_2(x_2)~\hat a^\dagger_R(-x_1) \hat c^\dagger_R(-x_3)
-  \hat b_2(x_2)~\hat a^\dagger_L(-x_1) \hat c^\dagger_L(-x_3)|p,s\rangle\Big),
\\\nn
\Delta T_f(x_{123})&\sim & \text{Re}\Big(
\langle p,s|\hat b_1(x_2)~\hat a^\dagger_R(-x_1)\hat c^\dagger_R(-x_3)
+ \hat b_1(x_2)~\hat a^\dagger_L(-x_1)\hat c^\dagger_L(-x_3)|p,s\rangle\Big).
\end{eqnarray}
That is, it represents the interference between a quark-anti-quark pair and a gluon.
\item Sector $(x_1<0,~x_2>0,~x_3>0)$:
\begin{eqnarray}\nn
T_f(x_{123})&\sim & \text{Im}\Big(
\langle p,s|\hat b_2(x_2)\hat a_R(x_3)~\hat a^\dagger_R(-x_1)
+ \hat b_{2}(x_2)\hat a_L(x_3)~ \hat a^\dagger_L(-x_1)|p,s\rangle\Big),
\\\nn
\Delta T_f(x_{123})&\sim & \text{Re}\Big(
\langle p,s|\hat b_2(x_2)\hat a_R(x_3)~\hat a^\dagger_R(-x_1)
- \hat b_1(x_2)\hat a_L(x_3)~\hat a^\dagger_L(-x_1)|p,s\rangle\Big).
\end{eqnarray}
That is, it represents the interference between a quark-gluon pair and a quark.
\end{itemize}
The remaining sectors are obtained by the C-conjugation of the amplitude. I.e. they change each particle to its antiparticle and flip the particle's emission(absorption) to absorption(emission). 

In this way, each sector of the hexagon corresponds to an interference between particular compositions of partonic fields inside the proton. This interference interpretation is demonstrated in fig.~\ref{fig:Hex_Interp} by parton-level diagrams within each sector.

The border lines $x_i=0$ correspond to the cases of an infinitely soft quark ($x_3=0$), gluon $(x_2=0)$ or anti-quark ($x_1=0$) field forming a pair with another parton. Naively, this pair corresponds to a leading particle with altered quantum numbers (color and spin) but the same momentum. It is expected that the twist-three distributions are continuous at these lines, which is also supported by the evolution equation.

In contrast to the twist-two PDFs, there are no separate quark and anti-quark twist-three distributions. Each distribution incorporates all configurations for a given flavor and continuously switches from one to another. However, one may think of the sectors with $(x_3>0, x_1<0)$ as "quark" sectors, because in these sectors the distribution is related to the interference of quark fields accompanied by a gluon. Similarly, the sectors with $(x_1>0, x_3<0)$ are "anti-quark" sectors. Finally, the sectors with $(x_1>0,x_3>0)$ and  $(x_1<0, x_3<0)$ are special, since they correspond to gluon/quark-pair interference. Often, these sectors contribute to both quark and anti-quark observables.

We would like to mention one other contrasting difference between twist-two and twist-three distributions. In the case of twist-two distributions, the quark PDFs are significantly larger than the anti-quark PDFs (at least in the valence region). This follows from the fact that the proton (at moderate values of $x$) consists of two $u$ and one $d$ quark, which dominate the corresponding number densities. Nevertheless, in the case of twist-three distributions, one should not expect any striking imbalance between "quark" and "anti-quark" sectors, nor between $u/d$ and $sea$ quarks, because quantum interference is driven merely by background color-field correlations rather than a simple count of the valence content. Of course, there are still some differences between flavors because the proton medium prefers particles to anti-particles and distinguishes flavors, but they are not as extremely pronounced. 

Also, there exist other interpretations of several specific integrals of twist-three distributions. They correspond to forces, transverse momentum, etc, and are analogous to quark-number or momentum sum rules in the case of twist-two PDFs. These interpretations are correspondingly discussed in sections \ref{sec:d2+force} and \ref{sec:burkardt}.

\subsection{Symmetries} 
\label{sec:sym}

Twist-three distributions obey a number of symmetries with respect to reflection and permutation of momentum fractions. These symmetries reflect the discrete symmetries of QCD.

The quark distributions in various bases satisfy
\begin{eqnarray}\label{sym:1}
T(x_{123})&=&T(-x_{321}),\qquad~ \Delta T(x_{123})=-\Delta T(-x_{321}),
\\\label{sym:2}
S^{\pm}(x_{123})&=&S^{\mp}(-x_{321}),
\qquad
\mathfrak{S}^{\pm}(x_{123})=\pm\mathfrak{S}^{\pm}(-x_{123}).
\end{eqnarray}
While the gluon distributions satisfy
\begin{eqnarray}\label{sym:3}
T^{\pm}_{3F}(x_{123})&=&\mp T^{\pm}_{3F}(-x_{123}),
\qquad  T^{\pm}_{3F}(x_{123})=\mp T^{\pm}_{3F}(x_{321}),
\\\label{sym:4}
\mathfrak{F}^{\pm}(x_{123})&=&\mp\mathfrak{F}^{\pm}(-x_{123}),\qquad  \mathfrak{F}^{\pm}(x_{123})=\mp \mathfrak{F}^{\pm}(x_{132}).
\end{eqnarray}
Here, we have used the notation convention,
$$
(-x_{ijk})\longleftrightarrow (-x_i, -x_j, -x_k).
$$
Note that gluon distributions exhibit a larger number of symmetry relations than the quark ones. Visually, these symmetries represent transformations of the hexagonal domain. Namely
\begin{eqnarray}\nn
(x_{123})\to (-x_{321})&:\qquad& \text{mirroring with respect to $x_2=0$ axis}
\\
\nn
(x_{123})\to (-x_{123})&:\qquad& \text{mirroring with respect to $x_{1,2,3}=0$ point}
\\
\nn
(x_{123})\to (x_{132})~~&:\qquad& \text{turning the hexagon by $\pi/3$ counter-clockwise.}
\end{eqnarray}
Correspondingly, distributions are symmetric or anti-symmetric with respect to these transformations.

The symmetry relations are employed in all expressions presented in this work. That is, some equations are presented in a form that is already simplified using (\ref{sym:1}-\ref{sym:4}). For that reason, it is important to check that the implementation of non-perturbative distributions satisfies the symmetry relations at each step.

It is also important to mention that the evolution code \cite{Rodini:2024usc}, presented in the following sections, does not check the input functions, but assumes that they satisfy the proper symmetry conditions. Thus, the boundary values of the PDFs must obey these conditions. Otherwise, the code may provide an incorrect result.

\subsection{Evolution} 
\label{sec:evol}

The leading-order (LO) evolution equations for twist-three distributions were derived already in the 1980s \cite{Bukhvostov:1985rn, Balitsky:1987bk}, and revised later in refs.\cite{Mueller:1997yk, Kang:2008ey, Braun:2009mi}. The evolution equations have a rather involved form and, contrary to the twist-two intuition, mix the $T$ and $\Delta T$ functions. 

In ref.~\cite{Braun:2009mi}, it is demonstrated that a convenient basis for the solution of the evolution equations is the basis of definite-C-parity functions $(\mathfrak{S}^\pm,\mathfrak{F}^\pm).$ In these terms, there is no mixing between different elements of the basis, and the evolution equations can be written in the following form:
\begin{eqnarray}\label{def:ev-ns-singlet}
\mu^2 \frac{\partial}{\partial \mu^2} \mathfrak{S}^\pm_{\text{NS}_i}&=&-a_s(\mu) \mathbb{H}_{\text{NS}}\otimes\mathfrak{S}^\pm_{\text{NS}_i},
\\\label{def:ev-singlet}
\mu^2 \frac{\partial}{\partial \mu^2} \(\begin{array}{c}
\mathfrak{S}^\pm_{\text{S}}\\
\mathfrak{F}^\pm
\end{array}\)&=&
-a_s(\mu) \(
\begin{array}{cc}
\mathbb{H}_{qq}     &  \mathbb{H}_{qg} \\
\mathbb{H}_{gq}     &  \mathbb{H}_{gg}
\end{array}\)
\otimes
\(\begin{array}{c}
\mathfrak{S}^\pm_{\text{S}}\\
\mathfrak{F}^\pm
\end{array}\),
\end{eqnarray}
where $a_s(\mu)=g^2/(4\pi)^2$, $\otimes$ denotes an integral convolution, and, for brevity, we have omitted the argument $(x_1,x_2,x_3;\mu)$ of distributions.  The functions $\mathfrak{S}^\pm_{\text{S}}$ and $\mathfrak{S}^\pm_{\text{NS}}$ are, respectively, the flavor-singlet and flavor-non-singlet combinations of quark distributions, discussed below. The kernels $\mathbb{H}$ have a complicated form, and their explicit expressions are summarized in ref.~\cite{Rodini:2024usc}.

The integral convolution between kernels $\mathbb{H}$ and the distributions relates a point $x_{123}$ to a line integral within the domain (\ref{def:domain}). Different kernels are defined along different lines, including those that do not contain the point $x_{123}$. To describe this structure, it is convenient to introduce a single variable that describes the distance of a point on the hexagon from the origin.
\begin{eqnarray}
\|x\|=\max(|x_1|,|x_2|,|x_3|).
\end{eqnarray}
For example, in fig.~\ref{fig:Hex_Interp}, the dashed hexagon has $\|x\|=x$. 

All convolution integrals posses a feature that they involve only points with $\|y\|\geqslant \|x\|$, where $y$ is the integration variable. Therefore, the solution of the evolution equation at a point $x_{123}$ incorporates the distribution in the region between $\|x\|$ and the boundary. It is analogous to the DGLAP evolution equation, where the evolution of a PDF at $x$ incorporates the distributions in the range $[x,1]$. In that sense the variable $\|x\|$ serves as an ordinary kinematic scale for the collinear distribution. The "small-x" regime of twist-three PDFs corresponds to $\|x\|\to 0$, and the "large-x" regime corresponds to $\|x\|\to1$, while individual components of $x_{123}$ can take any values in $[-1,1]$.

In contrast to twist-two PDFs, the twist-three distributions do not have "quark" and "anti-quark" components, but instead are defined for a quark of a given flavor and smoothly pass from one interpretation region to another (see section \ref{sec:interpret}). Therefore, in equations (\ref{def:ev-ns-singlet}, \ref{def:ev-singlet}), the singlet and non-singlet contributions are defined as follows
\begin{eqnarray}
\mathfrak{S}^\pm_{\text{S}}&=&\sum_{f}\mathfrak{S}^\pm_{f},
\\
\mathfrak{S}^\pm_{\text{NS}_1}&=&\mathfrak{S}^\pm_{u}-\mathfrak{S}^\pm_{d},
\\
\mathfrak{S}^\pm_{\text{NS}_2}&=&\mathfrak{S}^\pm_{u}+\mathfrak{S}^\pm_{d}-2\mathfrak{S}^\pm_{s},
\\
\mathfrak{S}^\pm_{\text{NS}_3}&=&\mathfrak{S}^\pm_{u}+\mathfrak{S}^\pm_{d}+\mathfrak{S}^\pm_{s}-3\mathfrak{S}^\pm_{c},
\\
\mathfrak{S}^\pm_{\text{NS}_4}&=&\mathfrak{S}^\pm_{u}+\mathfrak{S}^\pm_{d}+\mathfrak{S}^\pm_{s}+\mathfrak{S}^\pm_{c}-4\mathfrak{S}^\pm_{b},
\end{eqnarray}
where the summation goes over all active flavors. 

The number of active flavors depends on the scale. In the present implementation, we use the simplest Zero Mass Variable Flavor Number Scheme \cite{Collins:1986mp}, which consists in neglecting parton distributions below a certain threshold and treating quarks as massless. We consider two threshold scales
\begin{eqnarray}\label{def:mu-scales}
\mu_c=m_c=1.40\text{GeV},\qquad
\mu_b=m_b=4.75\text{GeV}.
\end{eqnarray}
So, for various regions of $\mu$ we consider
\begin{eqnarray}\nn
\mu<\mu_c &:& N_f=3,\qquad \text{included flavors =}\{g,u,d,s\},
\\
\mu_c<\mu<\mu_b &:& N_f=4,\qquad \text{included flavors =}\{g,u,d,s,c\},
\\\nn 
\mu_b<\mu &:& N_f=5,\qquad \text{included flavors =}\{g,u,d,s,c,b\}.
\end{eqnarray}
The solution of the evolution equation is always performed from a smaller $\mu_{i}$ to a larger $\mu_{f}$, which allows one to continuously include all flavors. The initial distributions for $c$ and $b$ flavors are zero, and they are generated solely through the evolution procedure. These values (\ref{def:mu-scales}) are taken equal to the corresponding values in the MSHT20 extraction of $a_s$ and unpolarized PDFs \cite{Bailey:2020ooq}, which are used as input functions for all procedures that require $a_s(\mu)$ or unpolarized PDFs (see sec.\ref{sec:TMD}). Note that the same scheme is implemented in the TMD-related part.

The evolution equations for genuine twist-three PDFs impose a certain analytical structure on their solution. In particular, a part of the kernel $\mathbb{H}_{qq}$\footnote{\label{foot:singular} More specifically, the kernel $\mathcal{H}_{13}^d$ is proportional to $x_1x_3/x_2^3$. This singularity can be seen explicitly in eqn.~(103) in \cite{Rodini:2024usc}, or eqn.~(A.16) in \cite{Braun:2009mi}, or eqn.~(240) in \cite{Ji:2014eta}. This kernel is unique and represents the contribution of the gluon equation of motion.} is singular at $\|x\|\to0$, irrespectively of the behavior of $\mathfrak{S}$ at this point. Therefore, the solution always has singular behavior at $\|x\|\to 0$, at least along the $x_1=x_3$ direction. For the moment, it is not clear whether there is some compensating mechanism for this feature within the factorization theorems, but in the present LO study this feature is present and plays an important role in the properties of twist-three observables. 

The numerical solution of the twist-three evolution equation is realized in the \texttt{snowflake}-code, which, for purposes of global fit, is merged with and distributed together with the \texttt{artemide}-code \cite{artemide}. We refer to sec.~\ref{sec:implementation}, as well as, to the original publication \cite{Rodini:2024usc} for further details on the evolution procedure and its numerical implementation.

\section{DIS Observables} 
\label{sec:DIS}

In this section we provide the details related to the structure function $\overline{g}_2(x)$ and its moment $d_2$. In the first part we provide theoretical expressions for these functions in terms of genuine twist-three distributions, and related sum rules. In the last subsection, we make a review of available data.

\subsection{$g_{2}$ structure function}
\label{sec:g2}

The structure function $g_2$ is measured in the polarized DIS alongside the helicity structure function $g_1$. It has been studied theoretically in many works, for review,  see \cite{Jaffe:1989xx, Jaffe:1996zw, Braun:2022gzl} and references therein. Traditionally, it is considered the main source of information about twist-three distributions, because it was the first object in which the contribution of twist-three distributions was studied. 

The structure function $g_2$ can be decomposed into two distinct terms
\begin{eqnarray}
g_2(x,Q)=g_2^{\text{WW}}(x,Q)+\overline{g}_2(x,Q),
\end{eqnarray}
where $Q$ is the DIS photon virtuality, and $x$ is the Björken variable. These terms are entirely independent and are made of twist-two PDFs ($g_2^{\text{WW}}$) and genuine twist-three PDFs ($\overline{g}_2$).

The term $g_2^{\text{WW}}$ is the, so-called, Wandzura-Wilczek (WW) contribution \cite{Wandzura:1977qf}, which is expressed exactly via the DIS structure function $g_1(x)$ as
\begin{eqnarray}\label{g2:WW}
g^{\text{WW}}_2(x,Q)=-g_1(x,Q)+\int_x^1 \frac{dy}{y}g_1(y,Q).
\end{eqnarray}
In turn, the structure function $g_1$ is given via the twist-two helicity PDF $\Delta f$ 
\begin{eqnarray}
g_1(x,Q)=\frac{x}{2}\sum_f e_f^2 \int_x^1 \frac{dy}{y} \[
\Delta C_{qq}\(\frac{x}{y},Q\)\Delta f_q(y,Q)
+
\Delta C_{qg}\(\frac{x}{y},Q\)\Delta f_g(y,Q)\],
\end{eqnarray}
where  $e_f$ is the electric charge of a quark with flavor $f$ and $\Delta C$ is the perturbative coefficient function \cite{Zijlstra:1993sh}. The WW contribution is sizable numerically, and produces the dominant part of the $g_2$ structure function. For the present work, the WW contribution represents a contamination of the twist-three signal.  We computed this term with the help of the APFEL package \cite{Bertone:2017gds}, using the MAPPDFpol1.0 extraction \cite{Bertone:2024taw}.

The term $\overline{g}_2$ is the term composed only from genuine twist-three distributions. It is convenient to present it in the following form \cite{Jaffe:1989xx, Ji:1990br, Ali:1991em, Kodaira:1994ge, Braun:2011aw}
\begin{eqnarray}\label{def:g2}
\overline{g}_2(x,Q)&=&\sum_f \frac{e_f^2}{2} \int_x^1 \frac{dy}{y}\(\Delta q_{T,f}(y,Q^2)+\Delta q_{T,f}(-y,Q^2)\),
\end{eqnarray}
where the function $\Delta q_{T,f}$ at LO reads  
\begin{eqnarray}\label{def:g2-tw3}
\Delta q_{T,f}(x,Q^2)=\int [dy]\(T_f(y_{123},Q)+\Delta T_f(y_{123},Q)\) \frac{d}{dy_3}\frac{\delta(x+y_1)-\delta(x-y_3)}{y_1+y_3}.
\end{eqnarray}
Here, the integration over $y$ is restricted by the domain (\ref{def:domain}) and the delta-functions. 

The expression (\ref{def:g2-tw3}), although being rather compact, is complicated to implement numerically. To make it suitable for the numerical computation we integrate it by parts (using that distributions vanish at the boundary) and, after algebraic simplifications, obtain the following expression
\begin{eqnarray}\label{expession:g2}
\overline{g}_{2}(x,Q)&=&\frac{1}{2}\sum_f e^2_f \int_{-1}^{1} dy_1 dy_3
\(\frac{\partial}{\partial y_3}\mathfrak{S}^+(y_1,-y_1-y_3,y_3;Q)\)
\\\nn &&\times 
\Big[
\frac{\theta(y_1>x,y_3<-x)}{y_1y_3}
+\frac{\theta(y_1<x,y_3<-x)}{y_3(y_1+y_3)}
+\frac{\theta(y_1>x,y_3>-x)}{y_1(y_1+y_3)}
\Big].
\end{eqnarray}
In contrast to (\ref{def:g2}), this expression looks asymmetric with respect to the quark/anti-quark regions. This is due to the symmetry of $\mathfrak{S}^\pm$ and the integral kernel, which allows one to unify both sectors, and eliminate the $\mathfrak{S}^-$ contribution. Note that the first term in the square brackets is regular, while the last two terms are singular along the line $y_1=-y_3=x$ ($y_2=0$), but cancel each other when added together.

The terms $\Delta q_T(y)$ and $\Delta q_T(-y)$ represent the quark and anti-quark contribution to $\overline{g}_2$ respectively. However, as it is discussed in sec.~\ref{sec:interpret}, in terms of genuine distributions, there is no clear separation between quark and anti-quark. Still, these terms have different domains of integrations. The term $\Delta q_T(y)$ is integrated over the red-shaded area in fig.\ref{fig:Hex_Interp}, while the term $\Delta q_T(-y)$ over the green-shaded area. Note that both functions receive contributions from the "quark-anti-quark" sector, and for $x<0.5$ some portion of this area contributes to both of the terms at the same time. In what follows, we also discuss particular flavor contributions to $\overline{g}_2$, which we define as
\begin{eqnarray}\label{def:g2f}
\overline{g}_{2}(x,Q)=\sum_f e^2_f \overline{g}_{2f}(x,Q),
\end{eqnarray}
i.e., as the coefficient in front of $e_f^2$ in eqn.~(\ref{expession:g2}).

The numerical implementation of equation (\ref{expession:g2}) is straightforward in the \texttt{snowflake} code, which is designed to evaluate such kinds of integrals over the 2D Chebyshev interpolation function. The only difference with other integrals, already implemented in \texttt{snowflake}, is the derivative of $\mathfrak{S}^+$. The simplest way to evaluate it is to apply the derivative matrix to $\mathfrak{S}^+$. However, we found that this method is not very stable --  due to the need to cancel each singular contribution at the $y_2=0$ line exactly, which becomes increasingly noisy in the small-$\|x\|$ regime. Instead, we directly perform the derivative of the interpolation function, which can be done analytically. The code has been tested against an independent realization in the \texttt{honeycomb} library by S.Rodini, and is in agreement with it.

\subsection{$d_2$ moment}
\label{sec:d2}

The $d_2$ moment is defined as \cite{Jaffe:1990qh}
\begin{eqnarray}\label{d2:via_g2}
d_{2f}(Q)=3\int_0^1 dx\,x^2\overline{g}_{2f}(x,Q),
\end{eqnarray}
where $\overline{g}_{2f}$ is the contribution to $\overline{g}_{2}$ of a quark and anti-quark of flavor $f$. We use this definition due to its common employment by experimental groups. Correspondingly, a hadron's $d_2$ moment is defined as
\begin{eqnarray}\label{d2=d2f}
d_2(Q)=\sum_f e^2_f d_{2f}(Q),
\end{eqnarray}
in complete analogy with eqn.~(\ref{def:g2f}).

On the other hand, $d_2$ represents the matrix element of a local operator of twist-three:  
\begin{eqnarray}\label{d2:operator}
\langle p,s|g \bar q(0)F^{\mu+}(0)\gamma^+ q(0)|p,s\rangle=4\epsilon^{\mu\nu}s_\nu p_+^2 M d_{2f}.
\end{eqnarray}
Comparing (\ref{d2:operator}) with the definition (\ref{def:T}), we find that
\begin{eqnarray}\label{def:d2}
d_{2f}(Q)&=&\frac{1}{2}\int [dy]T_f(y_{123},Q)=\frac{1}{4}\int [dy]\mathfrak{S}_f^+(y_{123},Q).
\end{eqnarray}
In this way, the moment $d_2$ represents the integral over the whole hexagonal domain, and provides the normalization of the twist-three PDF $T$ or $\mathfrak{S}^+$.

We have noted that there are several conventions for $d_2$ in the literature which are related to each other by a factor of $2$ in definition (\ref{d2:operator}) and/or of individual-flavor components (\ref{d2=d2f}) (see also discussion in ref.~\cite{Burger:2021knd}). We have checked that our set of relations is consistent and agrees with the definitions of experimental groups, whose data we use. Furthermore, we have checked that the computation of integrals (\ref{def:d2}) and (\ref{expession:g2}, \ref{d2:via_g2}) produces the same result, which provides an additional check of the numerical implementation. 

\subsection{Burkhardt-Cottingham sum rule}
\label{sec:sum-rules}

The Burkhardt-Cottingham sum rule \cite{Burkhardt:1970ti} states
\begin{eqnarray}\label{BC-sumrule}
\int_0^1 dx\overline{g}_2(x,Q) = 0.
\end{eqnarray}
Substituting (\ref{expession:g2}) into the integral (\ref{BC-sumrule}), one obtains
\begin{eqnarray}\label{BC-sumrule1}
\int_{-1}^1 dy_1dy_3
\(\frac{\partial}{\partial y_3}\mathfrak{S}^+(y_1,-y_1-y_3,y_3;Q)\)\frac{\theta(y_{1,3}>0)-\theta(y_{1,3}<0)}{y_1+y_3}=0.
\end{eqnarray}
Changing $y_1\to -y_2-y_3$, and integrating by parts, we obtain that this is equivalent to
\begin{eqnarray}
\int_{-1}^1 \frac{dy}{y}\mathfrak{S}^+(0,y,-y;Q)=0.
\end{eqnarray}
Therefore, the sum rule is automatically satisfied due to the symmetry of $\mathfrak{S}^+$ (\ref{sym:2}), if the integral is understood in the sense of the principal value.

There is, however, a very important detail. The Burkhardt-Cottingham sum rule \cite{Burkhardt:1970ti} is derived under the assumption that $\lim_{x\to0}xg_2(x)=0$. This assumption could be violated, in which case the integral (\ref{BC-sumrule}) is not well defined. To interpret it on the level of twist-three distributions, let us introduce the truncated sum rule
\begin{eqnarray}
\text{BC}_{x_0}(Q)=\int_{x_0}^1 dx\overline{g}_2(x,Q).
\end{eqnarray}
Evaluating the integral over (\ref{expession:g2}) we obtain
\begin{eqnarray}\label{BC-sumrule-cut}
\text{BC}_{x_0}(Q)&=&-x_0\overline{g}_2(x_0,Q)+\text{BC}^{(0)}_{x_0}(Q),
\end{eqnarray}
where
\begin{eqnarray}
\text{BC}^{(0)}_{x_0}(Q)&=&
\int_{-1}^1 dy_1 dy_3
\(\frac{\partial}{\partial y_3}\mathfrak{S}^+(y_1,-y_1-y_3,y_3;Q)\)
\\\nn &&\qquad\qquad\times 
\frac{\theta(y_1>x_0,y_3>-x_0)-\theta(y_1<x_0,y_3<-x_0)}{y_1+y_3}.
\end{eqnarray}
The term $\text{BC}^{(0)}_{x_0}$ is well defined, due to the symmetry of $\mathfrak{S}^+$. In the limit $x_0\to0$ it reduces to the expression  (\ref{BC-sumrule1}), and vanishes. Meanwhile, the term $-x_0\overline{g}_2(x_0)$ is not presented in (\ref{BC-sumrule1}) since setting $x_0=0$ nullifies it. However, in the limit $x_0\to0$ this term is essentially problematic due to the integral's singular behavior at the center of the hexagon. Using (\ref{expession:g2}), one can deduce that the most singular part is
\begin{eqnarray}
x_0\overline{g}_2(x_0,Q)&\sim &x_0\int_{x_0}^1 dy_1 \int_{-1}^{-x_0} dy_3
\frac{\partial}{\partial y_3}\mathfrak{S}^+(y_1,-y_1-y_3,y_3;Q)\frac{1}{y_1y_3}+\cdots~,
\end{eqnarray}
where the dots denote less singular contributions. The limit strongly depends on the behavior of $\mathfrak{S}^+$ at $\|x\|\to 0$. It must possess a zero of at least second order $\|x\|^{2+\epsilon}$ to satisfy the Burkhardt-Cottingham sum rule.

Therefore, the Burkhardt-Cottingham sum rule is the statement about the behavior of the genuine twist-three distributions at $\|x\|\to0$. If $\mathfrak{S}^+$ vanishes at $\|x\|\to0$ as $\|x\|^{2+\epsilon}$, the Burkhardt-Cottingham sum rule is exact. In the opposite case, the integral (\ref{BC-sumrule}) diverges. 

Meanwhile, the evolution equation of genuine twist-three PDFs induces the singularity\footref{foot:singular}  at $\|x\|\to0$ in sectors $(x_1>0,x_3>0)$ and $(x_1<0,x_3<0)$. In the present LO implementation, this effect does not have any compensation (if such mechanism exists at all). Therefore, the Burkhardt-Cottingham sum rule is violated by the effects of twist-three evolution equation. 

\subsection{Efremov-Leader-Teryaev sum rule}

The Efremov-Leader-Teryaev sum rule \cite{Anselmino:1994gn, Efremov:1996hd} states
\begin{equation}
\int_0^1 dx\,x\overline{g}^V_2(x,Q)=0,
\end{equation}
where the super-script $V$ indicates the valence quark contribution. The valence component of $\overline{g}_2$ should be interpreted via the expression (\ref{def:g2}), interpreting $\Delta q_{T}(y)$ as the quark term, and $\Delta q_{T}(-y)$ as the anti-quark term. Note that this interpretation does not hold at the genuine twist-three level, since both $\Delta q_{T}(\pm y)$ incorporate ``quark'' and ``anti-quark'' sectors of the hexagon domain.

Considering the valence part of the structure function $\overline{g}_{2f}$ in the sense of \cite{Efremov:1996hd}, we define
\begin{eqnarray}
\overline{g}^V_{2,f}(x,Q)&=&\frac{1}{2}\int_x^1 \frac{dy}{y}\(\Delta q_{Tf}(y,Q^2)-\Delta q_{Tf}(-y,Q^2)\).
\end{eqnarray}
The minus sign between the terms leads to a different kind of internal symmetry of the integrand, which reduces genuine twist-three PDFs $\mathfrak{S}^-$, instead of ordinary $\mathfrak{S}^+$. Namely, we found
\begin{eqnarray}\label{ELT-sr2}
\int_0^1 dx\,x\overline{g}^V_2(x,Q)=\int_{-1}^1 dy_1dy_3 \frac{\partial}{\partial y_3}\mathfrak{S}_f^-(y_1,-y_1-y_3,y_3)=0.
\end{eqnarray}
This expression is always valid, because $\mathfrak{S}^-$ vanishes at the boundary. Consequently, we conclude that the Efremov-Leader-Teryaev sum rule is a trivial statement (\ref{ELT-sr2}), which is always fulfilled.

It is interesting to observe that, if one ignores the ``valence'' label and utilizes expression (\ref{expession:g2}) directly, the sum rule turns into
\begin{eqnarray}\label{ELT-sr1}
\int_0^1 dx\,x\overline{g}_2(x,Q)=\sum_{f}\frac{e^{2}_{f}}{2}\int_{0}^{1} dy_{1} dy_3 \frac{y_{1}}{(y_{1}+y_{3})^2}\mathfrak{S}^{+}_{f}(y_{1},-y_{1}-y_{3},y_{3}).
\end{eqnarray}
This integral is non-zero for a generic function $\mathfrak{S}^+$.

\subsection{Review of experimental data} 
\label{sec:DIS-data}

\begin{table}[t]
\centering
\begin{tabular}{||l|c|c|c|p{4.5cm}||}
\hline
\multicolumn{5}{||c||}{$g_2$ structure function}
\\\Xhline{5\arrayrulewidth}
Data set & Ref. & Kinematic range & $N_{\text{pt}}$ & Comment
\\\Xhline{5\arrayrulewidth}
E142 & \cite{E142:1996thl} & 
\makecell[c]{$1.1<Q^{2}<5.5$GeV$^2$ \\ $0.03<x<0.6$} &
 &
 \makecell[l]{
Neutron (He$^{3}$) target. \\  Excluded from the fit due \\ to large uncertainties.}
\\\hline
E143(1996) & \cite{E143:1995xmc} & 
\makecell[c]{$1.49<Q^2<8.85$GeV$^2$ \\$0.029<x<0.799$} &
&
\makecell[l]{
Proton, deuteron targets. \\
Superseded in ref.~\cite{E143:1998hbs}.}
\\\hline
E143(1998) & \cite{E143:1998hbs} & 
\makecell[c]{$1.49<Q^2<8.85$GeV$^2$ \\ $0.029<x<0.799$ } &
22 &
\makecell[l]{
Proton, deuteron targets. \\ Data for $g_2^n$ are derived \\ from $g_2^p$ and $g_2^d$, and thus \\ are excluded from the fit.}
\\\hline
E154 & \cite{E154:1997eyc} & 
\makecell[c]{$1.2<Q^2<15.$GeV$^2$ \\ $0.014<x<0.7$} &
15 &
Neutron target.
\\\hline
E155(1999) & \cite{E155:1999eug} & 
\makecell[c]{$1.15<Q^2<29.25$GeV$^2$ \\ $0.022<x<0.839$ }&
36 &
\makecell[l]{Proton, deuteron targets.  \\ 
$E=38.8$GeV.}
\\\hline
E155(2003) & \cite{E155:2002iec} & 
\makecell[c]{$0.8<Q^2<18.40$GeV$^2$ \\ $0.021<x<0.78$} &
56 &
\makecell[l]{Proton, deuteron targets.  \\ 
$E=21.9\text{GeV} \& 31.3 \text{GeV}.$}
\\\hline
JLab(2004)& \cite{JeffersonLabHallA:2004tea}  & 
\makecell[c]{ $2.71<\langle Q^2\rangle<4.83$GeV$^2$ \\ $0.33<x<0.60$} & 
 &
\makecell[l]{
Neutron (He$^3$) target. \\
Superseded in ref.~\cite{JeffersonLabHallA:2016neg}. }
\\\hline
JLab(2016)& \cite{JeffersonLabHallA:2016neg}  & 
\makecell[c]{
$2<Q^2<6$GeV$^2$  \\ $0.25<x<0.9$} &
26 &
\makecell[l]{Neutron (He$^3$) target.}
\\\hline
HERMES& \cite{HERMES:2011xgd}  & 
\makecell[c]{
$0.38<\langle Q^2\rangle<10.35$GeV$^2$ \\ $0.009<x<0.68$} & 
13 &
\makecell[l]{Proton target}
\\\hline
SMC & \cite{SpinMuonSMC:1997mkb} &
\makecell[c]{$1.36<Q^{2}<17.07$GeV$^2$ \\ $0.006<x<0.6$} &  & 
\makecell[l]{Proton target.
\\ Excluded from the fit due \\ to large uncertainties.
}
\\\Xhline{5\arrayrulewidth}
\multicolumn{3}{||r|}{\textbf{Total:} } &
168 &\\
\hline
\end{tabular}
\caption{\label{tab:g2-data} Review of available high-energy data for the $g_2$ structure function. The column $N_{\text{pt}}$ indicates the number of data points that survive the kinematic restrictions and are included in the fit. The reason for dataset exclusion, as well as the type of target, is presented in the last column.}
\end{table}

The $g_2$ structure function has been measured in a variety of experiments. The synopsis of data suitable for analysis is given in table \ref{tab:g2-data}. This table does not include data measured at very low values of $Q$, such as \cite{Kramer:2005qe}, because they cannot be described within the factorization theorem approach. Moreover, there are some datasets presented in table \ref{tab:g2-data} which are not included in the fit because their data is either superseded by a later measurement, or has too large uncertainty to have a restrictive power. These reasons are pointed out in the right column of table  \ref{tab:g2-data}. Still, we present the comparison with excluded data set in the corresponding sections.

There are two principal sources of measurements for the $d_2$ moment. The synopsis of suitable data is given in table \ref{tab:d2-data}.

The first source is the integral of $\overline{g}_2$ (\ref{d2:via_g2}), which is determined from integrating the events in DIS measurements. Consequently, $d_2$ has better statistical precision in comparison to $g_2$, and some experiments report only $d_2$ without reporting $g_2$ data. It is important to mention that in practice, the integral of $g_2$ is taken over a limited range of $x$. In some cases, it is included into the systematic uncertainty, while in other cases, the group presents the cut integral. In the latter, we fit the data using the complete expression for $d_2$. In sec.~\ref{sec:d2+force}, we discuss the dependence on the $x$-cut.

The second source of information about $d_2$ are lattice simulations which measure the local operator (\ref{d2:operator}) directly for a particular flavor. Unfortunately, many lattice measurements are incomplete \cite{Gockeler:1995wg, Gockeler:2000ja, Crawford:2024wzx, Gao:2026wlz}, in the sense that they do not provide a complete analysis of systematic uncertainties. This makes them unsuitable for the analysis of a delicate observable such as $d_2$. Nonetheless, we present the comparison with some of them in their corresponding plots.

To justify the application of the factorization theorem, one must set a constraint on the value of $Q$. We employ the constraint
\begin{eqnarray}\label{data:Q2>2}
Q^2>2\text{GeV}^2.
\end{eqnarray}
On the one hand, this constraint is not very restrictive, and one could expect sizable contribution of power corrections to the factorization theorem. However, most of the data has large uncertainties that allow to accommodate a contamination due to power corrections. On the other hand, the majority of included data (119 out of 168 points) have the much safer scale constraint $Q^2>4$GeV$^2$. Nonetheless, it is instructive to mention that the restriction (\ref{data:Q2>2}) is more conservative than those used in other fits of the $g_{2}$ structure function (see for instance ref.\cite{Sato:2016tuz}).

At low energies one should expect a sizable contribution of the target mass corrections \cite{Blumlein:1998nv, Dong:2007iv} and higher-twist contamination effect. There are indications that these corrections could be as large as $25\%$ for $d_2$ at $Q^2= 2$GeV$^2$ \cite{Dong:2007iv}, but rapidly drop with the increase of the scale. In this work, we do not include any effects of this type for the sake of simplicity and better control over the details of this first study. However, these corrections should be included in the following fits.

After applying the kinematic restriction (\ref{data:Q2>2}), we obtained 168 points for $g_2$ and 13 points for $d_2$. This extends the number of points found in \cite{Vladimirov:2025qrh} due to the inclusion of additional datasets to the ones used in the earlier work. The distribution of the points in the plane ($x$, $Q$) is presented in fig.~\ref{fig:all-datapoints}.

\begin{table}[t]
\centering
\begin{tabular}{||l|c|c|c|p{5.5cm}||}
\hline
\multicolumn{5}{||c||}{$d_2$ moment}
\\\Xhline{5\arrayrulewidth}
Dataset & Ref. & Kinematic range & $N_{\text{pt}}$ & Comment
\\\Xhline{5\arrayrulewidth}
\multicolumn{5}{|c|}{Collider measurements}
\\\hline
E143 (1998) & \cite{E143:1998hbs} & 
$Q^2=5$GeV$^2$ &
2 &
\makecell[l]{
Proton, deuteron targets. \\ $d_2^n$ is derived \\ from $g_2^p$ and $g_2^d$, and thus, is  \\ excluded from the fit.}
\\\hline
E154 & \cite{E154:1997eyc} & 
$Q^2=3.6$GeV$^2$ &
 &
 \makecell[l]{
Neutron target.
\\
Excluded from the fit due\\ to large uncertainties.}
\\\hline
E155 (1999) & \cite{E155:1999eug} & 
$Q^2=5$GeV$^2$ &
2 &
\makecell[l]{
Proton, deuteron targets.
\\
$E=38.8$GeV}
\\\hline
E155 (2003) & \cite{E155:2002iec} & 
$Q^2=5$GeV$^2$ &
2 &
\makecell[l]{
Proton, deuteron targets.
\\
Combined from  \\ $E=21.9$GeV \& $31.3$GeV  \\ measurements.}
\\\hline
JLab(2004)& \cite{JeffersonLabHallA:2004tea}  & 
$Q^2=5$GeV$^2$ &
&
\makecell[l]{
Neutron (He$^3$) target. 
\\
Superseded in ref.~\cite{JeffersonLabHallA:2016neg}. }
\\\hline
JLab(2016)& \cite{JeffersonLabHallA:2014gzr, JeffersonLabHallA:2016neg}  & 
\makecell[c]{
$ Q^2=3.21$GeV$^2$
\\
$Q^2=4.32$GeV$^2$
}&
2 &
\makecell[l]{
Neutron (He$^3$) target. 
\\
The value from Ref.\cite{JeffersonLabHallA:2016neg} is taken
\\ since it provides a more detailed  \\ estimation of uncertainty. }
\\\hline
HERMES& \cite{HERMES:2011xgd}  & 
$\langle Q^2\rangle=5$GeV$^2$ & 
1 &
Proton target.
\\\hline
SANE & \cite{SANE:2018pwx}  & 
\makecell[c]{
$\langle Q^2\rangle=2.8$GeV$^2$
\\
$\langle Q^2\rangle=4.3$GeV$^2$
} &
2 &
\makecell[l]{
Proton target.
\\
The matrix element $d_2$ is defined \\ with opposite sign.
}
\\\hline
RSS & \cite{RSS:2006tbm, ResonanceSpinStructure:2008ceg} & 
\makecell[c]{
$\langle Q^2\rangle=1.28$GeV$^2$
\\
$\langle Q^2\rangle=1.3$GeV$^2$}
& 
&
\makecell[l]{Proton target. \\ Excluded because of \\ $Q^2$-cut.}
\\\Xhline{5\arrayrulewidth}
\multicolumn{5}{|c|}{Lattice measurements}
\\\hline
RQCD & \cite{Burger:2021knd} &
$\mu=2$GeV & 
2 & 
\makecell[l]{
$u$ and $d$ quarks. 
\\ Definition is $d_2^{[RQCD]}=2d_2$}
\\\Xhline{5\arrayrulewidth}
\multicolumn{3}{||r|}{\textbf{Total:} } &
13 &\\
\hline
\end{tabular}
\caption{\label{tab:d2-data} Review of available high-energy data for the $d_2$ moment. The column $N_{\text{pt}}$ indicates the number of data points that survive the kinematic restrictions and are included in the fit. The reason for data-set exclusion, as well as, other comments are presented in the last column.}
\end{table}

\section{TMD Observables} 
\label{sec:TMD}

In this section we provide the details related to the Sivers and worm-gear-T functions, which are accessible through the structure functions of Semi-Inclusive DIS (SIDIS) and the polarized Drell-Yan reaction. In the first part of this section, we provide the theoretical expressions for structure functions within the TMD factorization theorem, and the relation between TMDPDFs and genuine twist-three PDFs. In the last subsection, we make a review of available data. Note that in many aspects, the TMD approach of the present analysis repeats our earlier analysis of unpolarized distributions ART25 and ART23 \cite{Moos:2023yfa, Moos:2025sal}, to which we refer for additional details.

\subsection{Spin-asymmetries in TMD factorization (SIDIS)}

The SIDIS reaction is
$$
\ell(l)+h_1(P,S)\to \ell(l')+h_2(p_h)+X,
$$
where in brackets we specify the momentum of each particle, and $S$ is the spin-vector of the target hadron. The differential SIDIS cross-section has a rich angular structure which is encoded in the SIDIS structure functions \cite{Diehl:2005pc, Bacchetta:2006tn}. For the present discussion only three of them are needed:
\begin{eqnarray}
\frac{d\sigma}{dxdydz\,d\phi_Sd\phi_h\,dp_{h\perp}^2}&=&\frac{\alpha_{\text{em}}^2}{Q^2}\frac{y}{2(1-\varepsilon)}
\Big[F_{UU,T}+
|S_\perp|\sin(\phi_h-\phi_S)F_{UT,T}^{\sin(\phi_h-\phi_S)}
\\\nn &&
+
|S_\perp|\lambda_\ell \sqrt{1-\varepsilon^2}\cos(\phi_h-\phi_S)F_{LT}^{\cos(\phi_h-\phi_S)}+ \cdots\Big],
\end{eqnarray}
where $\lambda_\ell$ is the helicity of the lepton, and
\begin{eqnarray}
q^2=-Q^2,\quad x=\frac{Q^2}{2(P\cdot q)},\quad y=\frac{(P\cdot q)}{(P\cdot l)},\quad z=\frac{(P\cdot P_h)}{(P\cdot q)},\quad \varepsilon=\frac{1-y}{1-y+\frac{y^2}{4}},
\end{eqnarray}
with $q=l-l'$ being the momentum of the virtual photon. The variables $p_{h\perp}$ and $S_\perp$ are the transverse components of momentum $p_h$ and the target-hadron's spin vector $S$ with respect to the plane $(q,P)$. The angles $\phi_h$ and $\phi_S$ are the azimuthal angles of the $p_{h\perp}$ and $S_\perp$ (see \cite{Bacchetta:2004jz}).

On the theory side, the SIDIS cross-section can be described within the TMD factorization approach, which is valid at small values of $p_{h\perp}$ \cite{Collins:2011zzd, Bacchetta:2006tn, Boer:2011xd, Rodini:2023plb}. In this approach, the structure function is described through convolutions of non-perturbative TMD distributions. For the structure functions under consideration the expressions are
\begin{eqnarray}
F_{UU,T}&=&\Big|C_V\(\frac{Q^2}{\mu^2}\)\Big|^2 \mathcal{B}_0[f_1D_1]+\mathcal{O}\(\frac{p_{h\perp}^2}{z^2Q^2}\),
\\
F_{UT}^{\sin(\phi_h-\phi_S)}&=&-M\Big|C_V\(\frac{Q^2}{\mu^2}\)\Big|^2\mathcal{B}_1[f_{1T}^\perp D_1]
+\mathcal{O}\(\frac{p_{h\perp}^2}{z^2Q^2}\),
\\
F_{LT}^{\cos(\phi_h-\phi_S)}&=&M\Big|C_V\(\frac{Q^2}{\mu^2}\)\Big|^2\mathcal{B}_1[g_{1T}^\perp D_1]+\mathcal{O}\(\frac{p_{h\perp}^2}{z^2Q^2}\),
\end{eqnarray}
where $C_V$ is the hard coefficient function, and $M$ is the mass of the target hadron. The integral convolution is defined as
\begin{eqnarray}\label{def:TMD-convolution}
\mathcal{B}_n[fD]=\sum_f e_f^2\int_0^\infty \frac{b\,db}{2\pi} b^n J_n\(\frac{b|p_\perp|}{z}\)f_{f}(x_S,b;\mu,Q^2)D_{f}(z,b;\mu,Q^2),
\end{eqnarray}
where $J_n$ is the Bessel function of the first kind of order $n$. The TMD distributions $f_1$, $D_1$, $f_{1T}^\perp$ and $g_{1T}^\perp$ are the unpolarized TMDPDF, unpolarized TMDFF, Sivers TMDPDF and worm-gear-T (Kotzinian-Mulders) TMDPDF, respectively. Here, we utilize the symmetric point for the rapidity factorization scale-parameter, $\zeta=\bar \zeta=Q^2$, which is the standard choice in TMD phenomenology.

The variable $x_S$ is the leading-power approximation for the parton momentum fraction\footnote{\label{foot:eulc} By bold font, we conventionally denote the scalar products of transverse components in negative sign, $\vec p_{h\perp}^2=-p^2_{h\perp}>0$, and $\vec q_T^2=-q_T^2>0$.}
\begin{eqnarray}
x_S=-\frac{q^+}{P^+}=x\(1-\frac{\vec p_\perp^2}{z^2Q^2}\).
\end{eqnarray}
It deviates from $x$ due to the transformation between the laboratory frame (where the SIDIS cross-section is defined) and the factorization frame. Although this difference is power-suppressed, we include it in our definition, as was done in ART25. 

The experiment does not present the structure functions $F_{UT,T}^{\sin(\phi_h-\phi_S)}$, and $F_{LT}^{\cos(\phi_h-\phi_S)}$ directly, but rather provides the values for the single and double spin-asymmetries. These asymmetries are related to structure functions through
\begin{eqnarray}\label{A1}
A_{UT,T}^{\sin(\phi_h-\phi_S)}=\frac{F_{UT,T}^{\sin(\phi_h-\phi_S)}}{F_{UU,T}},
\\\label{A2}
A_{LT}^{\cos(\phi_h-\phi_S)}=\frac{F_{LT}^{\cos(\phi_h-\phi_S)}}{F_{UU,T}}.
\end{eqnarray}
In the limit of the TMD factorization theorem these asymmetries are
\begin{eqnarray}\label{A3}
A_{UT,T}^{\sin(\phi_h-\phi_S)}=-M\frac{\mathcal{B}_1[f_{1T}^\perp D_1]}{\mathcal{B}_0[f_{1} D_1]}
+\mathcal{O}\(\frac{p_{h\perp}^2}{z^2Q^2}\),
\\\label{A4}
A_{LT}^{\cos(\phi_h-\phi_S)}=M\frac{\mathcal{B}_1[g_{1T}^\perp D_1]}{\mathcal{B}_0[f_{1} D_1]}
+\mathcal{O}\(\frac{p_{h\perp}^2}{z^2Q^2}\).
\end{eqnarray}
These expressions are used to compare with the data. 

Let us mention that the definitions of the asymmetries (\ref{A1}, \ref{A2}) are incomplete, since they should include the structure functions $F_{UU,L}$ and $F_{UT,L}^{\sin(\phi_h-\phi_S)}$. The latter represent the contribution of a longitudinally polarized photon, and have the same angle-dependence as $F_{UU,T}$ and $F_{UT,T}^{\sin(\phi_h-\phi_S)}$, but accompanied by a factor $\varepsilon$. These structure functions are power suppressed in the TMD factorization, and thus expressions (\ref{A3}, \ref{A4}) remain valid. Still, one should keep in mind that the contribution of these terms can be sizable at lower energies, see studies in refs.~\cite{Piloneta:2025jjb, Balitsky:2026nux}.

The unpolarized TMD distributions have been studied in detail both theoretically and phenomenologically, and have been extracted in many studies. The most significant extractions are \cite{Scimemi:2017etj, Bacchetta:2017gcc, Bacchetta:2019sam, Bertone:2019nxa, Scimemi:2019cmh, Bacchetta:2022awv, Moos:2023yfa, Moos:2025sal, Bacchetta:2025ara, Barry:2025glq}. In this study we use unpolarized TMD distributions from the ART25 extraction \cite{Moos:2025sal}. The uncertainty of the ART25 unpolarized extraction is taken into account, as described in sec.~\ref{sec:def-uncert}. We have found that the uncertainty in the extraction due to the unpolarized parts is much smaller in comparison to others.

It should be noted that, although differences between the values of TMD distributions have been reduced in the most recent studies, they are still present (see the comparison in ref.~\cite{Moos:2025sal}). It has been found that the unpolarized TMDPDFs mostly agree between different groups, while unpolarized TMDFFs differ substantially. This difference is most probably generated by differences in the implementation of the SIDIS factorization theorem between different groups. Since the data are described almost equally well in each case, it is expected that the differences in the extracted TMDFFs are compensated by corresponding differences in the implementations, such that the final result is alike. In this work, we follow the same procedural rules and the same implementation of TMD factorization as in ART25, to mitigate possible tensions in the implementation. Furthermore, it is known that the influence of TMDFFs on the prediction is rather small, because TMDFFs enter equally into the numerator and denominator of the asymmetries (\ref{A3}, \ref{A4}). Summarizing these observations, we expect that the hidden systematic uncertainty due to differences in the unpolarized TMD distributions is small, and is covered by the included uncertainty of ART25.

\begin{figure}[t]
\centering
\includegraphics[width=0.65\linewidth]{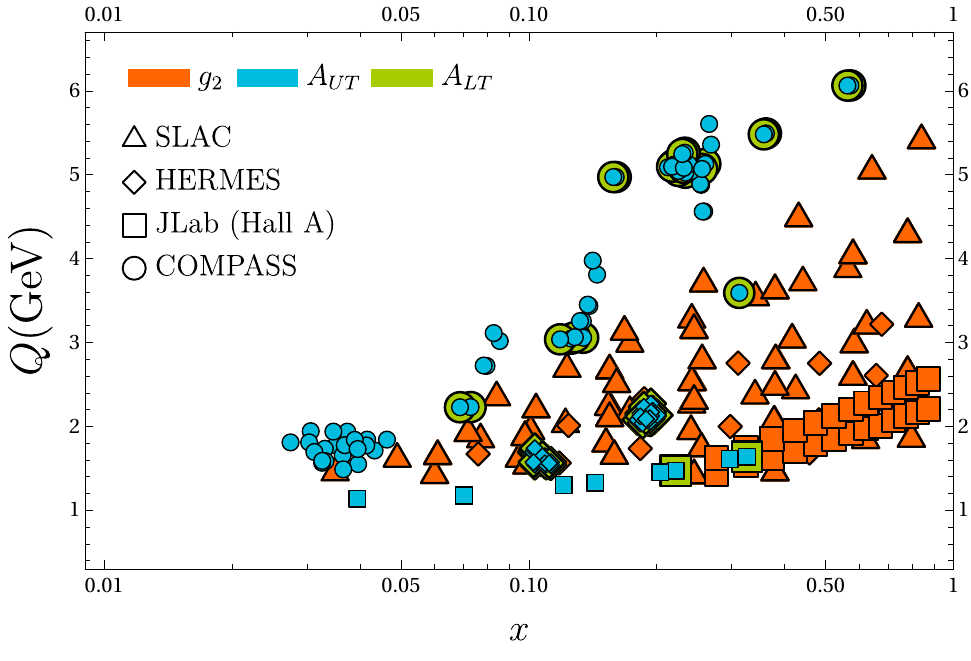}
\caption{\label{fig:all-datapoints} The distribution of the data points included in the fit in the $(x,Q)$ plane. Different colors correspond to different types of observables, and different shapes indicate different experiments, as indicated in the figure.}
\end{figure}

\subsection{Spin-asymmetries in TMD factorization (Drell-Yan reaction)}
\label{sec:TMD-DY}

The spin-dependent cross-section of the Drell-Yan reaction
$$
h_1(p_1,s_1)+h_2(p_2,s_2)\to B(q)(\to\ell\ell')+X,
$$
with $B$ being an electro-weak boson, can be written as \cite{Arnold:2008kf}
\begin{eqnarray}
\frac{d\sigma}{dQ^2d y d\varphi dq_T^2}=\frac{\alpha_{\text{em}}^2}{9sQ^2}\Big[F_{UU}^1+|S_{1T}|\sin(\varphi-\varphi_S)F_{TU}^1+\cdots \Big],
\end{eqnarray}
where $s=(p_1+p_2)^2$, $q^2=Q^2$, and dots denote other structure functions. The variables $\varphi$, $q_T$ and $y$ are the angle, transverse component of the momentum, and rapidity of the
electroweak boson measured in the center-of-mass frame respectively. The subscript $T$ denotes the transverse component of the vectors with respect to the ($p_1, p_2$) plane, and should not be confused with the $\perp$-label used in SIDIS.

Within the TMD factorization, the structure functions $F_{UU}^1$ and $F_{TU}^1$ can be written as
\begin{eqnarray}
F^1_{UU}&=&\Big|C_V\(\frac{-Q^2}{\mu^2}\)\Big|^2 \mathcal{B}^{\text{DY}}_0[f_1f_1]+\mathcal{O}\(\frac{q_T^2}{Q^2}\),
\\
F^1_{UT}&=&-M\Big|C_V\(\frac{-Q^2}{\mu^2}\)\Big|^2 \mathcal{B}^{\text{DY}}_1[f_{1T}^\perp f_1]+\mathcal{O}\(\frac{q_T^2}{Q^2}\),
\end{eqnarray}
where the coefficient function $C_V$ is the same as in the SIDIS case, but evaluated for time-like kinematics. The integral convolution is given by 
\begin{eqnarray}\label{def:TMD-convolution3}
&&\mathcal{B}_n^{\text{DY}}[f_1f_2]=
\\\nn
&&
\sum_{\text{ch.}}\sum_f z_l^{\text{ch.}}z_q^{\text{ch.}}\Delta^{\text{ch.}}
\int_0^\infty \frac{b\,db}{2\pi} b^n J_n(b|q_T|)
f_{1f}(x_1,b;\mu,Q^2)\overline{f}_{2f}(x_2,b;\mu,Q^2),
\end{eqnarray}
with $\overline{f}$ being a TMDPDF for the antiquark flavor. The variables $x_{1,2}$ are\footref{foot:eulc}
$$
x_1=\sqrt{\frac{Q^2+\vec q_T^2}{s}}e^y,\qquad x_2=\sqrt{\frac{Q^2+\vec q_T^2}{s}}e^{-y}.
$$
The factors $z_f^{\text{ch.}}$ are a combination of the electro-weak coupling constants, and $\Delta^{\text{ch}}$ is the corresponding propagator (multiplied by $Q^4$). These variables depend on the channel of the reaction. For example, at low energies, there is only a photon channel. In this case: $z^{\gamma\gamma}_l=1$, $z_f^{\gamma\gamma}=e_f^2$ and $\Delta^{\gamma\gamma}=1$. In the cases involving Z-boson and W-boson production, the expressions are more involved and include the $WW$, $ZZ$ and $Z\gamma$ channels (see equations (2.26)-(2.27) in ref.~\cite{Bury:2021sue}).

In this work we consider only the asymmetry $A_N$, measured by the STAR collaboration \cite{STAR:2023jwh}. In this case, it is defined as
\begin{eqnarray}\label{def:AN_DY}
A_N=\frac{F_{TU}^1}{F_{UU}^1}=-M \frac{\mathcal{B}^{\text{DY}}_1[f_{1T}^\perp f_1]}{\mathcal{B}^{\text{DY}}_0[f_1f_1]}+\mathcal{O}\(\frac{q_T^2}{Q^2}\).
\end{eqnarray}

One of the fundamental predictions of the TMD factorization theorem is the relative sign difference between the Sivers function involved in the Drell-Yan reaction and the one in SIDIS. This effect arises due to the differences in the gauge-structure of QCD operators involved in these processes \cite{Collins:2002kn, Boer:2003cm}. In the case of Drell-Yan, the Sivers function should be taken with the opposite sign:
\begin{eqnarray}\label{siv:sign-change}
f_{1T}^\perp(x,b)\Big|_{\text{SIDIS}}=-f_{1T}^\perp(x,b)\Big|_{\text{DY}}.
\end{eqnarray}
Everywhere (apart from this section), we take the definition of the Sivers function as that in the SIDIS case. Therefore, to reconstruct $A_N$ using our accepted SIDIS definition one should compute as $A_N=+M \mathcal{B}_1[f_{1T}^\perp f_1]/\mathcal{B}_0[f_1 f_1]$.

\subsection{TMD evolution}

The scales $\mu$ and $\zeta$ that are present in the arguments of TMD distributions (\ref{def:TMD-convolution}) are, respectively, the hard and rapidity factorization scales. The dependence on $\mu$ and $\zeta$ is dictated by the TMD evolution equations \cite{Aybat:2011zv}
\begin{eqnarray}\label{def:TMDev1}
\mu^2 \frac{d}{d\mu^2} F(x,b;\mu,\zeta)&=&\frac{\gamma_F(\mu,\zeta)}{2}F(x,b;\mu,\zeta),
\\\label{def:TMDev2}
\zeta \frac{d}{d\zeta} F(x,b;\mu,\zeta)&=&-\mathcal{D}(\mu,\zeta)F(x,b;\mu,\zeta),
\end{eqnarray}
where $F$ is any TMD distribution (including the TMDFF $D_1$). The TMD anomalous dimension $\gamma_F$ is perturbative. The Collins-Soper kernel $\mathcal{D}$ is a fundamental non-perturbative function, that describes the propagation of the soft-gluons through the QCD vacuum \cite{Vladimirov:2020umg}. 

As an implementation of the TMD evolution, we employ the $\zeta$-prescription \cite{Scimemi:2018xaf}. For the $\zeta$-prescription, the solution of evolution equations (\ref{def:TMDev1}, \ref{def:TMDev2}) and the dependence of TMD distribution on the scale is given by 
\begin{eqnarray}
F(x,b;\mu,\zeta)=\(\frac{\zeta}{\zeta_\mu(b)}\)^{-\mathcal{D}(\mu,b)}F(x,b),
\end{eqnarray}
where $F(x,b)$ is the so-called optimal TMD distribution that is scale-independent, universal and does not contain any part of the Collins-Soper kernel $\mathcal{D}$. The dependence on the latter is entirely encapsulated in the prefactor, with $\zeta_\mu(b)$ being the special equi-evolution line that is a functional of $\mathcal{D}$ \cite{Vladimirov:2019bfa}. Therefore, the integral convolution (\ref{def:TMD-convolution}) takes the form
\begin{eqnarray}\label{def:TMD-convolution2}
\mathcal{B}_n[fD]=\sum_f e_f^2\int_0^\infty \frac{b\,db}{2\pi} b^n J_n\(\frac{b|p_\perp|}{z}\)
\(\frac{Q^2}{\zeta_\mu(b)}\)^{-2\mathcal{D}(\mu,b)}
f_{f}(x,b)D_{f}(z,b),
\end{eqnarray}
and analogously for the Drell-Yan case (\ref{def:TMD-convolution3}). Here, both TMD distributions $f$ and $D$ are optimal distributions. 

The main advantage of the $\zeta$-prescription is that it entirely disentangles non-perturbative effects of TMD parton dynamics from the TMD evolution. Therefore, the perturbative series related to the TMD evolution, and the other part of the TMD distributions are independent. This allows us to use the maximum known perturbative precision for the TMD evolution, and the TMD hard coefficient function $C_V$, without internal conflict with the order of matching to collinear distribution. 

Let us mention that not all implementations of the TMD factorization allow for such separation. For instance, the implementations based on the $b^*$-prescription couple the order of collinear matching to the TMD evolution, see for example \cite{Bacchetta:2022awv, Barry:2023qqh,Bacchetta:2024qre}. In this case, it is compulsory to utilize the same order of the TMD evolution in the fits of polarized data (at the moment it is LO), such as in refs.~\cite{Bacchetta:2020gko, Cammarota:2020qcw, Cocuzza:2023vqs}. Therefore, the $\zeta$-prescription (or an analogous one) is crucial to reach the maximum possible precision in the global fit, since the coupled fits of unpolarized data are all based on the high-precision data and highest-order of perturbative input. 

The present setup is based on the ART25 extraction \cite{Moos:2025sal}, and thus inherits all its perturbative definitions, apart from the polarized TMD PDFs, which are discussed in the following section. Consequently, we work with the TMD anomalous dimension at N$^4$LO \cite{Moch:2018wjh, Herzog:2018kwj, Lee:2022nhh}, and the perturbative part of the Collins-Soper kernel at N$^3$LO \cite{Moult:2022xzt, Duhr:2022yyp}. This corresponds to the N$^4$LL precision of the TMD factorization approach. Collinear matching of unpolarized TMD PDFs and TMD FF is taken at N$^3$LO. We refer to \cite{Moos:2023yfa, Moos:2025sal} for further details of the perturbative input.

\subsection{Small-$b$ matching of TMD distributions}
\label{sec:TMD->tw3}

The TMD distributions are nonperturbative functions of $(x,b)$. There are no first-principles restrictions on these functions, except that, for asymptotically small values of $b$, they turn into collinear distributions. The asymptotic expansion in terms of collinear distributions can be determined by the operator product expansion (OPE). Therefore, knowing a collinear distribution, one can significantly restrict the corresponding TMD distribution and vice versa.

The matching relation between a TMD distribution $F$ at small values of $b$ has the form
\begin{eqnarray}
F(x,b)=\sum_{f}C_{F/f}\otimes f(x,b;\mu_{\text{OPE}})+\mathcal{O}(b^2),
\end{eqnarray}
where $f$ are various collinear distributions that contribute to the OPE,  $C$ is a perturbative coefficient function, and $\mu_{\text{OPE}}$ is the scale of the OPE. Power corrections can be computed (see e.g. \cite{Moos:2020wvd}), but involve unknown collinear functions of higher twist. Therefore, it is more efficient to replace these corrections by fitting an ansatz. In this way, the standard phenomenological expression for TMD distributions has the form
\begin{eqnarray}\label{TMD:ansatz-structure}
F(x,b)=\sum_{f}C_{F/f}\otimes f(x,b)\cdot f_{\text{NP}}(x,b),
\end{eqnarray}
where $f_{\text{NP}}$ is a function which satisfies $f(x,0)\sim 1+\mathcal{O}(b^2)$.

The OPE scale $\mu_{\text{OPE}}$ enters the expression for the coefficient function $C_{F/f}$ via logarithms 
\begin{eqnarray}\label{def:L}
\mathbf{L}=\ln\(\frac{|b|^2\mu_{\text{OPE}}^2}{4e^{-2\gamma_E}}\),
\end{eqnarray}
as the argument of $\alpha_s(\mu_{\text{OPE}})$. Generally speaking, the dependence on $\mu_{\text{OPE}}$ cancels between the evolution of the collinear distribution and that of the coefficient function. Thus, one can set $\mu_{\text{OPE}}$ to any reasonable value. However, in practice there is a residual dependence due to missing higher-order terms of the coefficient function which scales as $\sim \alpha_s^{n+1}\mathbf{L}^n$, where $n$ is the last included order. In our fit we use the general scheme of the ART25 extraction and set
\begin{eqnarray}\label{def:muOPE}
\mu_{\text{OPE}}=\frac{2e^{-\gamma_E}}{|b|}+5\text{GeV}.
\end{eqnarray}
This choice allows one to nullify logarithm contributions at small-$b$ and freezes the value of $\mu_{\text{OPE}}$ at large-$b$. In this way, the ansatz smoothly interpolates between the perturbative and non-perturbative regimes, while avoiding problems with the low-energy definition of the perturbative expression (Landau pole, quark masses).

In the case for the unpolarized distributions, the corresponding collinear distributions are unpolarized PDFs and FFs. They are known with extreme precision, and thus provide a significant restriction to the unpolarized TMD distributions. The inclusion of these matching relations is required to successfully describe precise data from the LHC (see for instance \cite{Hautmann:2020cyp, Bertone:2019nxa, Bacchetta:2022awv}). Although, even in this case, one can find some impact of TMD extractions on PDFs \cite{Barry:2025glq}. Here, we use a N$^3$LO matching, with coefficient functions taken from refs.~\cite{Luo:2019szz, Luo:2019hmp, Ebert:2020yqt, Ebert:2020qef}. ART25 used unpolarized PDFs from the MSHT20 N$^2$LO set \cite{Bailey:2020ooq}, and unpolarized FFs from the MAPFF N$^2$LO set \cite{Khalek:2021gxf, AbdulKhalek:2022laj}.

In the case of polarized distributions, the situation with collinear matching changes completely. The collinear counter-parts of TMD distributions are unknown or almost unknown, and thus, we use the TMD distributions to restrict the collinear part. The expressions for the corresponding coefficient functions are known at NLO \cite{Scimemi:2019gge, Rein:2022odl}. However, we restrict ourselves to the LO matching because it is the maximum available order of evolution of twist-three distributions (see sec.~\ref{sec:evol}).

The LO collinear matching for the Sivers function of quarks reads \cite{Scimemi:2019gge, Ji:2006ub, Kanazawa:2015ajw, Scimemi:2018mmi, Rein:2022odl}
\begin{eqnarray}\label{TMD:sivers-smallb-1}
f_{1T,q}^\perp(x,b)=-\pi T(-x,0,x;\mu_{\text{OPE}})
+\mathcal{O}(\alpha_s)+\mathcal{O}(b^2).
\end{eqnarray}
For the anti-quark Sivers function it is
\begin{eqnarray}\label{TMD:sivers-smallb-2}
f_{1T,\overline{q}}^\perp(x,b)=-\pi T(x,0,-x;\mu_{\text{OPE}})
+\mathcal{O}(\alpha_s)+\mathcal{O}(b^2).
\end{eqnarray}
Both expressions are given for the SIDIS definition of the Sivers function. Therefore, the Sivers function gives access to the genuine twist-three distribution $T$ at the line $(-x,0,x)$ (shown as the red line in fig.~\ref{fig:Hex_Interp}). This function is also known as the Qiu-Sterman function \cite{Qiu:1991pp}, or sometimes as the Efremov-Teryaev-Qiu-Sterman function \cite{Efremov:1983eb}.

In contrast to the $\overline{g}_2$ structure function, the Sivers function is sensitive to both C-even and C-odd twist-three PDFs. Using (\ref{T->Sigma},\ref{sym:2}), one can find that
\begin{eqnarray}
f_{1T,q}^\perp(x,b) \pm f_{1T,\bar q}^\perp(x,b)=-\pi \mathfrak{S}^\pm(-x,0,x;\mu_{\text{OPE}})
+\mathcal{O}(\alpha_s)+\mathcal{O}(b^2).
\end{eqnarray}
So, different flavor configurations are sensitive to various components of $\mathfrak{S}^\pm$. It is somewhat unfortunate because it requires including an independent function $\mathfrak{S}^-$ in the fit.

The worm-gear-T function has a more involved OPE structure. It incorporates two parts:
\begin{eqnarray}\label{wgt:tw2+tw3}
g_{1T}^\perp(x,b)&=&g^{\text{tw2}}_{1T}(x,b)+g^{\text{tw3}}_{1T}(x,b)+\mathcal{O}(b^2).
\end{eqnarray}
The first term represents a matching to the twist-two helicity distribution analogous to the Wandzura-Wilczek part of $g_2$ (\ref{g2:WW}), while the second incorporates only genuine twist-three distributions.

The twist-two part of the worm-gear-T function is given by
\begin{eqnarray}\label{wgt:tw2}
g^{\text{tw2}}_{1T}(x,b)&=&\sum_f x\int_x^1 \frac{dy}{y} C^{\text{tw2}}_{q/f}\(\frac{x}{y},\mu_{\text{OPE}}\) g_{1,f}(y;\mu_{\text{OPE}}),
\end{eqnarray}
where $g_1$ is the helicity PDF. The NLO coefficient functions in the $\zeta$-prescription are
\begin{eqnarray}\label{gT:WW-quark}
C^{\text{tw2}}_{q/q}(x,\mu)&=&1+a_s(\mu)C_F\Big[\mathbf{L}(2\ln x-4\ln(1-x)-1-2x)
\\\nn &&\qquad\qquad-2(1-x)-2\ln x-\frac{\pi^2}{6}\Big]+\mathcal{O}(a_s^2),
\\\label{gT:WW-gluon}
C^{\text{tw2}}_{q/g}(x,\mu)&=&a_s(\mu)\Big[-\mathbf{L}(\ln x+2-2x)+1-x+\frac{1}{2}\ln x\Big]+\mathcal{O}(a_s^2),
\end{eqnarray}
where $\mathbf{L}$ is defined in (\ref{def:L}). One can significantly improve the perturbative convergence of the coefficient function by resumming large-x logarithms $\ln(1-x)$ \cite{delRio:2025qgz}. In this case, the quark contribution (\ref{gT:WW-quark}) changes to 
\begin{eqnarray}
C^{\text{tw2}}_{q/q}(x,\mu)&=&\frac{e^{\overline{\mathcal{E}}}}{(1-x)^{\alpha_f}}+a_s(\mu)C_F(2\mathbf{L}-1)(1-x+\ln x)+\mathcal{O}(a_s^2), 
\end{eqnarray}
where $\overline{\mathcal{E}}$ and $\alpha_f$ are known up to N$^3$LO. Their explicit expressions are given in appendix A of ref.\cite{delRio:2025qgz}. In this work we use the resummed version of the twist-two matching, along the helicity distribution extracted at NNLO using the MAPPDFpol1.0 set \cite{Bertone:2024taw}.

The twist-three part of the worm-gear-T function reads \cite{Rein:2022odl, Scimemi:2018mmi}
\begin{eqnarray}\label{wgt:tw3}
g^{\text{tw3}}_{1T}(x,b)=
2x\int [dy]\int_0^1 d\alpha \delta(x-\alpha y_3)\(\frac{\Delta T(y_{123})}{y_2^2}+\frac{T(y_{123})-\Delta T(y_{123})}{2y_2y_3}\)+\mathcal{O}(a_s).
\end{eqnarray}
For numerical evaluation it is convenient to integrate over $\alpha$ and extract the apparent pole explicitly. The result is better looking in terms of the function $S^+$:
\begin{eqnarray}
&&g^{\text{tw3}}_{1T,q}(x,b)=
2x\int [dy]\Big[
\(\frac{\theta(y_1<-x,y_3>x)}{y_1y_3^2}-\frac{\theta(y_1>-x,y_3>x)}{y_2y_3^2}\)S^+(y_{123})
\\\nn &&
\qquad
+\(\frac{\theta(y_1<-x;y_3<x)}{y_1y^2_2}
+\frac{\theta(y_1>-x;y_3>x)}{y^2_2y_3}\)\(S^+(y_{123})-S^+(-x,0,x)\)\Big],
\end{eqnarray}
which is explicitly finite at $y_2\to0$. The anti-quark part is obtained by changing $x\to -x$, and explicitly reads
\begin{eqnarray}
&&g^{\text{tw3}}_{1T,\bar q}(x,b)=
-2x\int [dy]\Big[
\(\frac{\theta(y_1>x,y_3<-x)}{y_1y_3^2}-\frac{\theta(y_1<x,y_3<-x)}{y_2y_3^2}\)S^+(y_{123})
\\\nn && 
\qquad
+\(\frac{\theta(y_1>x;y_3>-x)}{y_1y^2_2}
+\frac{\theta(y_1<x;y_3<-x)}{y^2_2y_3}\)\(S^+(y_{123})-S^+(x,0,-x)\)\Big].
\end{eqnarray}
Note that these expressions are similar to the expression (\ref{expession:g2}) for the $\overline{g}_2$ function. In particular, they have the same range of integration, given by the green (for the quark case) and red (for the anti-quark case) areas in fig.~\ref{fig:Hex_Interp}. However, the kernel of integration is different, and thus one cannot symmetrize the function $S^+$ to $\mathfrak{S}^+$. This means that the worm-gear-T function has sensitivity to both $\mathfrak{S}^+$ and $\mathfrak{S}^-$ components, like the Sivers function.

Apart from the known small-$b$ limit, there are no other fundamental constraints on the Sivers function and worm-gear-T function, although there are extra relations that they could satisfy, namely, the Burkardt sum rule \cite{Burkardt:2004ur} and the positivity constraint \cite{Bacchetta:1999kz}. These relations are based on the naive-parton model expectation, and do not have a fundamental origin, which, in particular, implies that they could be violated by evolution effects. Therefore, we do not include these constraints in our fit procedure, but rather test them afterwards, see discussions concerning the Burkardt sum-rule in sec.~\ref{sec:burkardt}, and concerning the positivity bound in sec.~\ref{sec:positivity}.

\subsection{Review of experimental data} 
\label{sec:SIDIS-data}

\begin{table}[t]
\centering
\begin{tabular}{||l|c|c|c|p{4.5cm}||}
\hline
Dataset & Ref. & Kinematic range & $N_{\text{pt}}$ & Comment
\\\Xhline{5\arrayrulewidth}
\multicolumn{5}{||c||}{$A_{UT}^{\sin(\phi_h-\phi_S)}$ asymmetry in SIDIS}
\\\hline
COMPASS (2008) & \cite{COMPASS:2008isr} & 
\makecell[c]{
$1<Q<16$GeV \\ $0.03<x<1$ \\ $0.2<z<1$ \\ $W>5$GeV} &
27 &
\makecell[l]{
$\pi^\pm$, $K^\pm$ on \\ the deuteron target
}
\\
\hline 
COMPASS (2016) & \cite{COMPASS:2016led} & 
\makecell[c]{
$1<Q<9$GeV \\ $0.003<x<0.9$ \\ $0.1<z<1$ \\ $W>\sqrt{10}$GeV} &
28 &
\makecell[l]{
$h^\pm$ on \\ the proton target
}
\\
\hline 
COMPASS (2023) & \cite{COMPASS:2023vhr} & 
\makecell[c]{
$1<Q$GeV \\ $0.003<x<0.7$ \\ $0.2<z<1$ \\ $W>5$GeV} &
12 &
\makecell[l]{
$h^\pm$ on \\ the deuteron target
}
\\
\hline 
HERMES & \cite{HERMES:2020ifk} & 
\makecell[c]{
$1<Q$GeV \\ $0.023<x<0.6$ \\ $0.2<z<0.7$ \\ $W>\sqrt{10}$GeV} &
44 &
\makecell[l]{
$\pi^\pm$, $K^\pm$ on \\ the proton target
}
\\
\hline 
JLab & \cite{JeffersonLabHallA:2011ayy, JeffersonLabHallA:2014yxb} & 
\makecell[c]{
$1<Q$GeV \\ $W>2.3$GeV} &
6 &
\makecell[l]{
$\pi^\pm$ , $K^\pm$ on \\ the neutron target
}
\\\Xhline{5\arrayrulewidth}
\multicolumn{5}{||c||}{$A_{UT}$ asymmetry in the Drell-Yan reaction}
\\
\hline 
STAR (Z-boson) & \cite{STAR:2023jwh}& 
\makecell[c]{
$73<Q<114$GeV \\ $-1<y<1$ \\ $p_T<10$GeV} &
 &
\makecell[l]{
not included in the fit \\ (see text)
}
\\
\hline 
STAR (W-boson) & \cite{STAR:inprep}& 
\makecell[c]{
$50<Q<110$GeV \\ $-0.7<y<0.7$ \\ $p_T<10$GeV} &
 &
\makecell[l]{
not included in the fit \\ (see text)
}
\\\Xhline{5\arrayrulewidth}
\multicolumn{3}{||r|}{\textbf{Total:} } &
117 &\\
\hline
\end{tabular}
\caption{\label{tab:AUT-data} Review of available high-energy data for the $F_{UT}^{\sin(\phi_h-\phi_S)}$ structure function in SIDIS. The column $N_{\text{pt}}$ indicates the number of data points that survive the kinematic restrictions and are included in the fit.}
\end{table}

The SIDIS reaction has been measured at the HERMES, COMPASS and JLab experiments. The measurement of the Sivers asymmetry, which is a single-spin asymmetry, is simpler and thus the data is more accurate and incorporates more points. The measurement for the $A_{LT}$ asymmetry, which is a double spin-asymmetry, is more involved and thus the data possesses larger uncertainty and large-kinematic bins. The synopses of included data are presented in tables \ref{tab:AUT-data} and \ref{tab:ALT-data}.

In addition to SIDIS measurements there are measurements of the $A_{UT}$ asymmetry in the Drell-Yan reaction obtained by the STAR collaboration \cite{STAR:2015vmv, STAR:2023jwh}. These are unique data because they are performed at a very-high energy ($Z$ and $W$-boson production). Therefore, they provide information about the evolution of the Sivers function, and also allow us to test the sign-change relation (\ref{siv:sign-change}). However, we did not include these data in the fit. There are several reasons for it. The main reason is that this data set contains only 7 points in total, with rather large uncertainties. At the same time, the computation of these 7 points requires significant computer time, since one should compute the evolution of twist-three distributions up to much higher values, and integrate over the weak-boson peak. It slows down the computation procedure by about 20\% (which is significant given the present state of numerical computation), with almost zero impact on extracted values. Nonetheless, we discuss the description of these points and their impact to test the sign-change relation (\ref{siv:sign-change}) in sec.~\ref{sec:result-DY+sign}.

The TMD factorization is valid up to power corrections in $p_\perp/Q$. Therefore, in addition to the general restriction for $Q$ (\ref{data:Q2>2}), we additionally restrict the TMD data with respect to $p_\perp$. We use the following constraint:
\begin{eqnarray}\label{data:Q2>2+TMD}
Q^2>2\text{GeV}^2,\qquad \frac{p_\perp^2}{z^2Q^2}<0.35.
\end{eqnarray}
This is the same constraint that was used in the works \cite{Bury:2021sue, Horstmann:2022xkk}. In the case of some measurements (namely, JLab and COMPASS measurements \cite{JeffersonLabHallA:2011ayy, JeffersonLabHallA:2011vwy, JeffersonLabHallA:2014yxb, COMPASS:2008isr, COMPASS:2023vhr}), the $p_T$ bins are rather large, and thus the cut rule (\ref{data:Q2>2+TMD}) leaves only a single point in each series. For these measurements we have included more $p_T$ points, as long as they obey $p_T/z/Q<0.45$, to reduce the bias. 

\begin{table}[t]
\centering
\begin{tabular}{||l|c|c|c|p{5.5cm}||}
\hline
\multicolumn{5}{||c||}{$A_{LT}^{\cos(\phi_h-\phi_S)}$ asymmetry in SIDIS}
\\\Xhline{5\arrayrulewidth}
Dataset & Ref. & Kinematic range & $N_{\text{pt}}$ & Comment
\\\Xhline{5\arrayrulewidth}
\hline 
COMPASS (2016) & \cite{COMPASS:2016led} & 
\makecell[c]{
$1<Q<9$GeV \\ $0.003<x<0.9$ \\ $0.1<z<1$ \\ $W>\sqrt{10}$GeV} &
26 &
\makecell[l]{
$h^\pm$ on the proton target
}
\\
\hline 
HERMES & \cite{HERMES:2020ifk} & 
\makecell[c]{
$1<Q$GeV \\ $0.023<x<0.6$ \\ $0.2<z<0.7$ \\ $W>\sqrt{10}$GeV} &
44 &
\makecell[l]{
$\pi^\pm$, $K^\pm$ on the proton target
}
\\
\hline 
JLab & \cite{JeffersonLabHallA:2011vwy} & 
\makecell[c]{
$1.4<Q^2<2.7$GeV$^2$\\ $0.16<x<0.35$ \\ $W>2.3$GeV} &
4 &
\makecell[l]{
$\pi^\pm$ on the neutron target
}
\\\Xhline{5\arrayrulewidth}
\multicolumn{3}{||r|}{\textbf{Total:} } &
74 &\\
\hline
\end{tabular}
\caption{\label{tab:ALT-data} Review of available high-energy data for the $F_{UT}^{\cos(\phi_h-\phi_S)}$ structure function in SIDIS. The column $N_{\text{pt}}$ indicates the number of data points that survive the kinematic restrictions and are included in the fit.}
\end{table}

In total, the applied kinematic cuts are milder than the typical constraints used in unpolarized phenomenology, because \textit{(i)} the data has significantly larger uncertainty and thus allows a less restrictive consideration, and \textit{(ii)} it is expected that power corrections partially cancel in the ratio, and thus the TMD factorization has somewhat of a larger application range. This is confirmed in the final plots, where we observe that the theoretical curves continue to describe the data significantly beyond their restrictions.

Let us mention that the COMPASS measurements \cite{COMPASS:2008isr, COMPASS:2016led, COMPASS:2023vhr} are made for integrated kinematics with binning, vs.$x$, vs.$z$ and vs.$p_\perp$. We consider the complete binning simultaneously, because it provides us with important information about each kinematic variable and allows us to restrict the $x$ and $p_\perp$ dependence at the same time. In principle, it implies that some portion of the data is double-counted, however, the double-counting is partially mitigated by the application of the cut rule (\ref{data:Q2>2+TMD}).

Altogether, we have 117 points to extract the Sivers function, and 74 points to extract the worm-gear-T function. This data set is a bit larger than that of our previous studies \cite{Bury:2020vhj, Bury:2021sue, Horstmann:2022xkk} thanks to the adding-an-extra-point rule, and the recent measurement \cite{COMPASS:2023vhr}. The distribution of data points in the ($x,Q$) plane is shown in fig.~\ref{fig:all-datapoints} together with points for the $g_2$ structure function.

\section{Extraction procedure} 
\label{sec:fit}

This section is devoted to the clarification of various technical details concerning the extraction procedure. Most importantly, it includes the description of the fitting ansatz, and the propagation of various uncertainties.

\subsection{Interpretation of experimental data}

Compared to experimental data, we are performing several additional approximations.

\textbf{Bin treatment.} The experimental measurement is made for the cross-section averaged over a certain kinematic bin. Since the value of the cross-section changes within a bin, the theoretical prediction at the central (or average) kinematics is an approximation of the measurement. The complete comparison should be done integrating the theoretical prediction over the bin, i.e. simulating the result of the measurement. However, such an approach is numerically much more expensive, because it requires evaluating multi-dimensional integrals for each data point. Therefore, in our fit we have performed the computation with the average-bin value of the kinematic parameters that is provided by experimental groups. The only exceptions are the Drell-Yan measurements by STAR, for which we averaged the cross-section properly. The reason is that the Z/W production cross-section has a very rapidly changing shape with respect to $Q$, and thus could not be approximated sufficiently well. Note that some experiments (e.g. HERMES \cite{HERMES:2020ifk}) include the uncertainty due to the bin-averaging into the data uncertainty.

\textbf{Flavor decomposition.} The ansatz for genuine twist-three distributions is made for the proton, and incorporates the $\{u,d,s,c,b,g\}$ flavors. Meanwhile, the experiments provide us with measurements at proton, deuteron and neutron targets. To address it, we approximate the neutron target as a proton with exchanged $u$ and $d$ flavors, keeping the sea contributions intact, i.e. it has the content $\{d,u,s,c,b,g\}$. At the same time, the deuteron target is considered an average of a proton and neutron target, thus we obtain its distributions by computing $\{(u+d)/2,(u+d)/2,s,c,b,g\}$.

Some of the COMPASS measurements are given for charged hadron $h^\pm$. In this case, we approximate its fragmentation function as the sum of the $\pi$ and $K$ fragmentation functions, as it was done in ART25. This is a good approximation, since $\pi^\pm+K^\pm$ provides a 95\% of the hadron content at these energies (see e.g.\cite{COMPASS:2023vhr}). Also, all SIDIS measurements are asymmetries, thus their corresponding corrections should be added to the denominator and the numerator, and most likely compensate each other.

\subsection{Ansatz} 
\label{sec:ansatz}

At the present stage of the investigation, there is no preferred parametrization of genuine twist-three distributions. Apart from our proof-of-concept study \cite{Vladimirov:2025qrh}, there have been no other extractions and no experience determining these objects. As a consequence, currently it is not clear which shape one should expect, and which features should be allowed by the parametrization. Therefore, we intend to construct a sufficiently flexible ansatz, while keeping in mind that it should not incorporate too many parameters. We adopt the following general criteria for said ansatz:
\begin{itemize}
\item The ansatz should satisfy the symmetry relations (\ref{sym:1}-\ref{sym:4}).
\item The ansatz should vanish at the boundary $\|x\|\to1$.
\item It should be continuous and smooth in the complete domain (except maybe at the origin $\|x\|=0$).
\item It should be made for $u$, $d$ and $s$ quarks, and allow a non-zero gluon contribution.
\end{itemize}
It appears that it is not simple to create such functions with a limited number of parameters.

The structure of twist-three evolution equations adds specific features to the twist-three PDFs. Most importantly, it induces a singularity, of a yet unclear type, at $\|x\|=0$. This behavior was observed already in ref.~\cite{Rodini:2024usc}. Therefore, it is important to perform at least a few iterations of the evolution procedure before comparing with data. Thus, we impose the boundary ansatz at the scale $\mu_{i}=1$GeV, so that it is later iterated about 25 times before $Q=2$GeV (we employ the Runge-Kutta method with a logarithmic step of $\ln \mu\sim 0.03$), where we start to compare with the data.

The ansatz is made in terms of the functions $\mathfrak{S}^\pm$ and $T_{3F}^\pm$. All flavors incorporate the following function
\begin{equation}\label{ansatz1}
h(x_{123}) = \frac{(1-x_1^2)^{a_1} (1-x_2^2)^{a_2} (1-x_3^2)^{a_3}}{(x_1^2 + x_2^2 + x_3^2)^{a_0}},
\end{equation}
where $a_{0,1,2,3}$ are parameters to fit. Here, the parameter $a_0$ controls the behavior at $\|x\|\to0$, which could be singular (if $a_0>0$), or regular (if $a_0<0$). The parameters $a_{i}$ control the behavior at $|x_i|\to1$, and they are restricted to $a_i>0$. 

For the quark sector we multiply the function $h$ by a general polynomial of second order that preserves the symmetries $\mathfrak{S}^\pm(x_{123})=\pm \mathfrak{S}^\pm(-x_{123})$. We obtain
\begin{eqnarray}
\mathfrak{S}_f^+(x_{123};1\text{GeV})&=&h(x_{123}) \(\alpha_0^f+\alpha_{11}^f x_1^2+\alpha_{13}^fx_1x_3+\alpha_{33}^fx_3^2\),
\\
\mathfrak{S}_f^-(x_{123};1\text{GeV})&=&h(x_{123}) \(\alpha_1^fx_1+\alpha_3^f x_3\),
\end{eqnarray}
Therefore, each flavor incorporates $6$ extra parameters.

The gluon distribution possesses a larger symmetry (\ref{sym:3}), and simultaneously is less restricted from the experimental side. Within our datasets there is no single observable that would directly access any gluon distribution. Thus, the gluon part is restricted only through perturbative mixing. Therefore, we employ a minimalist ansatz for gluon input
\begin{eqnarray}
T_{3F}^+(x_{123})&=&h_g(x_{123})\beta_1(x_1-x_3),
\\
T_{3F}^-(x_{123})&=&h_g(x_{123})\beta_0,
\end{eqnarray}
where 
\begin{eqnarray}\label{ansatz6}
h_g(x_{123}) = \frac{(1-x_1^2)^{\frac{a_1+a_3}{2}} (1-x_2^2)^{a_2} (1-x_3^2)^{\frac{a_1+a_3}{2}}}{(x_1^2 + x_2^2 + x_3^2)^{a_0}}.
\end{eqnarray}
The equality of powers in $h_g$ is required by the symmetry relation (\ref{sym:4}). Therefore, the gluon distributions add another two parameters in the fit.

Altogether, we end up with an ansatz that incorporates 24 free parameters:
\begin{itemize}
\item $\{a_0,a_1,a_2,a_3\}$, common for all flavors,
\item $\{\alpha_0^f,\alpha_1^f,\alpha_3^f,\alpha_{11}^f,\alpha_{13}^f,\alpha_{33}^f\}$, specific for $\{u, d, s\}$ quark flavors,
\item $\{\beta_0,\beta_1\}$, specific for gluon distributions.
\end{itemize}
This ansatz is richer than that of ref.~\cite{Vladimirov:2025qrh}, where we set $\alpha_{11}=\alpha_{33}=0$, and $a_1=a_3$, which imposed unnecessary symmetries between $x_1$ and $x_3$. The extra flexibility of the ansatz is restricted by the increased number of data points (372 vs 243 in \cite{Vladimirov:2025qrh}). Still, we found that the results of fits are compatible with each other within uncertainties, see discussion in sec.~\ref{sec:comparison-with-others}.

Furthermore, the determination of the Sivers and worm-gear-T functions requires a parametrization of $f_{\text{NP}}$, which characterizes the non-perturbative transverse momentum part (\ref{TMD:ansatz-structure}). Based on our previous experiences \cite{Bury:2021sue, Horstmann:2022xkk}, we do not expect any significant sensitivity to this function. Therefore, we utilize the same flavor-independent function for both Sivers and worm-gear-T distributions. We use
\begin{eqnarray}\label{ansatz-fNP}
f_{\text{NP}}(x,b)=\frac{1}{\cosh(\lambda |b|)}.
\end{eqnarray}
This type of ansatz demonstrated a very good agreement with data in many earlier fits of TMD distributions (see e.g. \cite{Horstmann:2022xkk, Moos:2023yfa, Moos:2025sal}), including ART25 which plays an essential role in this analysis. 

However, the available data does not allow us to fix the parameter $\lambda$ with sufficient precision. We have found that it allowed any value in the range $0.1-2.$GeV without contradicting the data. This variation is then compensated by that of the parameters $\alpha_i$, which could obtain a very large value. This implies that the inclusion of $\lambda$ in the fit leads to the overfit problem, and makes the optimization procedure unstable. Let us note that similar conclusions were reached in previous studies of spin-asymmetries \cite{Bury:2021sue, Horstmann:2022xkk}, where similar parameters were found to be $0.94_{-0.93}^{+0.71}$ and $0.52^{+0.37}_{-0.42}$. Therefore, we fix the parameter $\lambda$ to
\begin{eqnarray}
\lambda=0.5\text{GeV}.
\end{eqnarray}
In sections \ref{sec:result-Sivers} and \ref{sec:result-wgt} we test the boundaries of this approximation for each TMD distribution.

\subsection{Definition of $\chi^2$}
\label{sec:def-chi2}

The fitting procedure and the method of estimation of uncertainties are inherited from our earlier works on the determination of TMD distributions \cite{Bertone:2019nxa, Scimemi:2019cmh, Vladimirov:2019bfa, Bury:2021sue, Bury:2022czx, Moos:2023yfa, Moos:2025sal}. In turn, it is based on the procedures carried by the NNPDF collaboration to estimate the goodness of their extractions, \cite{Ball:2008by, Ball:2012wy}. Since the present extraction incorporates observables of different kinds, we made minor modifications in the treatment of the global data, described below.

We use the standard definition of the $\chi^2$-test function adopted from fits of collinear PDFs in refs.~\cite{Ball:2008by, Ball:2012wy}, and widely utilized (for instance in \cite{Bacchetta:2022awv, Bacchetta:2019sam, Bacchetta:2024qre}). It is defined as
\begin{eqnarray}\label{def:chi2-0}
\chi^2_{test}=\sum_{i,j\in \text{data}}(m_i-t_i)V^{-1}_{ij}(m_j-t_j),
\end{eqnarray}
where $i$ and $j$ run over all data points included in the dataset, $m_i$ and $t_i$ are the experimental value and theoretical prediction for the point $i$, respectively, and $V^{-1}_{ij}$ is the inverse of the covariance matrix. The covariance matrix is defined as
\begin{eqnarray}\label{def:V}
V_{ij}=\delta_{ij}\Delta_{i,\text{uncorr}.}^2+\sum_{l}\Delta_{i,\text{corr}.}^{(l)}\Delta_{j,\text{corr}.}^{(l)},
\end{eqnarray}
where $\Delta_{i,\text{uncorr}.}$ is the uncorrelated uncertainty of the measurement $i$, and $\Delta_{i,\text{corr}.}^{(l)}$ is its $l$-th correlated uncertainty. In the present analysis, most of the data presents only uncorrelated uncertainties, and correlated uncertainty is present mainly due to the uncertainty in the target's polarization. 

It is expected that a statistically satisfying description of the data is reached if $\chi^2\approx N_{\text{pt}}$, where $N_{\text{pt}}$ is the number of data points included in the fit. This criterion is based on the assumption that if the number of data points is large (much larger than the number of parameters to fit), then the uncertainties are Gaussian and self-consistent (i.e. there are not undeclared correlations, nor tensions, etc.). Naturally, these criteria are not entirely valid for all datasets, because many of them were obtained using older technologies, some of them seemingly contradict each other, etc.  Altogether, this makes the estimation of the goodness of the global fit an ill-defined problem. 

In our case, we have found two principal issues. The first one is that some data points possess very large uncertainties, and thus result in negligible contributions to the total $\chi^2$ (for example, we have excluded from the fit the data by the E142 collaboration \cite{E142:1996thl}, since its uncertainty is much larger than the rest of data). The second is that we compare data for different observables which are measured using different methods, and thus have innate differences in quality. For example, double-spin asymmetries have naturally larger uncertainties, and a smaller number of data points in comparison to single-spin asymmetries. At the same time, all these data are important because they provide independent restrictions to the multi-dimensional distribution. 

As a consequence of these issues, a traditional and equal treatment of all independent data inputs is not a reliable strategy and easily leads to the overfit problem. If one considers all data in a single take, a part of the data would drive the fit, ignoring the rest. Particularly, the $d_2$ data set contains only 13 points, and it would be immediately shadowed by the rest of the measurements. To mitigate this problem, we consider several definitions of the $\chi^2$. First of all, we define $\chi^2$ functions for each of the principal processes
\begin{eqnarray}
\chi^2_{d2}&=&\sum_{i,j\in \text{tab.\ref{tab:d2-data}}}\chi_{ij}^2,
\qquad
\chi^2_{g2}=\sum_{i,j\in \text{tab.\ref{tab:g2-data}}}\chi_{ij}^2,
\\\nn
\chi^2_{UT}&=&\sum_{i,j\in \text{tab.\ref{tab:AUT-data}}}\chi_{ij}^2,
\qquad
\chi^2_{LT}=\sum_{i,j\in \text{tab.\ref{tab:ALT-data}}}\chi_{ij}^2,
\end{eqnarray}
where $\chi_{ij}^2$ is the element of the sum given in (\ref{def:chi2-0}). Each of these functions test the quality of the description for each process, and control a particular projection onto the hexagon domain, which is considered to be good if $\chi^2_{k}/N_k\approx 1$ (with $N_i$ being number of data points in each subset, i.e. $N_{d2}=13$, $N_{g2}=168$, $N_{UT}=117$ and $N_{LT}=74$). 

To test the global quality of the data description we study the sum of each measurement
\begin{eqnarray}\label{def:chi2_tot}
\chi^2_{\text{tot}}=\chi^2_{d2}+\chi^2_{g2}+\chi^2_{UT}+\chi^2_{LT}.
\end{eqnarray}
It should be compared with $N_{\text{tot}}=372$.

The minimization procedure, however, is made for the weighted $\chi^2$ function
\begin{eqnarray}
X^2=\frac{\chi^2_{d2}}{N_{d2}}+\frac{\chi^2_{g2}}{N_{g2}}+\frac{\chi^2_{UT}}{N_{UT}}+\frac{\chi^2_{LT}}{N_{LT}}.
\end{eqnarray}
This function treats each process subset equally. If the minima of these subsets are the same, it coincides with the minimum of $X^2$. However, if they are not the same, $X^2$ has an equilibrated minimum. In this way, we avoid the problem of overfitting large subsets (such as the $g_2$ subset) at the expense of the rest. In sec.~\ref{sec:validation} we demonstrate that this definition allows one to reach a good agreement between all parts of data and theory.

We stress that $X^2$ is used exclusively as the internal function of the minimization procedure (sec.~\ref{sec:def-uncert}). All results presented in the text, such as $\chi^2$ values, the uncertainty bands, significance statements, and the Hessian-based uncertainty estimation of sec.~\ref{sec:lambdas}, are computed from the standard definition of $\chi^2_{\text{tot}}$ (\ref{def:chi2_tot}). In principle, the minima of $\chi^2_{\text{tot}}$ and $X^2$ could differ. However, as shown in sec.~\ref{sec:validation}, the partial contributions to $X^2$ are close to unity, and $\chi_{\text{tot}}^2/N_{\text{pt}}$ is also close to unity, indicating that their minima are in agreement.

\subsection{Estimation of uncertainties}
\label{sec:def-uncert}

The search of the data-preferred distribution consists in obtaining the minimum value of $X^2$ with respect to the (24-dimensional) vector of parameters of the ansatz $\overrightarrow{\alpha}=\{a_0,a_1,...,\beta_1\}$. However, the prediction depends on a number of other theoretical elements, and the agreement with the data depends on their uncertainties. Therefore, the minimal value of $X^2$ is only an element of an ensemble of possible physical values (in the following we refer to it as the central value of the fit and denote it as $\overrightarrow{\alpha}_0$). Consequently, the central aim of the study is to find the parameters of this ensemble, which gives us the mean value and the uncertainty band of the physical distribution.

To determine the distribution $\overrightarrow{\alpha}$ we employ the parametric bootstrap method \cite{efron1994introduction}. It consists in the randomization of all uncertain inputs, in accordance to their proper distribution, and minimizing $X^2$ for each randomized setup. Each minimization procedure yields a value $\overrightarrow{\alpha}_i$, and the complete set of $\{\overrightarrow{\alpha}_i\}$ provides a statistical ensemble of the distribution of parameters that encode all inherited correlations, and other features.

There are two types of uncertainties:
\begin{itemize}
\item \textbf{Uncertainties due to the theoretical input.} These are uncertainties that appear due to other non-perturbative elements that are incorporated. These are the helicity distribution $g_1$ \cite{Bertone:2024taw}, unpolarized TMDPDF $f_1$ and unpolarized TMDFF $D_1$ \cite{Moos:2025sal}. The uncertainties of these distributions are given as statistical ensembles of distributions (replicas). To account for these uncertainties, we randomly selected replicas of these distributions for each minimization setup.
\item \textbf{Uncertainties due to the data input.} These are uncertainties presented in the data, which are given as uncorrelated and correlated uncertainties for each data point. To account for these uncertainties, we generated a set of pseudo-data following the rules given in ref.~\cite{Ball:2008by} for each minimization setup.
\end{itemize}
Furthermore, in order to avoid bias to the initial search point, we start each run with a random initial value. This is a very conservative approach to uncertainties, which allows us to be confident in the reliability of our result.

Using the ensemble $\{\overrightarrow{\alpha}_i\}$, we can find any derived quantity (the $g_2$ function, TMD distributions, cross-sections, values of parameters, etc) and their associated uncertainties, while preserving the correlation structure. For a quantity $F$, the procedure is as follows: We evaluate $F$ for each element of $\{\overrightarrow{\alpha}_i\}$, obtaining the ensemble $F_i=F[\overrightarrow{\alpha}_i]$. The mean value is then the mean $\langle F_i\rangle$, and the 68\% confidence interval (CI) uncertainty band is found by the resampling method by computing the 16\% and 84\% quantiles of $\{F_i\}$. The result is then presented as
\begin{eqnarray}
F=m^{+a}_{-b},
\end{eqnarray}
where $m$ is the mean value $\langle F_i\rangle$, and $(m-b,m+a)$ is the interval of the 68\%CI, i.e. $a$ and $b$ are the sizes of the uncertainty band in the positive and negative directions.

The procedure described above allows one to correctly propagate all correlations. Notice that $\langle F[\overrightarrow{\alpha}_i]\rangle \neq F[\overrightarrow{\alpha}_0]$, due to the correlations between members of $\{\overrightarrow{\alpha}_i\}$, asymmetries of distributions, and other effects. Nonetheless, in some tests we employ central values assuming that they are close enough to the mean values. The reason is that the computation of the central value requires only one minimization, while the computation of the mean requires multiple minimizations.

The computation of the ensemble $\{\overrightarrow{\alpha}_i\}$ is the most time-consuming part of the procedure. It requires minimizing $X^2$ multiple times. We have computed 300 replicas for our main fit and 150 replicas for test cases (TMD-only and DIS-only fits, discussed below).

\subsection{Technical implementation}
\label{sec:implementation}

\begin{figure}[t]
\centering
\input{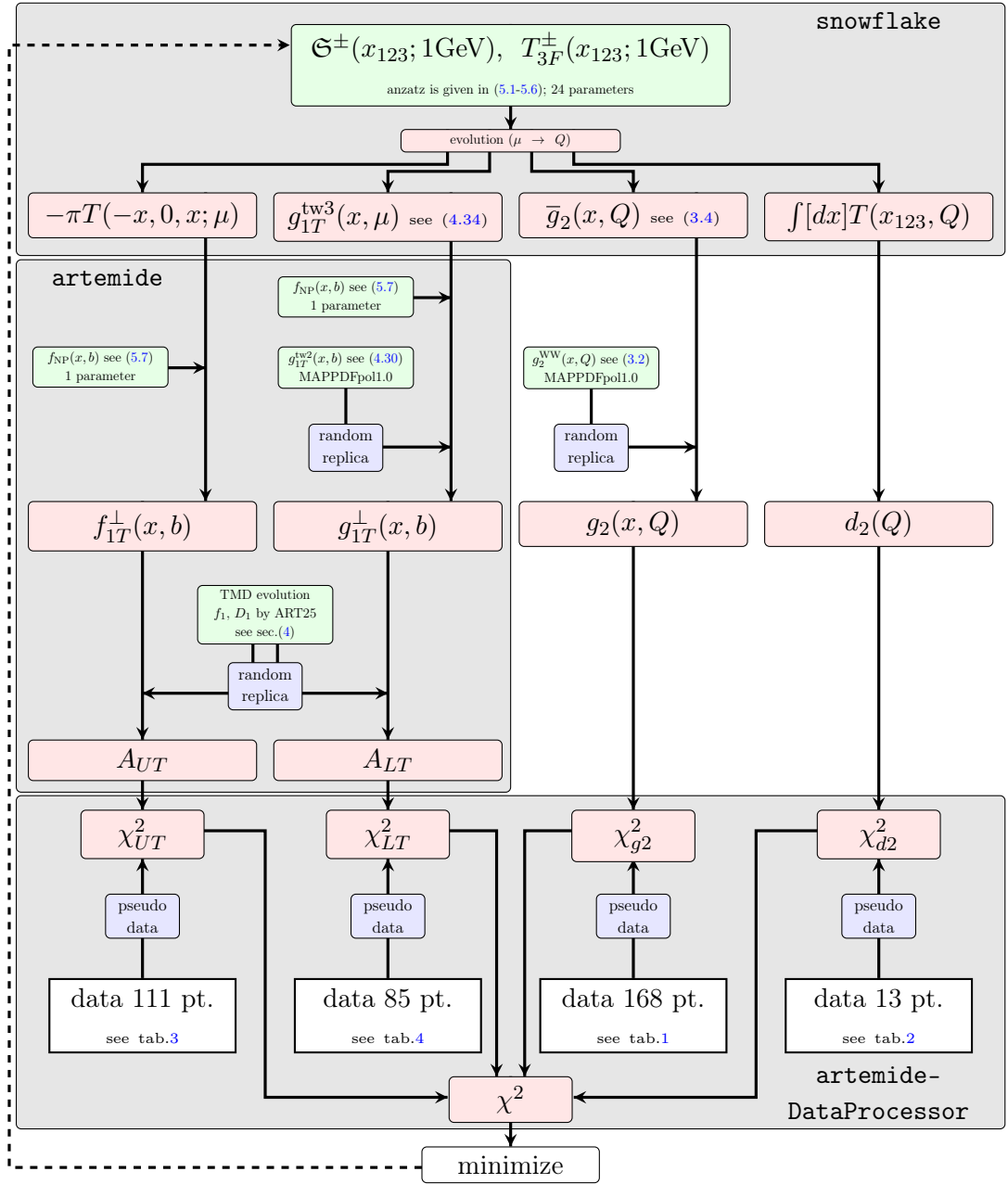}
\caption{\label{fig:scheme} The block-scheme that describes the relation between various elements of the present analysis, starting from the initial ansatz until the computation of the $\chi^2$-function. The green boxes show the elements that incorporate the non-perturbative elements (both external, and determined in this work). The red boxes show the computation parts. The blue boxes indicate the parts that were randomized in order to estimate the uncertainty band. The gray boxes demonstrate the package which is responsible for a particular part of the computation.}
\end{figure}

The computational base is taken from our earlier studies. There are several libraries incorporated in this computation, with the three main ones developed by our group. In this section, we provide references to all used libraries and give some details of their setup. Note that the complete working composition of this study can be downloaded from the repository \cite{artemide}.

The novel element of the analysis is the LO processing of genuine twist-three distributions. This is done with the code \texttt{snowflake} \cite{snowflake} (this repository contains a standalone version of \texttt{snowflake} with a limited functionality). This code performs the evolution of genuine twist-three distributions, and further computation of the required integrals in the $x_{123}$ plane. The algorithm and details of implementation are given in ref.~\cite{Rodini:2024usc}. In short, it approximates a twist-three function by a two-dimensional grid, and performs all necessary operations in a discrete manner, with a subsequent interpolation to any required point. 

In this study, we have used the (24$\times$12) grids for each sector of the hexagon. It provides a relative interpolation precision smaller than $10^{-5}$, and an accumulated error due to evolution smaller than $10^{-4}$, which is more than sufficient for the present data. The initial ansatz is given at $\mu_i=$1GeV and then evolved to $\mu_f\sim 110$GeV (preserving all intermediate values). Such values of $\mu_f$ are required by the TMD-part of the computation that evaluates the TMD distributions down to a very small value of $b$, and consequently, via (\ref{def:muOPE}), the twist-three distributions up to very high $\mu$. The inclusion of high-energy data requires the computation of a table up to even higher values of $\mu$, which significantly increases the computation time, and is the reason to exclude the STAR data from the fit. 

The computation of the TMD evolution, and the convolution of the TMD distributions into cross-sections are done with the help of \texttt{artemide} \cite{artemide}. \texttt{Artemide} is a multi-purpose library for computations within the TMD factorization framework. The details of its implementation can be found in the publications \cite{Moos:2023yfa, Moos:2025sal}. For the purposes of the present study, the codes of \texttt{snowflake} and \texttt{artemide} were merged into a single package that can be compiled in a single run. The package is available in the main repository \cite{artemide}.

The \texttt{artemide+snowflake} package provides one with the theoretical prediction for a given process and kinematics. The comparison with the data, as well as the statistical processing, was performed in the \texttt{artemide-Data-Processor} \cite{DataProcessor}\footnote{Particularly, all programs related to this work can be found in /FittingPrograms/Tw3\_2026/ of the repository \cite{DataProcessor}.}. This is a python library which provides a simple interface to \texttt{artemide}. The provided build incorporates all data described in this paper, as well as tools to compute the $\chi^2$ function, generate pseudo-replicas of data, etc.

The results of the fit and its analysis are collected in a separate repository \cite{REP}. It contains the grids of the extracted distributions, computed predictions, figures, and other information partially described in sec.~\ref{sec:results}.

Finally, we have also used external codes. Namely, the \texttt{APFEL++} package \cite{Bertone:2017gds} to evaluate the WW contribution for $g_2$ function, and \texttt{iminuit} \cite{iminuit} to perform the minimization procedure for the $\chi^2$ function.

The summary of the computation process is showcased in the form of a block-scheme in fig.~\ref{fig:scheme}. It describes the relation between various elements of the present analysis, starting from the initial ansatz until the computation of the $\chi^2$ function.

\section{Validation of the results}
\label{sec:validation}

The fit of twist-three distributions is a novel direction of studies. Therefore, it is not clear what to expect from such a fit, especially in view of a combined description of various experiments. In this section, we present the extracted values of our model's parameters, and discuss their relation to experimental data. Moreover, we discuss several validation tests of our analysis, namely, partial fits, a zero-value test, and a comparison with available fits.

\subsection{Extracted parameters and agreement with the data}

\begin{figure}[t]
\centering
\includegraphics[width=0.9\linewidth]{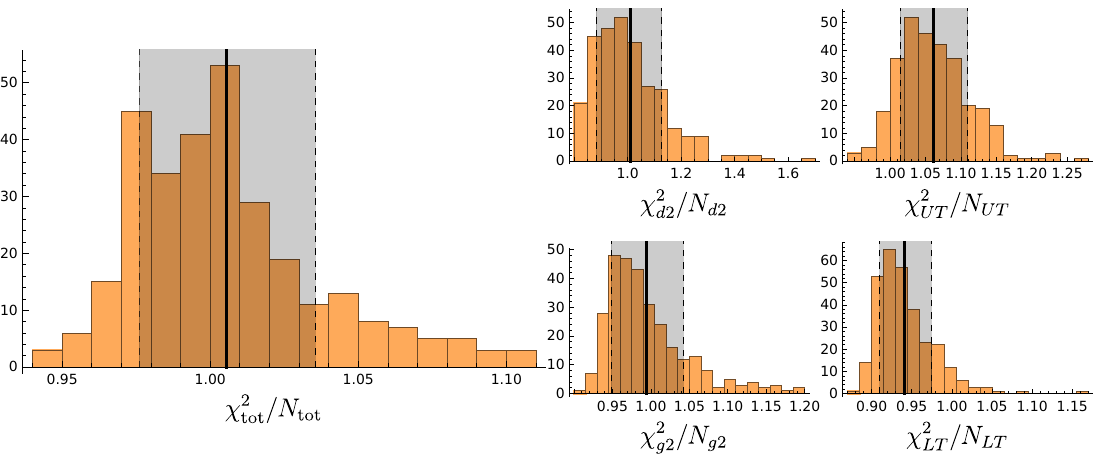}
\caption{\label{fig:chi2_main} Distribution of $\chi^2/N$ for 300 replicas extracted in the main fit. The gray rectangle represents the 68\%CI, and the black line the mean value of the distribution.}
\end{figure}

The fit converges very well and demonstrates a well-defined minimum. The set of 300 replicas, obtained through the procedure described in secs. \ref{sec:def-chi2} and \ref{sec:def-uncert}, has the following values of the $\chi^2$ distributions:
\begin{eqnarray}\label{chi2:main}
\frac{\chi^2_{d2}}{N_{d2}}=1.01^{+0.12}_{-0.13},\quad
\frac{\chi^2_{g2}}{N_{g2}}=0.99^{+0.05}_{-0.05},\quad
\frac{\chi^2_{UT}}{N_{UT}}=1.06^{+0.05}_{-0.05},\quad
\frac{\chi^2_{LT}}{N_{LT}}=0.94^{+0.03}_{-0.03},
\end{eqnarray}
with
\begin{eqnarray}\label{chi2:tot}
\frac{\chi^2_{\text{tot}}}{N_{\text{tot}}}=1.01_{-0.03}^{+0.03}.
\end{eqnarray}
The distributions of $\chi^2$ are shown in fig.~\ref{fig:chi2_main}. Note that these values represent the distribution of $\chi^2$ computed for each replica. Meanwhile, the value of $\chi^2$ for the mean vector of the fitted parameters is smaller, as described below.

Contributions of particular datasets to the global $\chi^2$ are presented in table \ref{tab:chi2}. The contributions are rather homogeneous, i.e. there are no sudden excesses or inadequate drops, and all datasets demonstrate a very good agreement with the theory. It should be noted that many data subsets have $\chi^2/N_{\text{pt}}<1$, which is usually due to an excessive size of the uncertainty of these measurements (especially for older experiments, such as E143). Note that the measurement of the spin-asymmetry in Z/W boson production measured at STAR \cite{STAR:2023jwh, STAR:inprep} was not included in the fit, although it is presented in the table. The discussion concerning these data is provided in sec.~\ref{sec:result-DY+sign}.

The worst agreement is observed with respect to the HERMES $K^\pm$ production data, for both the single and double spin-asymmetry cases ($\chi^2/N_{\text{pt}}=1.71$ and $1.31$, respectively). In these cases, the main disagreement comes specifically from the $K^+$ production: 2.01 (for $A_{UT}$) and 1.53 (for $A_{LT}$), while the $K^-$ channel demonstrates better agreement: 1.10 (for $A_{UT}$) and 1.40 (for $A_{LT}$). This is obviously due to the large scattering of the data points, see fig.~\ref{fig:HERMES2} in the appendix.

The obtained values for the parameters are presented in the following table:
\begin{center}
\renewcommand{\arraystretch}{2.}
\renewcommand\theadset{\def\arraystretch{2.}}
\begin{tabular}{||l|c|c||}
\hline
\multicolumn{3}{||c||}{Values of the parameters in the main fit}
\\\Xhline{5\arrayrulewidth}
\hline 
common & \multicolumn{2}{|c||
}{
$\begin{aligned}
&a_0&=&-1.71^{+0.32}_{-0.33} ~{\scriptstyle(\pm 0.04)}\quad &a_1=8.36^{+0.89}_{-0.79} ~{\scriptstyle(\pm 0.13)}
\\
&a_2&=&1.65^{+0.66}_{-0.74} ~{\scriptstyle(\pm 0.09)}\qquad &a_3=8.34^{+0.89}_{-0.68} ~{\scriptstyle(\pm 0.09)}
\end{aligned}$
 } 
\\\Xhline{5\arrayrulewidth}
& $\mathfrak{S}^+$ & $\mathfrak{S}^-$\\
\hline
$u$-quark & 
$\begin{aligned}
&\alpha_0^u=1.75^{+0.57}_{-0.62}~{\scriptstyle(\pm0.03)}
\\
&\alpha_{11}^u=7.6^{+3.4}_{-4.1}~{\scriptstyle(\pm0.7)}
\\ 
&\alpha_{13}^u=6.2^{+4.5}_{-4.6}~{\scriptstyle(\pm0.8)}
\\
&\alpha_{33}^u=1.6^{+3.4}_{-3.0}~{\scriptstyle(\pm0.3)}
\end{aligned}$
&
$\begin{aligned}
&\alpha_1^u=2.4^{+3.0}_{-2.8}~{\scriptstyle(\pm1.9)}
\\
&\alpha_{3}^u=7.3^{+3.4}_{-3.1}~{\scriptstyle(\pm1.3)}
\end{aligned}$
\\\hline
$d$-quark & 
$\begin{aligned}
&\alpha_0^d=-1.73^{+0.70}_{-0.64}~{\scriptstyle(\pm0.04)}
\\
&\alpha_{11}^d=-5.6^{+4.3}_{-4.3}~{\scriptstyle(\pm0.4)}
\\
&\alpha_{13}^d=-6.3^{+6.9}_{-7.3}~{\scriptstyle(\pm0.8)}
\\
&\alpha_{33}^d=-2.6^{+2.5}_{-2.4}~{\scriptstyle(\pm0.2)}
\end{aligned}$
&
$\begin{aligned}
&\alpha_1^d=-2.4^{+7.1}_{-7.8}~{\scriptstyle(\pm5.3)}
\\
&\alpha_{3}^d=-16.4^{+9.3}_{-9.7}~{\scriptstyle(\pm6.6)}
\end{aligned}$
\\\hline
$s$-quark & 
$\begin{aligned}
&\alpha_0^s=1.95^{+0.90}_{-1.08}~{\scriptstyle(\pm0.05)}
\\
&\alpha_{11}^s=10.^{+12.}_{-11.}~{\scriptstyle(\pm1.)}
\\
&\alpha_{13}^s=7.3^{+8.6}_{-9.2}~{\scriptstyle(\pm2.3)}
\\
&\alpha_{33}^s=7.3^{+8.1}_{-7.9}~{\scriptstyle(\pm0.9)}
\end{aligned}$
&
$\begin{aligned}
&\alpha_1^s=6.6^{+4.3}_{-5.6}~{\scriptstyle(\pm0.6)}
\\
&\alpha_{3}^s=-2.9^{+3.0}_{-3.2}~{\scriptstyle(\pm0.6)}
\end{aligned}$
\\\Xhline{5\arrayrulewidth}
& $T_{3F}^+$ & $T_{3F}^-$ \\\hline
gluon & $\beta_1=5.9^{+2.9}_{-3.6}~{\scriptstyle(\pm0.4)}$
&
$\beta_0=4.2^{+5.4}_{-5.4}~{\scriptstyle(\pm0.7)}$
\\\hline
\end{tabular}
\end{center}
The corresponding distributions are presented in figure \ref{fig:parameter_distr} of the appendix, where one can see that they present well-defined peaks. The uncertainty given in brackets corresponds to the estimation of induced systematic uncertainty due to fixing value of parameters $\lambda$, explained in the following section.

The mean-parameter replica (i.e. the replica corresponding to the mean values of the parameters) has a smaller $\chi^2/N$ in comparison to (\ref{chi2:main}, \ref{chi2:tot}) , $\chi^2_{\text{tot}}/N_{\text{tot}}=0.95$. It indicates that the correlation between parameters is not negligible. The correlation matrix is shown in fig.\ref{fig:parameter_correlation} of the appendix. Mostly, the parameters are correlated inside blocks of a given flavor. The largest correlations are between parameters $(\alpha_3^d, \alpha_3^s)$, $(\alpha_3^u,a_0)$, $(\alpha_3^d,\alpha_3^u)$; they are $(0.77, -0.72, -0.70)$, respectively. However, most of the correlations are rather small, as 85\% of elements of the correlation matrix are smaller than $|0.5|$, and 52\% are smaller than $|0.25|$.

In table \ref{tab:chi2}, we also present the values of $\chi^2/N$ for the central fit, i.e., the fit obtained by minimizing $\chi^2$ on the unmodified central data. Although it does not play a statistical role in the analysis, it is useful for validating the fit procedure independently of replica-averaging effects. The values of $\chi^2$ for this fit are very close to the replica-averaged values. The central fit yields the following values of the model parameters:
\begin{gather*}
a_0=-1.78, \quad a_1= 8.68, \quad a_2=1.55, \quad a_3=8.84\,,
\\
\alpha_0^u=1.87, \quad \alpha_{11}^u=9.0, \quad \alpha_{13}^u=7.5, \quad \alpha_{33}^u=1.4, \quad \alpha_1^u=1.9, \quad \alpha_3^u=6.9\,, 
\\
\alpha_{0}^d=-1.91, \quad \alpha_{11}^d=-7.5, \quad \alpha_{13}^d=-9.0, \quad \alpha_{33}^d=-3.5, \quad \alpha_{1}^d=-3.6, \quad \alpha_{3}^d=-22.3\,,
\\
\alpha_{0}^s=2.1, \quad \alpha_{11}^s=14., \quad \alpha_{13}^s=11.1, \quad \alpha_{33}^s=9.1, \quad \alpha_{1}^s=7.6, \quad \alpha_{3}^s=-2.8\,, 
\\
\beta_1=5.6, \quad \beta_0=4.1\,.
\end{gather*}
As one can see, all these values lie within one standard deviation of the main extraction.

\begin{table}[t]
\begin{center}
\scalebox{0.96}{
\renewcommand{\arraystretch}{1.1}
\begin{tabular}{||l|c|c||c|c|c|c||}
\hline
Set of data & $N_{\text{pt}}$ & \makecell{$\chi^2/N_{\text{pt}}$ \\ (main)}
& \makecell{ $\chi^2/N_{\text{pt}}$ \\ (central)}
  & \makecell{ $\chi^2/N_{\text{pt}}$ \\ (DIS)} & \makecell{ $\chi^2/N_{\text{pt}}$ \\ (TMD)} & \makecell{$\chi^2/N_{\text{pt}}$ \\ (null)}
\\\Xhline{5\arrayrulewidth}
\multicolumn{7}{|c||}{$d_2$ moment} 
\\\hline
E143, E155 & 6 & 0.45 & 0.44 & 0.33 & 1.07 & 1.94
\\\hline
JLab, HERMES, SANE & 5 & 1.52 & 1.43 & 2.05 & 1.77 & 3.47
\\\hline
RQCD (lattice) & 2 & 0.40 & 0.47 & 0.58 & 0.50 & 2.00
\\\Xhline{5\arrayrulewidth}
\textbf{Total $d_2$} & 13 & 0.85 & 0.84 & 1.03 & 1.25 & 2.54
\\\Xhline{5\arrayrulewidth}
\multicolumn{7}{|c||}{Structure function $\bar g_2$} 
\\\hline
E143 & 22 & 0.43 & 0.43 & 0.41 & 0.49 & 0.52   
\\\hline
E154 & 15 & 0.99 & 1.00 & 0.99 & 1.11 & 1.02
\\\hline
E155 & 92 & 1.15 & 1.16 & 1.07 & 4.46 & 1.34
\\\hline
HERMES & 13 & 0.99 & 0.98 & 0.95 & 1.18 & 0.90
\\\hline
Hall A & 26 & 0.66 & 0.65 & 0.73 & 6.45 & 1.23
\\\Xhline{5\arrayrulewidth}
\textbf{Total $\bar g_2$} & 168 & 0.96 & 0.96 & 0.92 & 3.69 & 1.15
\\\Xhline{5\arrayrulewidth}
\multicolumn{7}{|c||}{$A_{UT}^{\sin(\phi_h-\phi_S)}$ asymmetry} 
\\\hline
COMPASS(2008) & 27 & 0.64 & 0.63 & 0.63 & 0.64 & 0.69
\\\hline
COMPASS(2016) & 28 & 0.98 & 0.93 & 2.36 & 1.24 & 1.95
\\\hline
COMPASS(2023) & 12 & 0.45 & 0.44 & 0.48 & 0.39 & 0.49
\\\hline
HERMES($\pi^\pm$) & 22 & 1.10 & 1.13 & 1.34 & 1.00 & 1.34
\\\hline
HERMES($K^\pm$) & 22 & 1.71 & 1.71 & 1.93 & 1.55 & 1.98
\\\hline
JLab & 6 & 0.87 & 0.78 & 1.00 & 0.74 & 0.84 
\\\hline
\red{$^*$}STAR ($W^\pm/Z$ in DY) & 7 & 0.61 & 0.89 & 0.67 & 0.61 & 0.67
\\\Xhline{5\arrayrulewidth}
\textbf{Total $A_{UT}^{\sin(\phi_h-\phi_S)}$} & 117 & 1.00 & 0.99 & 1.42 & 1.00 & 1.34
\\\Xhline{5\arrayrulewidth}
\multicolumn{7}{|c||}{$A_{LT}^{\cos(\phi_h-\phi_S)}$ asymmetry} 
\\\hline
COMPASS(2016) & 26 & 0.47 & 0.47 & 1.06 & 0.80 & 1.20
\\\hline
HERMES($\pi^\pm$) & 22 & 0.98 & 0.98 & 0.99 & 0.99 & 0.99
\\\hline
HERMES($K^\pm$) & 22 & 1.32 & 1.31 & 1.32 & 1.34 & 1.34
\\\hline
JLab & 4 & 1.09 & 1.10 & 1.19 & 1.03 & 1.13
\\\Xhline{5\arrayrulewidth}
\textbf{Total $A_{LT}^{\cos(\phi_h-\phi_S)}$} & 74 & 0.91 & 0.91 & 1.03 &  0.89 & 1.11
\\\Xhline{8\arrayrulewidth}
\textbf{Total} & 372 & 0.95 & 0.95 & 1.03 & 2.20 & 1.25
\\\hline
\end{tabular}
}

\vspace{8pt}
\small{
\red{$^*$}The STAR measurement was not included in the fitting procedure, but tested against results of the fit. For that reason, its contribution is not included in the total $\chi^2$'s.}
\end{center}
\caption{\label{tab:chi2} Drop-down of the contributions to the total $\chi^2$ by different datasets. The value of $\chi^2$ is computed for the mean replica, and thus it differs a bit from the mean value of the $\chi^2$-distribution (\ref{chi2:main}, \ref{chi2:tot}). The column $\chi^2$(central) reports the values obtained in a single minimization of unmodified central data. The three last columns show the results of the fits for reduced datasets (see sec. \ref{sec:DISvsTMD}) and the results of the null-hypothesis (see sec.\ref{sec:null}).}
\end{table}

\begin{figure}[ht]
\centering
\includegraphics[width=0.98\linewidth]{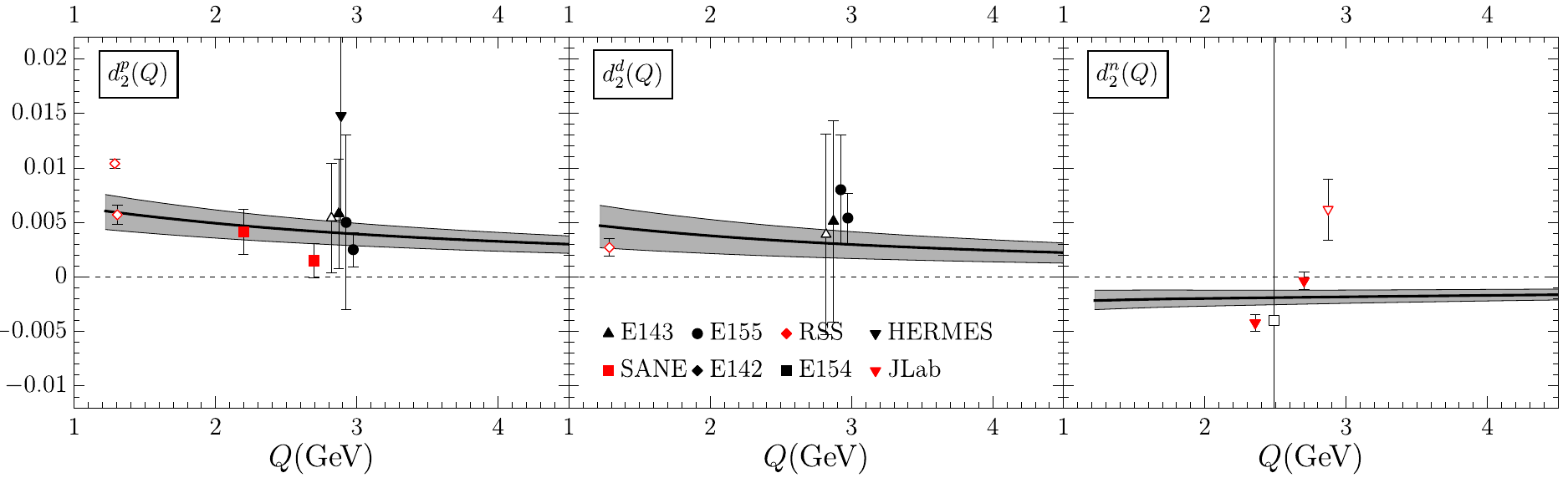}
\includegraphics[width=0.98\linewidth]{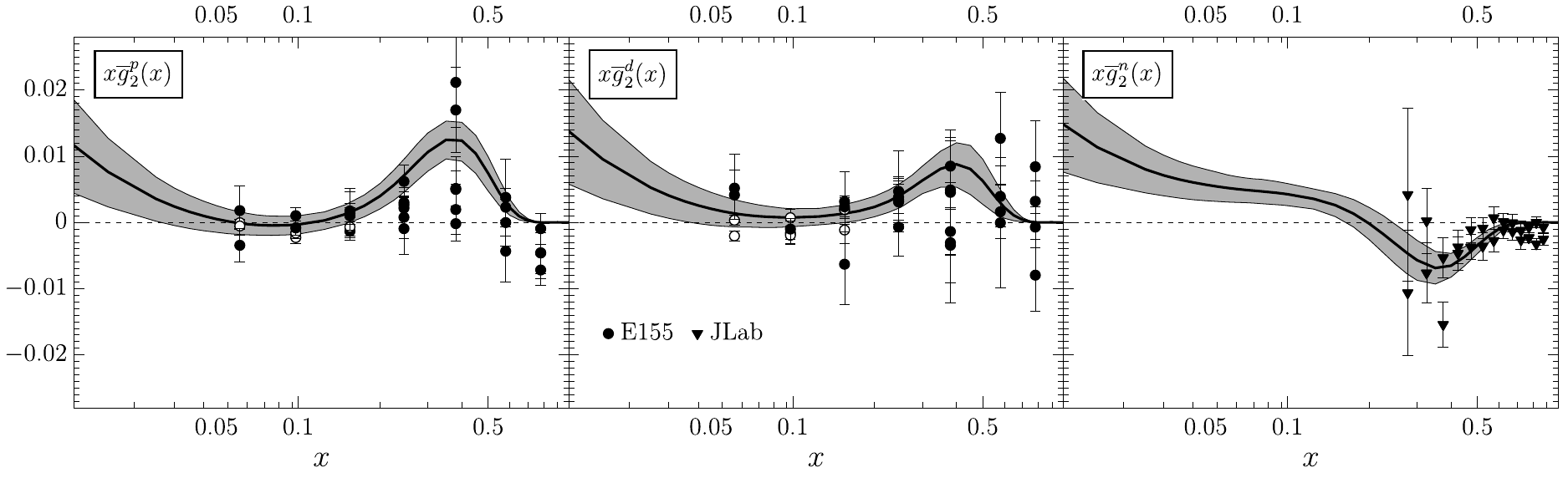}
\includegraphics[width=0.98\linewidth]{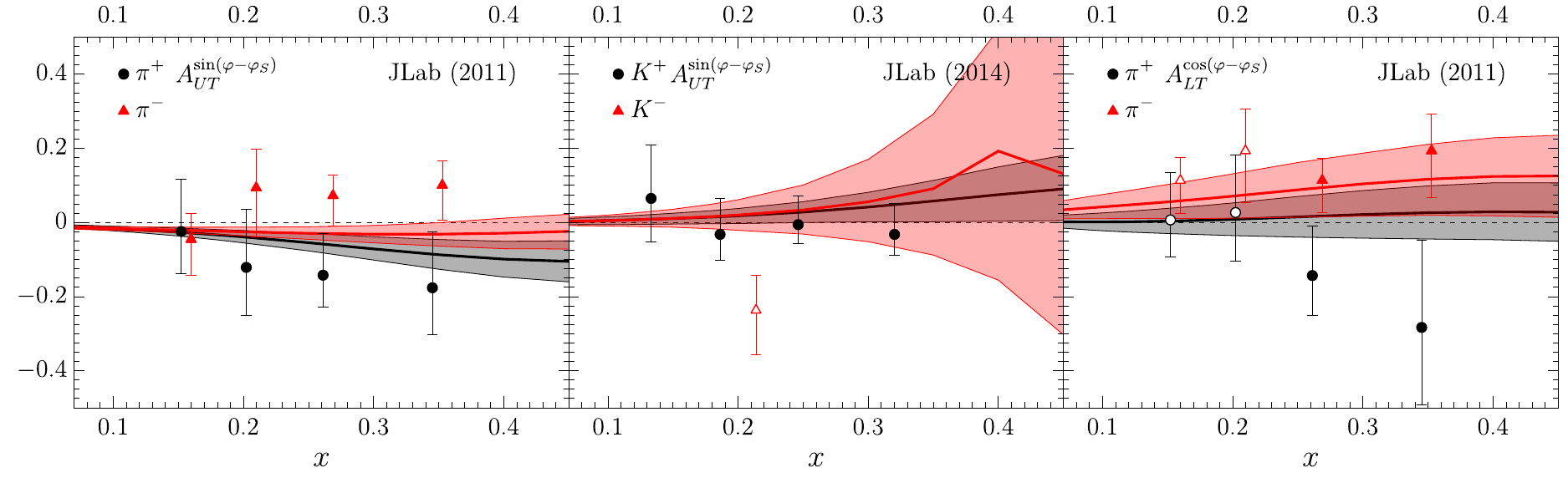}
\caption{\label{fig:d2_exp} Comparison of the theory prediction with experimental data. In all plots, filled (empty) points were included (excluded) in the fit. The upper panel shows the moment $d_2$ vs. $Q$ for different targets. The middle panel shows the structure function $\overline{g}_2$ vs.$x$ for different targets. The lower panel shows comparison of various spin-asymmetries with the data of JLab \cite{JeffersonLabHallA:2011ayy, JeffersonLabHallA:2014yxb, JeffersonLabHallA:2011vwy} vs. $x$.  Comparison with other experiments is shown in appendix fig.~\ref{fig:g2_all} - \ref{fig:HERMES4}. }  
\end{figure}

In fig.\ref{fig:d2_exp} we demonstrate the comparison of $\bar g_2$ and $d_2$ computed with our extraction against the experimental data. The lower panel of this figure demonstrates the asymmetries $A_{UT}^{\sin(\phi_h-\phi_S)}$ and $A_{LT}^{\cos(\phi_h-\phi_S)}$ in comparison with the measurements carried out by JLab. The agreement is reasonably good, and visually corresponds to the declared values of $\chi^2$. In the appendix we present the rest of the figures comparing data and theory, see fig.~\ref{fig:g2_all} - \ref{fig:HERMES4}, which include all considered datasets. 

It is interesting to observe that theoretical curves based on our extraction continue to describe the data even outside of the region of the TMD factorization theorem (i.e. for larger $p_\perp/z$). This feature has been also observed in other analyses of SIDIS data, see e.g.\cite{Moos:2025sal, Bacchetta:2022awv}. However, in all cases, the agreement becomes worse for lower $Q$. 

\subsection{Estimation of uncertainty due to $\lambda$ parameters}
\label{sec:lambdas}

The present data for TMD observables have very little sensitivity to the $p_\perp$-shape of the TMD distributions. This is because the majority of the data has effectively one-dimensional binning, integrating over two out of three main variables in a wide range. Due to this, there is only a small number of points that constrain the $p_\perp$ dynamics (e.g. 5 surviving points in COMPASS (2008) and (2016), 2 points in COMPASS (2023), etc.). Furthermore, the $p_\perp$-binned measurements usually have the largest uncertainties. For that reason, it is practically inefficient to fit parameters $\lambda_{UT}$ and $\lambda_{LT}$ together with the rest, and we have made the main fit fixing $\lambda_{UT}=\lambda_{LT}=0.5$GeV. In this section, we estimate the impact of this decision.

To study the dependence on $\lambda$'s we performed a series of central fits in a grid of fixed $(\lambda_{UT},\lambda_{LT})$'s, $(0.3,0.55) \times (0.3,0.55)$. We have found that the $\chi^2$ practically does not change in this region. The maximum difference is $\Delta \chi^2=0.01$ between the minimum and the point $(0.55,0.3)$. Fitting these results with a quadratic form, we found the Hessian matrix $H$, and the covariance matrix $\Sigma=2 H^{-1}/N_{\text{pt}}$. That allows us to estimate the uncertainty for parameters $\lambda$. We have found that
\begin{eqnarray}\label{siv:lambda_uncert}
\lambda_{UT}=0.34\pm 0.35 \text{GeV},\qquad\lambda_{LT}=0.44\pm 0.39 \text{GeV},
\end{eqnarray}
with a negligible correlation between these parameters $\text{corr}(\lambda_{UT},\lambda_{LT})=-0.004$. Note that the uncertainty band for $\lambda_{UT}$ is so large that it can be zero within it. This should be considered with caution since, close to these non-physical regimes, the profile of $\chi^2$ is most probably non-quadratic, while our estimation is done in the vicinity of the minimum.

In general, it demonstrates that the selection of $\lambda_{UT}=\lambda_{LT}=0.5$GeV does not affect the quality of our fit, although this freedom induces a systematic uncertainty to the parameters of genuine twist-three distributions. To estimate this uncertainty, we have fit parameters $a_i$, $\alpha_i^f$, $\beta_i$ as functions of $(\lambda_{UT}, \lambda_{LT})$, and estimated the systematic uncertainty as
\begin{eqnarray}
\delta \alpha_{\text{sys.}}=\sqrt{\Sigma_{11}\(\frac{\partial \alpha}{\partial \lambda_{UT}}\)^2 +\Sigma_{22}\(\frac{\partial \alpha}{\partial \lambda_{LT}}\)^2 
+2 \Sigma_{12}\frac{\partial \alpha}{\partial \lambda_{UT}}\frac{\partial \alpha}{\partial \lambda_{LT}}}.
\end{eqnarray}
The corresponding numbers are presented in the table of parameters obtained in the main fit, in brackets. This additional uncertainty appears to be rather small. With a few exceptions, it is less than $15\%$ of the uncertainty band. The main exceptions are the parameters for $\mathfrak{S}^-$ ($u$ and $d$ quarks). In these cases, the systematic uncertainty due to $\lambda$'s is very large, 60-70\%. The main contribution to it is the $\lambda_{LT}$ direction. It is natural, because among the TMD observables, only $A_{LT}$ receives a contribution from $\mathfrak{S}^-$ at leading order.

\subsection{Null-hypothesis test}
\label{sec:null}

Given the poor state of the data, and the general smallness of the signal, it is of vital importance to validate the observation of genuine twist-three distributions. To do so, we compare our results with the assumption of the absence of genuine twist-three contributions (a.k.a. the null-hypothesis). I.e. we compute the theory prediction with all $\alpha_i^f=0$ and $\beta_i=0$. In this case, we have observed the total $\chi^2/N_{\text{pt}}=1.25$, which corresponds to $\Delta \chi^2=111$ with 24 fitting parameters. The contribution of particular datasets is presented in the last column of table \ref{tab:chi2}.

The value $\chi^2/N_{\text{pt}}=1.25$ corresponds to a 3.4$\sigma$ deviation from the data, since for $N_{\text{pt}}=372$ the $1\sigma$-deviation is equivalent to an increment of $\chi^2/N_{\text{pt}}$ by $\approx 0.07$. Alternatively, treating $\Delta\chi^2=111$ as following a $\chi^2$-distribution with 24 degrees of freedom, the significance of the twist-three signal is $\approx7.2\sigma$. We consider the latter estimate too optimistic, since part of the data contains unknown systematic uncertainty, and thus $\chi^2$ is not fully reliable.

It is relevant to take into account that genuine twist-three distributions are not sign-definite, and thus, the null-hypothesis is correct for certain kinematic ranges and flavor combinations. Indeed, a large portion of sets demonstrate a very good agreement with the null-hypothesis and with the results of the fit, e.g. for $\overline{g}_2$ E143 ($0.52\to0.43$), E154 ($1.02\to 0.99$); for $A_{UT}$ COMPASS(2008) ($0.69\to 0.60$), JLab ($0.84\to 0.87$), etc.

To further probe the significance of the twist-three signal, we consider a kinematic sub-division of the data into two parts: $x>0.2$ (220 points) and $x<0.2$ (152 points). This separation is motivated by the observation that the observables $\overline{g}_2$ and $f_{1T}^\perp$ are small or vanishing for $x<0.2$. For these sub-sets we obtain
\begin{eqnarray}
x>0.2~ (N_{\text{pt}}=220):\qquad \frac{\chi^2_{\text{main}}}{N_{\text{pt}}}=1.01,\qquad \frac{\chi^2_{\text{null}}}{N_{\text{pt}}}=1.42,\qquad \Delta \chi^2=90.1,
\\\nn
x<0.2~ (N_{\text{pt}}=152):\qquad \frac{\chi^2_{\text{main}}}{N_{\text{pt}}}=0.91,\qquad \frac{\chi^2_{\text{null}}}{N_{\text{pt}}}=1.01,\qquad \Delta \chi^2=21.0.
\end{eqnarray}
Consequently, for $x>0.2$ the significance of the twist-three signal is of the order of $4.3\sigma$ (using the $\sqrt{2/N_{\text{pt}}}$ estimate as above), while for $x<0.2$ the signal is insignificant. 

We note that these numbers are only estimates, and a rigorous null-hypothesis test would require a dedicated re-fit. Still, we conclude that the signal of genuine twist-three distributions is statistically significant.

\subsection{Fit of DIS and TMD subsets independently}
\label{sec:DISvsTMD}

Other important cross-checks of the fit are independent fits of DIS (i.e. $d_2$ and $\overline{g}_2$) data, and TMD data (i.e. $A_{UT}^{\sin(\phi_h-\phi_s)}$ and $A_{LT}^{\cos(\phi_h-\phi_s)}$). Such a fit reflects how important the inclusion of each process in the fit is, and what impact one observable has on the others.

To make these tests we have performed two additional extractions (with 150 replicas each):
\begin{itemize}
\item \textbf{DIS only fit}. This fit includes only data for $d_2$ and $\overline{g}_2$ (with a total of 181 data points). In this case, one has no sensitivity to the functions $\mathfrak{S}^-$ and $T_{3F}^-$, since they do not contribute to DIS. Therefore, we neglected these functions entirely, by setting parameters $\alpha_1^f=\alpha_3^f=\beta_0=0$.
\item \textbf{TMD only fit}. This fit includes only data related to SIDIS, i.e. $A_{UT}^{\sin(\phi_h-\phi_s)}$ and $A_{LT}^{\cos(\phi_h-\phi_s)}$ (with 191 data points).
\end{itemize}

\begin{figure}[t]
\centering
\includegraphics[width=0.75\linewidth]{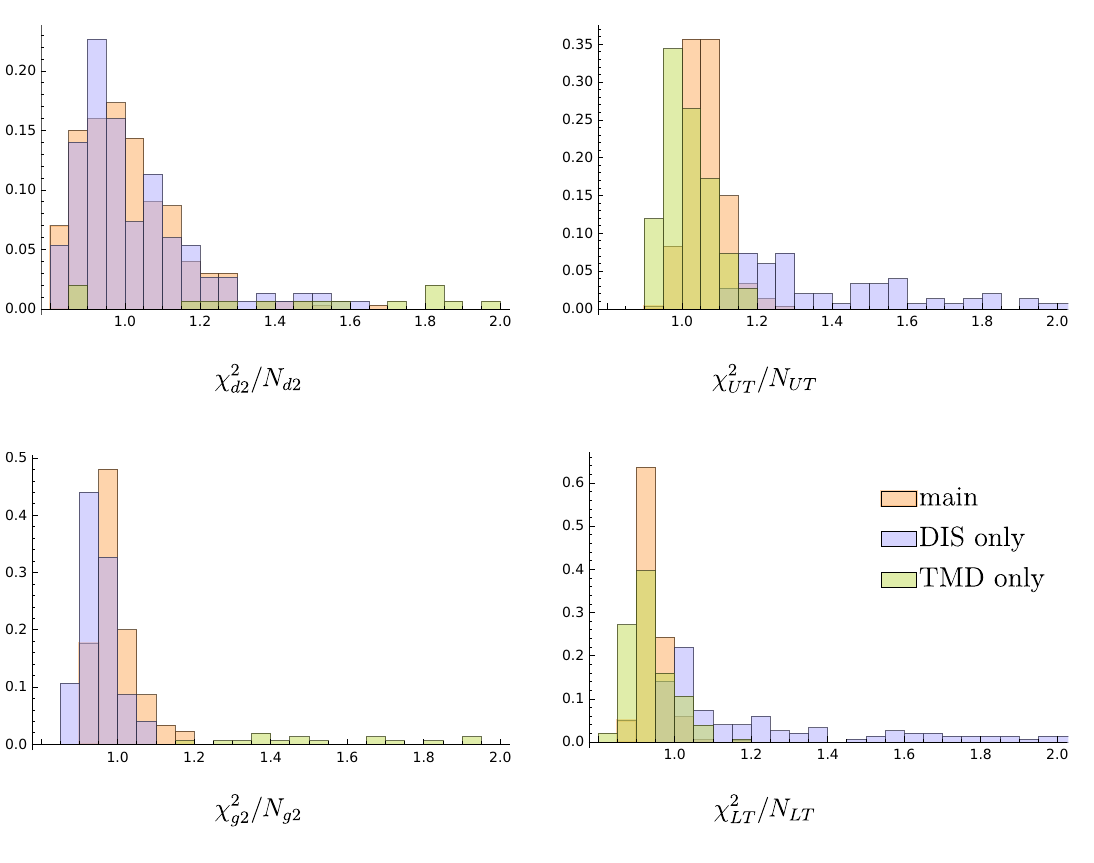}
\caption{\label{fig:chi2_together} Distribution of $\chi^2/N$ for particular observables for different kinds of fits. All distributions are normalized to unity.}
\end{figure}

The distributions of $\chi^2$ obtained in these fits are shown in fig.~\ref{fig:chi2_together}. These distributions have the following characteristics:
\begin{eqnarray}\label{chi2:disonly}
&&\textbf{DIS only:}
\\&&\nn
\frac{\chi^2_{d2}}{N_{d2}}=1.02^{+0.13}_{-0.13},\quad
\frac{\chi^2_{g2}}{N_{g2}}=0.95^{+0.04}_{-0.04},\quad
\frac{\chi^2_{UT}}{N_{UT}}=8.17^{+1.42}_{-6.93},\quad
\frac{\chi^2_{LT}}{N_{LT}}=1.78^{+0.29}_{-0.78},
\\\label{chi2:tmdonly}
&&\textbf{TMD only:}
\\\nn &&
\frac{\chi^2_{d2}}{N_{d2}}=47.2^{+39.7}_{-44.7},\quad
\frac{\chi^2_{g2}}{N_{g2}}=32.1^{+21.1}_{-29.2},\quad
\frac{\chi^2_{UT}}{N_{UT}}=1.01^{+0.06}_{-0.06},\quad
\frac{\chi^2_{LT}}{N_{LT}}=0.94^{+0.06}_{-0.05}.
\end{eqnarray}
Evidently, the fit of DIS/TMD data only produces a good result in its own sector, but leaves other processes with rather undetermined values of $\chi^2$. It is important to observe that the central values of the parameters are not that significantly different in both fits, see tables below. Therefore, the $\chi^2$ computed at the central replica (see table \ref{tab:chi2}) produces reasonable values for both types of measurements.

The values of parameters extracted in both fits are given in the following tables;
\begin{center}
\renewcommand{\arraystretch}{2.}
\renewcommand\theadset{\def\arraystretch{2.}}
\begin{tabular}{||l|c|c||}
\hline
\multicolumn{3}{||c||}{Values of parameters in the DIS-only fit}
\\\Xhline{5\arrayrulewidth}
\hline 
common & \multicolumn{2}{|c||
}{$a_0=-2.26^{+1.01}_{-1.14}\quad a_1=6.7^{+2.4}_{-2.9}\quad a_2=1.64^{+0.82}_{-1.01}\quad a_3=6.2^{+2.2}_{-2.7}$ } 
\\\Xhline{5\arrayrulewidth}
& $\mathfrak{S}^+$ & $\mathfrak{S}^-$\\
\hline
$u$-quark & \makecell[c]{
$\alpha_0^u=1.1^{+1.2}_{-1.5}\quad \alpha_{11}^u=4.^{+15.}_{-16.}$\\ $\alpha_{13}^u=19.^{+17.}_{-14.}\quad \alpha_{33}^u=-1.0^{+8.9}_{-10.5}$}
&
$\alpha_1^u=0.\quad \alpha_{3}^u=0.$
\\\hline
$d$-quark & \makecell[c]{
$\alpha_0^d=-0.6^{+2.1}_{-1.8}\quad \alpha_{11}^d=-3.^{+14.}_{-12.}$\\ $\alpha_{13}^d=-12.^{+12.}_{-16.}\quad \alpha_{33}^d=-5.3^{+6.3}_{-9.7}$}
&
$\alpha_1^d=0.\quad \alpha_{3}^d=0.$
\\\hline
$s$-quark & \makecell[c]{
$\alpha_0^s=0.3^{+2.7}_{-3.3}\quad \alpha_{11}^s=16.^{+24.}_{-20.}$\\ $\alpha_{13}^s=-1.^{+25.}_{-21.}\quad \alpha_{33}^s=13.^{+23.}_{-16.}$}
&
$\alpha_1^s=0.\quad \alpha_{3}^s=0.$
\\\Xhline{5\arrayrulewidth}
& $T_{3F}^+$ & $T_{3F}^-$ \\\hline
gluon & $\beta_1=5.9^{+6.2}_{-5.6}$
&
$\beta_0=0.$
\\\hline
\end{tabular}
\end{center}

\begin{center}
\renewcommand{\arraystretch}{2.}
\renewcommand\theadset{\def\arraystretch{2.}}
\begin{tabular}{||l|c|c||}
\hline
\multicolumn{3}{||c||}{Values of parameters in the TMD-only fit}
\\\Xhline{5\arrayrulewidth}
\hline 
common & \multicolumn{2}{|c||
}{$a_0=-1.39^{+0.84}_{-0.60}\quad a_1=8.0^{+2.0}_{-2.8}\quad a_2=1.16^{+0.36}_{-1.05}\quad a_3=8.1^{+1.9}_{-2.7}$ } 
\\\Xhline{5\arrayrulewidth}
& $\mathfrak{S}^+$ & $\mathfrak{S}^-$\\
\hline
$u$-quark & \makecell[c]{
$\alpha_0^u=0.7^{+1.5}_{-1.8}\quad \alpha_{11}^u=10.^{+17.}_{-16.}$\\ $\alpha_{13}^u=9.^{+16.}_{-11.}\quad \alpha_{33}^u=9.^{+16.}_{-11.}$}
&
$\alpha_1^u=-1.0^{+2.6}_{-2.8}\quad \alpha_{3}^u=3.0^{+4.1}_{-3.9}$
\\\hline
$d$-quark & \makecell[c]{
$\alpha_0^d=-2.4^{+2.8}_{-2.9}\quad \alpha_{11}^d=-22.^{+22.}_{-25.}$\\ $\alpha_{13}^d=0.^{+22.}_{-25.}\quad \alpha_{33}^d=28.^{+20.}_{-22.}$}
&
$\alpha_1^d=2.7^{+3.6}_{-3.9}\quad \alpha_{3}^d=-8.9^{+8.8}_{-10.6}$
\\\hline
$s$-quark & \makecell[c]{
$\alpha_0^s=1.1^{+2.9}_{-2.4}\quad \alpha_{11}^s=0.^{+32.}_{-31.}$\\ $\alpha_{13}^s=5.^{+32.}_{-33.}\quad \alpha_{33}^s=-4.^{+28.}_{-30.}$}
&
$\alpha_1^s=8.3^{+10.4}_{-7.6}\quad \alpha_{3}^s=1.3^{+7.8}_{-7.1}$
\\\Xhline{5\arrayrulewidth}
& $T_{3F}^+$ & $T_{3F}^-$ \\\hline
gluon & $\beta_1=-4.7^{+6.7}_{-7.1}$
&
$\beta_0=2.3^{+3.8}_{-3.7}$
\\\hline
\end{tabular}
\end{center} 

The visual comparison of the parameters is presented in fig.\ref{fig:parameter_comparison} in the appendix. As one can see, the fits agree with each other very well, except for the parameters $\alpha_{33}^d$, and $\beta_1$ (their disagreement is of the order of 1$\sigma$). This again strongly confirms the universality of twist-three functions and the factorization approach simultaneously. 

It implies that one can determine both the Sivers and worm-gear-T distributions rather precisely by considering only DIS data (see the upper panels of fig.~\ref{fig:sivers} and fig.~\ref{fig:wgt_comparison}). However, such a determination has a larger uncertainty, as the central values are sufficiently close to each other. Note, however, that such determination misses the contributions of $\Delta T$ and $T_{3F}^-$. Conversely, one can consider only TMD data and reasonably reconstruct the structure function $\bar g_2$ and the $d_2$ moment, see fig.~\ref{fig:g2_collection}(upper panel). 

The fact that the distribution of $\chi^2$ is very wide for the non-fitted data, together with a similar value for the parameters, implies that the correlation between these parameters is very different for different cases. This is exactly the theoretically expected behavior. Namely, both types of observables are given by different integrals of a single function, therefore, the average value of the function is the same, but the correlations are fit differently. 

From this test it is evident that both types of processes do not contradict each other and allow for a simultaneous and consistent description within the framework of twist-three distributions. It directly confirms the universality of the QCD factorization approach and the twist-three distributions. Moreover, both types of processes complement each other and help stabilize the values of twist-three distributions by adding internal correlations between parameters. 

This is, probably, one of the main conclusions of this work, and it invites us to extend this direction of studies by adding more kinds of observables, and increasing the precision of the extracted twist-three distributions.

\subsection{Comparison with other extractions}
\label{sec:comparison-with-others}

The extraction presented in this work is the first of its kind. Nonetheless, we can compare with the extractions of particular one-dimensional distributions considered in earlier works.

\begin{figure}[t]
\centering
\includegraphics[width=0.8\linewidth]{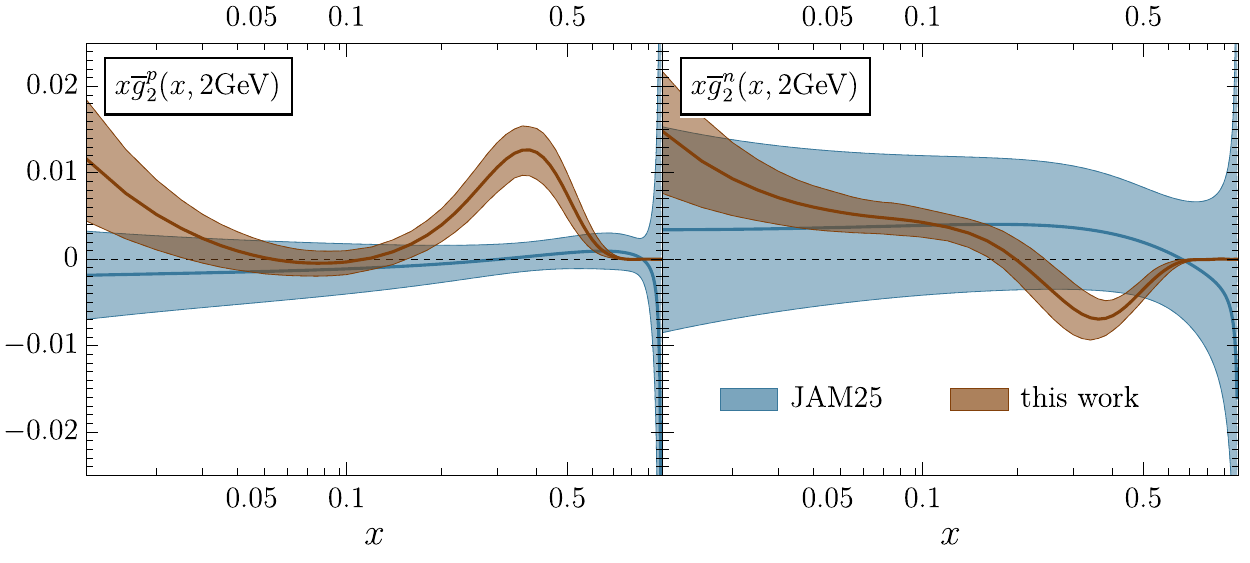}
\caption{\label{fig:g2_JAM} Comparison of the $\bar g_2$ function extracted in this work and the "higher-twist" contribution of $g_2$ determined in ref.~\cite{Cocuzza:2025qvf}.}
\end{figure}

In fig.\ref{fig:g2_JAM} we compare the value of the $\bar g_2$ structure function obtained in this work with the results of the analysis in ref.~\cite{Cocuzza:2025qvf}. Note that in ref.~\cite{Cocuzza:2025qvf} the structure function $g_2$ is approximated by its WW term (\ref{g2:WW}), and used to determine the helicity PDFs. The curve demonstrated in fig.~\ref{fig:g2_JAM} represents the uncertainty of the missed "higher-twist" contribution (i.e. the $\bar g_2$ part). In these circumstances, the disagreement between our extraction and ref.~\cite{Cocuzza:2025qvf} is reasonable.

\begin{figure}
\centering
\includegraphics[width=0.45\linewidth]{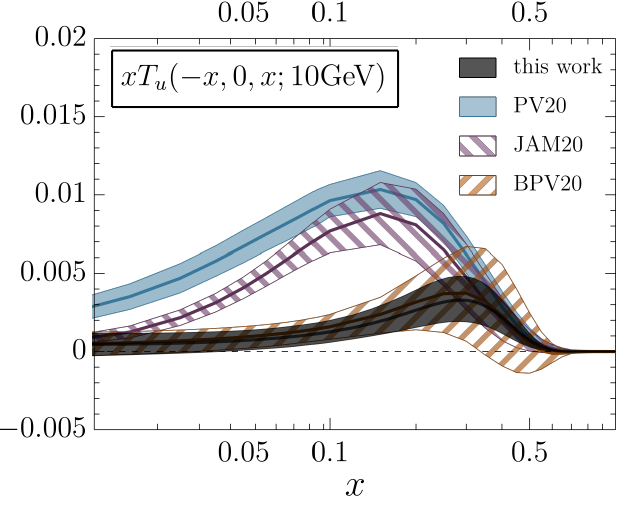}
\includegraphics[width=0.45\linewidth]{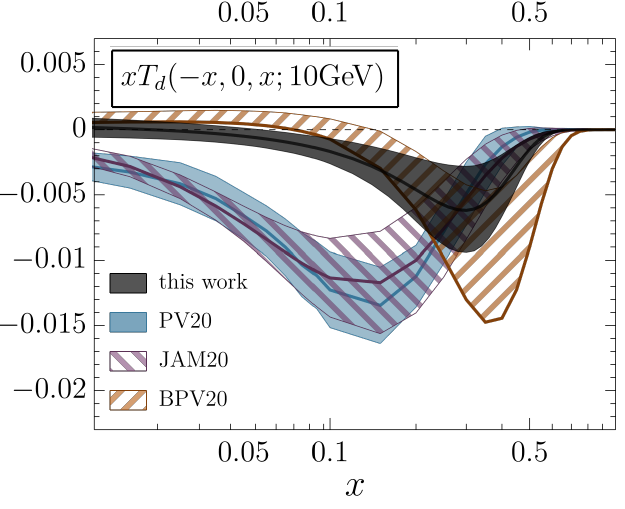}
\caption{\label{fig:QS_comparison} Comparison of the $T(-x,0,x)$ distribution extracted in this work with extractions made by the fit of the Sivers function in PV20 \cite{Bacchetta:2020gko}, JAM20 \cite{Cammarota:2020qcw}, BPV20 \cite{Bury:2021sue}. The comparison is done for $u$- and $d$-flavors, which correspond to the $x>0$ part, and at $\mu=10$GeV.}
\end{figure}

Another popular twist-three matrix element is the Qiu-Sterman function. It is determined in the fits of the Sivers function via the relation (\ref{TMD:sivers-smallb-1}), such as in refs.~\cite{Bacchetta:2011gx, Echevarria:2014xaa, Boglione:2018dqd, Bacchetta:2020gko, Cammarota:2020qcw, Bury:2020vhj, Bury:2021sue}. In fig.~\ref{fig:QS_comparison} we compare our extraction to the results of the most recent works \cite{Cammarota:2020qcw,  Bacchetta:2020gko, Bury:2021sue}.

Note that different groups use different notations. In order to produce the plot \ref{fig:QS_comparison}, we utilize the following relations. They are deduced by comparing our definitions with eqn.~(1) of ref.\cite{Cammarota:2020qcw} and  eqn.~(19) of ref.~\cite{Bacchetta:2020gko}:
\begin{eqnarray}
f_{1T,q}^{\perp(1)}(x,\mu)\Big|_{\text{PV20}}&=&-\pi T_f(-x,0,x;\mu),
\\
f_{1T,q}^{\perp(1)}(x,\mu)\Big|_{\text{JAM}}&=&-\frac{\pi}{2} T_f(-x,0,x;\mu).
\end{eqnarray}
The definition of the Qiu-Sterman function in ref.~\cite{Bury:2021sue} coincides with ours. The comparison shows a spectacular agreement with the results of BPV20 \cite{Bury:2020vhj, Bury:2021sue}. Note that the fit of BPV20 uses a different fitting ansatz, a different unpolarized input, does not implement the twist-three evolution, and is based only on the analysis of the Sivers asymmetry. Agreement with PV20 and JAM20 is worse. Most probably it is because in these works the Qiu-Sterman function is assumed to be proportional to the unpolarized PDF, and to evolve by the same kernel, which is incorrect.  Even so, for the $x\sim 0.3-0.8$ range, where most of the data is presented, we find a good agreement.

The comparison for the worm-gear-T function is not that straightforward, and we present it in sec.~\ref{sec:result-wgt}, together with additional details on the procedure of comparison. The comparison is done with earlier results from refs.~\cite{Bhattacharya:2021twu, Horstmann:2022xkk}, and demonstrates a very good agreement, see fig.~\ref{fig:wgt_comparison}(lower panel).

It is also instructive to compare with the preliminary results of the current extraction presented in ref.~\cite{Vladimirov:2025qrh}. In the preliminary study, we considered a smaller dataset and a simpler ansatz (the parameters $\alpha_{11}^f$ and $\alpha_{33}^f$ were absent and $a_1=a_3$). This led to a somewhat worse description of the data, and a more biased shape. Still, we found that the extracted functions are in a very good agreement. The main changes happen in the "anti-quark" region (which is smaller in this analysis), in the parameter $a_2$ (which is larger in this analysis), with smaller uncertainty bands, and better data description.

\section{Results} 
\label{sec:results}

This section is devoted to the discussion of the values and shape of the extracted genuine twist-three distributions, and related observables. The data used to make the plots presented in this article, such as predictions for observables, actual values of twist-three distributions at various scales, etc, is also collected in the public repository \cite{REP}.

The section is organized as follows. In sec.~\ref{sec:genuineTW3} we present the three-dimensional shape of the extracted twist-three PDFs and discuss their main features. Sec.~\ref{sec:d2+force} discusses the $d_2$ moment and its interpretation as an average color force, while sec.~\ref{sec:g2-result} examines the $\overline{g}_2$ structure function, including its comparison with the Wandzura-Wilczek contribution and implications for future experiments. The Sivers function is discussed in secs.~\ref{sec:result-Sivers} and \ref{sec:result-DY+sign}, the latter focusing on the sign-change test using Drell-Yan data from STAR. The worm-gear-T function is presented in sec.~\ref{sec:result-wgt}. Finally, secs.~\ref{sec:nucleon-tomography}--\ref{sec:burkardt} discuss derived quantities: the tomographic picture of the nucleon, the positivity bound, and the average transverse momentum of partons together with the Burkardt sum rule.

\begin{figure}[t]
\centering
\includegraphics[width=0.8\linewidth]{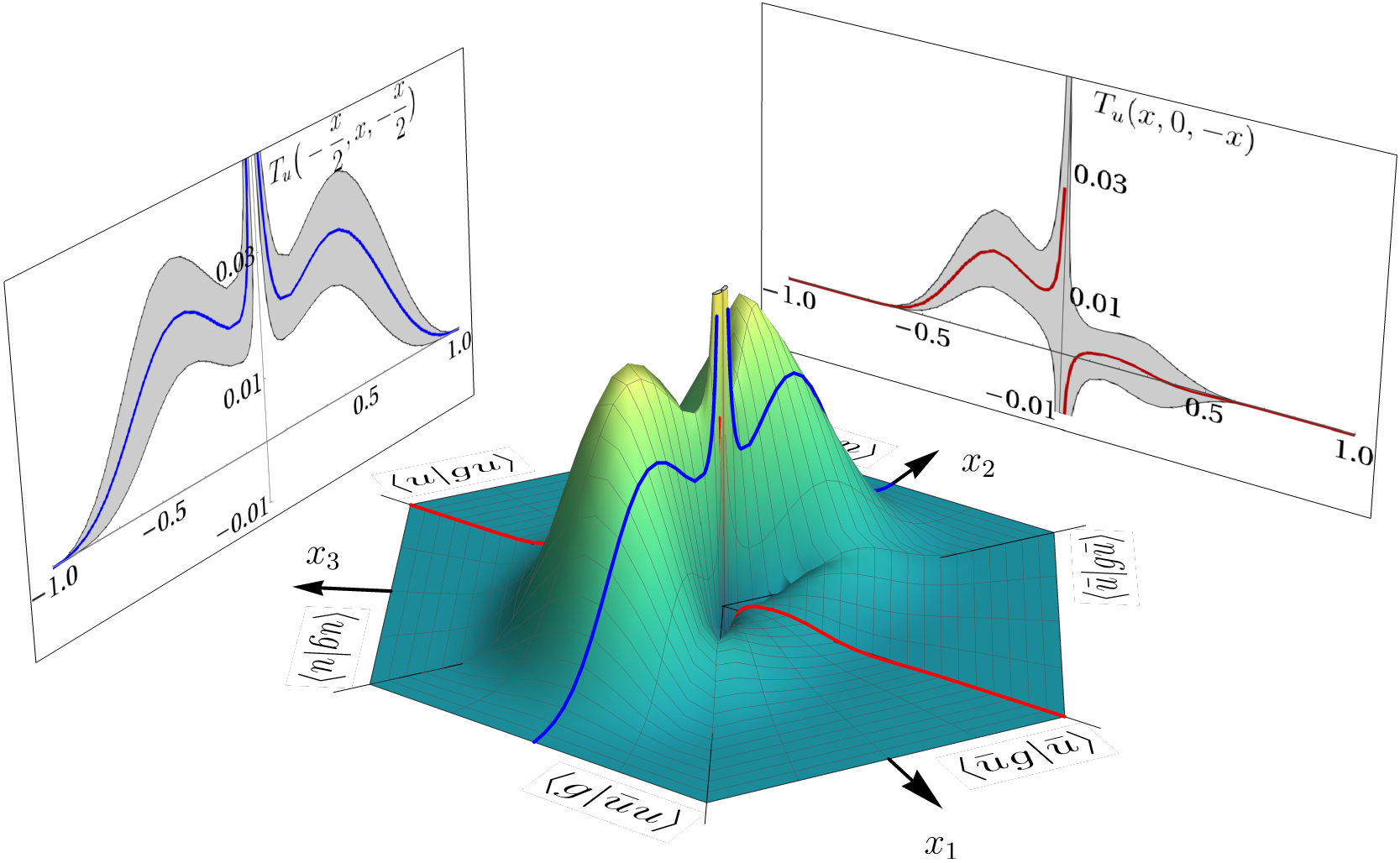}
\includegraphics[width=0.8\linewidth]{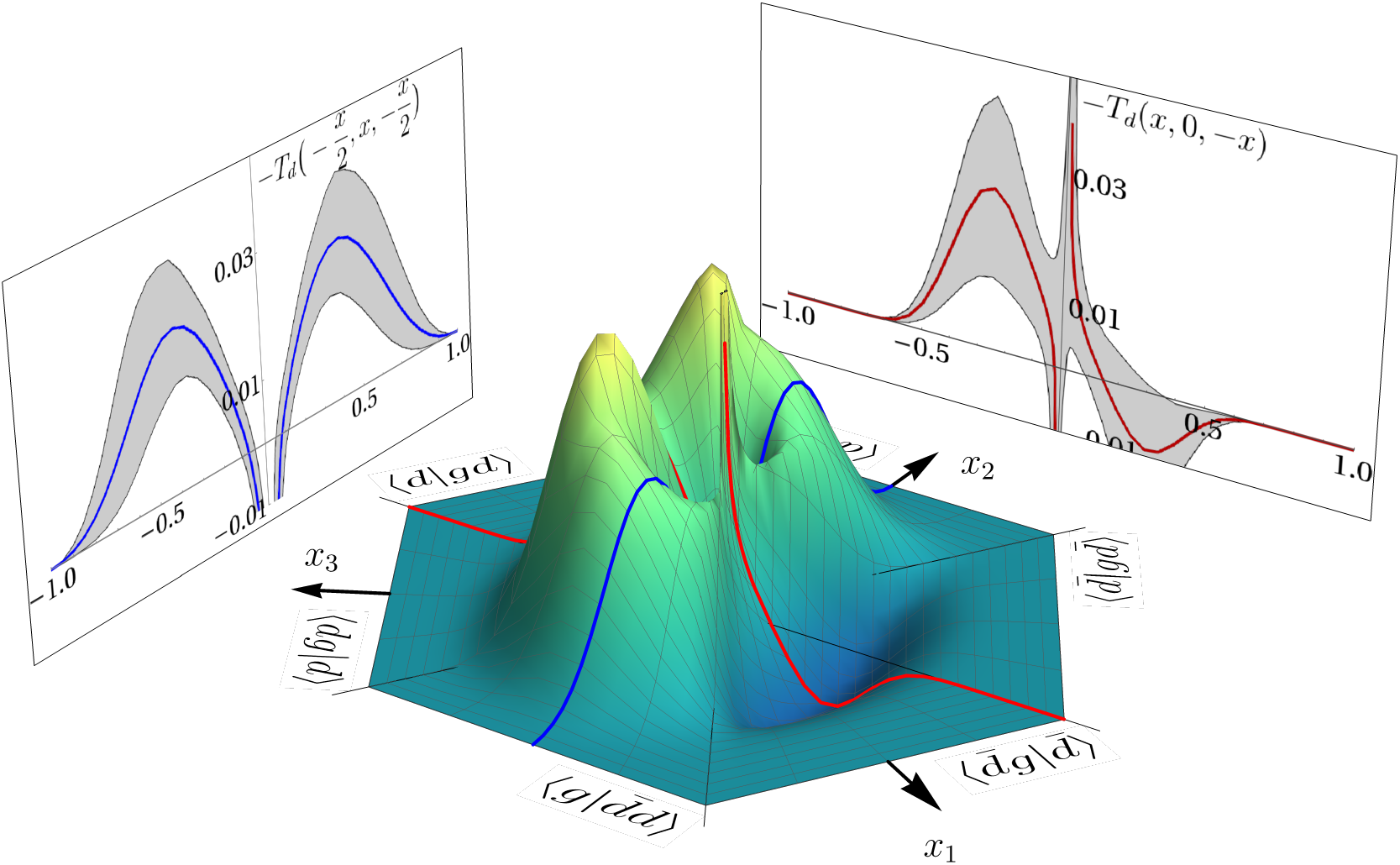}
\caption{\label{fig:T_main} Genuine twist-three PDFs $T_u(x_1,x_2,x_3)$ and $T_d(x_1,x_2,x_3)$ at $\mu=4$GeV. The three-dimensional plots showcase the mean value. The uncertainties can be understood from the sections $x_2=0$ (the Qiu-Sterman function) and $x_1=x_2$ shown in their corresponding panels. The function $T_d$ is multiplied by $(-1)$ for clearer interpretation.}
\end{figure}

\subsection{Genuine twist-three distributions}
\label{sec:genuineTW3}

In fig.~\ref{fig:T_main}, we present the value of the extracted genuine twist-three distributions $T$ for $u$ and $d$ quarks. Similar plots for the $s$-quark, gluon and functions $\Delta T$ are given in fig.~\ref{fig:AllDistributions} in the appendix. We also provide the plots of the C-definite distributions $\mathfrak{S}^\pm$ in fig.~\ref{fig:AllDistributions_C}.

These functions demonstrate a non-trivial structure. In particular, one can see that the functions $T$ are essentially asymmetric with respect to the $x_2$ axis. The "quark" part (towards $x_3>0$) is much larger and has a definite-sign preference, while the "anti-quark" part (towards $x_1>0$) is compatible with zero within the 68\%CI. Note that the dominance of the "quark" over the "anti-quark" part is a feature observed in the model computation of ref.~\cite{Braun:2011aw}. We have also found that $T_u\approx -T_d$, with $T_u$ being positive over the "quark" region. This is a well-known feature observed in various computational models, see for instance \cite{Bacchetta:2003rz, Yuan:2003wk, Bacchetta:2008af, Pasquini:2008ax, Pasquini:2010af, Braun:2011aw, Maji:2017wwd, He:2019fzn}. Nonetheless, apart from ref.~\cite{Braun:2011aw}, all model computations were done for the Qiu-Sterman matrix element, and do not provide information about the two-dimensional distribution.

We also observe that the $T_s$ distribution shows a significant asymmetry, resembling, in general, the shape of the $d$ distribution (reverted by $x_1\leftrightarrow x_3$). So, the "anti-quark" part of $T_s$ is larger and, generally, the same size as the "quark" part of $T_u$ and $T_d$, while its "quark" part is compatible with zero. For the moment, we do not draw any definitive conclusions from this interesting behavior, since it could be a result of tension between data. Particularly, it may emerge from a large value of $\chi^2$ in $K^\pm$ production measured by HERMES.

All distributions $T$ possess a singularity at $\|x\|\to0$, which, loosely speaking, corresponds to the small-$x$ limit. This singularity is not sign-definite, and depends on the direction of the limit. Along the $x_2=0$ line, the sign of the singularity depends on the flavor. Meanwhile, along the $x_1=x_3$ line, it goes to positive infinity for all distributions. 

The appearance of a singularity is a feature of the evolution equations, because the boundary ansatz has a zero at $\|x\|=0$, and this zero is of order $\sim\|x\|^{3.4}$. We have found that an increase of the power of said zero (up to $a_0\sim 5$) does not eliminate the singularity, although the symmetrization of the ansatz does reduce it. The power of singularity seems to be stable with growth of the scale, meanwhile the coefficient in-front of it slowly grows with the scale, see fig.~\ref{fig:T_evolution}. In some sense, this singularity presents a universal feature of genuine twist-three distributions, which is weakly dependent on the initial ansatz. Nonetheless, since most of the data included in the present fit constrains the large-$x$ region, and the singularity becomes numerically relevant only at high energies (being generated by evolution), the extracted distributions are only weakly sensitive to this effect in practice. This singularity should be studied in more detail in future works.

\begin{figure}[t]
\centering
\includegraphics[width=0.4\linewidth]{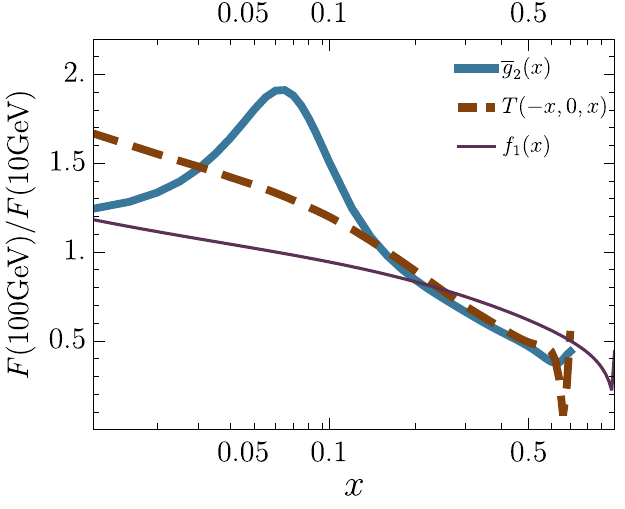}
\caption{\label{fig:evolution_comparison} Comparison of the evolution effects for $f_1$, $\overline{g}_2$, and the Qiu-Sterman function $T(-x,0,x)$. Curves demonstrate the ratio of functions at $\mu=100$GeV and $\mu=10$GeV for $u$-quark distributions (central values). At large $x$ the curves are truncated because, in this regime, the ratio is unstable due to vanishingly small distributions.}
\end{figure}

The evolution effects play an important role in the description of twist-three distributions. In contrast to twist-two PDFs, the effect of the evolution is more pronounced and strongly depends on the three-dimensional shape of the distributions. In fig.~\ref{fig:evolution_comparison}, we compare the ratios of evolved distributions at $100$ and $10$ GeV for the $u$ quark for the unpolarized PDF $f_1$, $g_2$, and the Qiu-Sterman function. It is important to emphasize that, the evolution of the Qiu-Sterman function is very different from the evolution of the unpolarized PDF, contrary to the common assumption, like that used in refs.~\cite{Bacchetta:2020gko, Cammarota:2020qcw, Anselmino:2008sga}. Furthermore, we include the explicit plot that demonstrates the evolution effects for the $\overline{g}_2$, Qiu-Sterman and worm-gear-T functions in fig.~\ref{fig:g2_QS_evolution} of the appendix.

The gluon distributions $T_{3F}^\pm$ are very poorly determined because there are no data that constrain them directly. The present determination is indirect via the coupling to quark distributions through the evolution equations. Still, we can determine the sign of these distributions and estimate their size, which appeared to be of the same order as the quark distributions. Note that for the $T_{3F}^+$ distribution we observe an unrealistically small uncertainty band. We associate this behavior to the over-constrained parametrization of the gluon distributions, in particular, to the equality in the small-$x$ behavior between the quark and gluon cases.

\begin{figure}[t]
\centering
\includegraphics[width=0.98\linewidth]{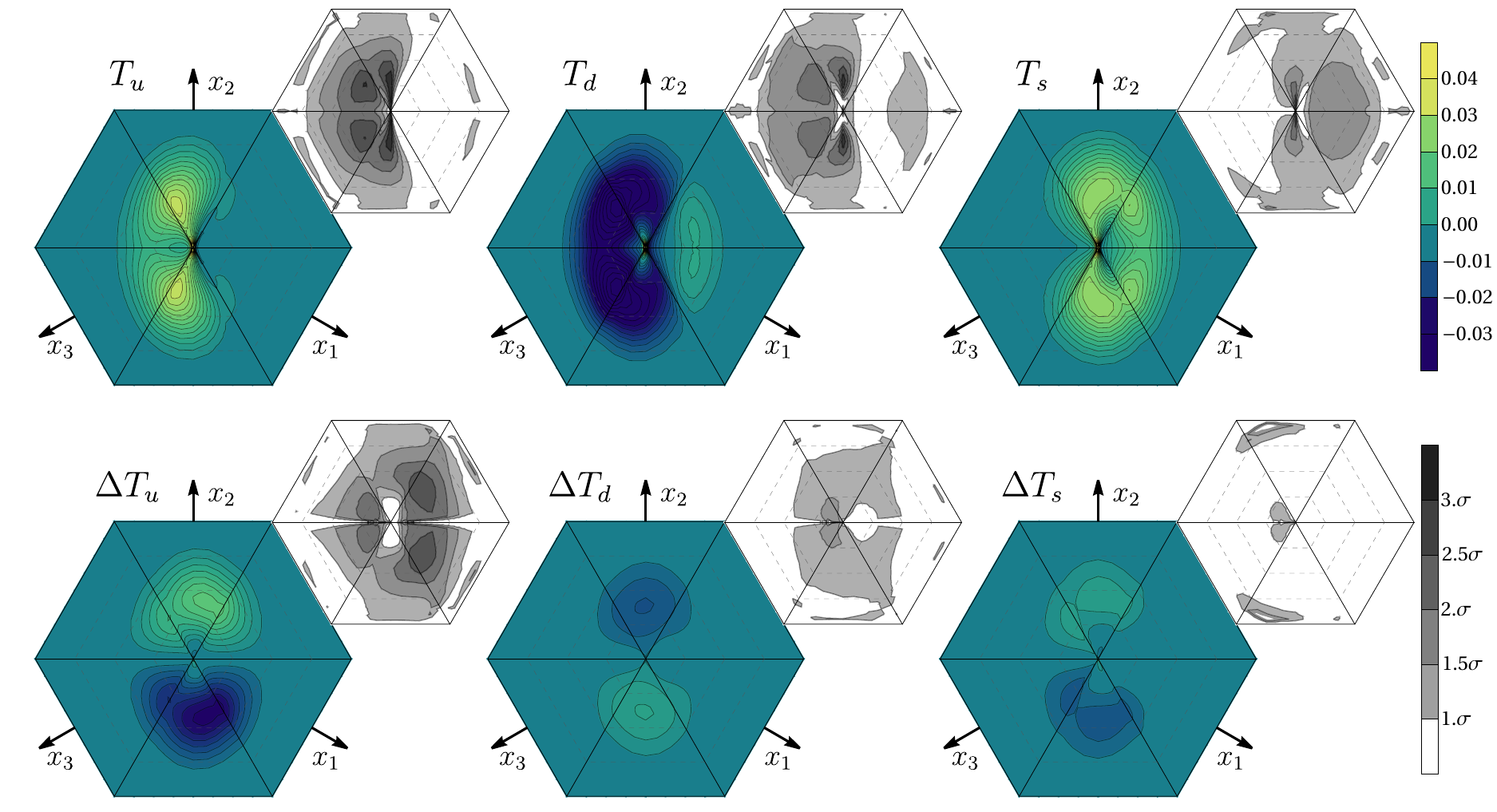}
\caption{\label{fig:hexagons} Contour plots of genuine twist-three PDFs $T$ and $\Delta T$ at $\mu=4$GeV for $u$, $d$ and $s$ quarks. The smaller hexagon demonstrates the contour plot of the uncertainty band, namely the $Z$-score of the distribution with respect to the zero-value (\ref{def:Z-score}).}
\end{figure}

In fig.~\ref{fig:hexagons} we show the density plots of the $T$ and $\Delta T$ distributions. In the smaller hexagons we show the size of the uncertainty. Namely, we plot the density of the $Z$-score with respect to the zero-value. For a function $t$ it is defined as
\begin{eqnarray}\label{def:Z-score}
Z(t)=\frac{2|t_0|}{t_{\text{max}}-t_{\text{min}}},
\end{eqnarray}
where $t_0$ is the mean value of the distribution, and $t_{\text{max}}$ and $t_{\text{min}}$ are the upper and lower boundaries of its 68\%CI respectively. I.e. this value indicates how many $\sigma$'s the value of the distribution deviates from zero. The most precise signals are found for the $T_u$ and $T_d$ distributions in the "quark" part, and for $\Delta T_u$ and $T_s$ in the "anti-quark" part. At some points the deviation from zero reaches a $3\sigma$ level, but in general it is of the order of $1-2\sigma$'s.

\subsection{The $d_2$ moment and the color force acting on quark}
\label{sec:d2+force}

Values of the $d_2$ moment and their dependence on $Q$ are shown in fig.~\ref{fig:d2_exp} together with experimental data. While there is good agreement with data, the agreement with lattice measurements is questionable, see fig.~\ref{fig:d2_lattice}.

Experimentally, the $d_2$ moment is obtained by integrating the structure function $g_2$ (\ref{d2:via_g2}) over the finite range of $x$,
\begin{eqnarray}
d_2^{\text{cut}}(Q)=3\int_{x_{\text{min}}}^{x_{\text{max}}}dx x^2\overline{g}_2(x,Q),
\end{eqnarray}
and adding the corresponding systematic uncertainty. For example, HERMES \cite{HERMES:2011xgd} uses a lower cut $x_{\text{min}}=0.023$, E155 \cite{E155:2002iec} has $x_{\text{min}}=0.02$, JLab and SANE \cite{JeffersonLabHallA:2014gzr, SANE:2018pwx} have $x_{\text{min}}\simeq0.25$. Also, many experiments have upper cuts around $x_{\text{max}}\simeq 0.8-0.9$. Having at our disposal the function $\overline{g}_2$ in a wide range, we can estimate the deviation from the complete integral due to these cuts. In the following table we present the typical values:
\begin{center}
\renewcommand{\arraystretch}{1.5}
\begin{tabular}{|c|c|c|c|c|c|c|} \hline
 $(x_{\text{min}}, x_{\text{max}})$    &  $(0.02,1)$ &  $(0.1,1)$ & $(0.2,1)$  &  $(0,0.9)$ &  $(0,0.8)$ & $(0.02,0.9)$ \\ \hline
$\Ds 1-d_2^{\text{cut}}/d_2$ & 0.19\% & 0.30\% & 2.09\% & -0.012\% & 0.010\% & 0.18\% \\\hline
\end{tabular}
\end{center}
Here, the computation is done with the central replica for the proton at $Q=2$GeV. Clearly, the impact of cuts is very small, because most of the integrand $x^2g_2$ is concentrated within the range $0.2\lesssim x\lesssim 0.5$.

The $d_2$ moment has a semi-classical interpretation as the average transverse color Lorentz force acting on the struck quark \cite{Burkardt:2008ps}, which is defined as\footnote{
Although the operator definition of the color-force is the same across the literature, there is a discrepancy in its relation to $d_2$. Here we use the definition based on the normalization (\ref{d2:operator}).
}
\begin{eqnarray}
\langle \vec f^i \rangle=-\frac{1}{\sqrt{2}P^+}\langle P,S|g\bar q(0)\gamma^+F^{+i}(0)q(0)|P,S\rangle =2\sqrt{2}M P^+ \epsilon^{ij}_TS_j d_2,
\end{eqnarray}
for the proton with spin-vector $S_\mu$. For a slow proton $P_+\simeq M/\sqrt{2}$ and the force is
\begin{eqnarray}\label{def:force}
\langle \vec f^i \rangle=2\epsilon^{ij}_TS_j M^2 d_2.
\end{eqnarray}

In table~\ref{tab:force}, we summarize the values obtained for $d_2$ and $\langle \vec f^x \rangle$ at various energies and for several quark flavors. As mentioned, the values of $d_2$ agree well with the data. However, it seems generally smaller than model computations \cite{Liu:2025ypg}. The forces acting on the $u$ and $d$ quarks are almost equal to each other and opposite in sign, in agreement with expectations \cite{Burkardt:2008ps}.

\begin{table}[h]
\begin{center}    
\renewcommand{\arraystretch}{1.5}
\begin{tabular}{||c|c||c||c||}
\hline
\multicolumn{4}{||c||}{$d_2$ moment ($\times 10^3$)} 
\\\Xhline{5\arrayrulewidth}
parton & $\mu=2$GeV    & $\mu=10$GeV & $\mu=50$GeV 
\\\hline
$u$     & $9.1_{-2.8}^{+2.9}$ & $5.3_{-1.8}^{+1.7}$ & $3.7_{-1.3}^{+1.3}$
\\\hline
$d$     & $-9.4_{-2.6}^{+2.6}$ & $-6.0_{-1.7}^{+1.6}$ & $-4.4_{-1.2}^{+1.2}$
\\\hline
$s$     & $12.1_{-7.2}^{+8.2}$ & $7.1_{-4.3}^{+4.9}$ & $5.0_{-3.0}^{+3.5}$
\\\hline
proton     & $4.3_{-1.2}^{+1.1}$ & $2.3_{-0.6}^{+0.6}$ & $1.5_{-0.4}^{+0.4}$
\\\hline
neutron     & $-1.9_{-0.6}^{+0.7}$& $-1.4_{-0.4}^{+0.4}$ & $-1.2_{-0.3}^{+0.3}$
\\\Xhline{5\arrayrulewidth}
\multicolumn{4}{||c||}{Average color force $\langle \vec f^x \rangle$ (MeV/fm)} 
\\\Xhline{5\arrayrulewidth}
$u$     & $81_{-25}^{+25}$ & $47_{-16}^{+15}$ & $33_{-11}^{+11}$
\\\hline
$d$     & $-83_{-23}^{+23}$ & $-53_{-15}^{+15}$ & $-39_{-11}^{+11}$
\\\hline
$s$     & $108_{-64}^{+73}$ & $63_{-38}^{+44}$ & $45_{-27}^{+31}$
\\\hline
\end{tabular}
\end{center}
\caption{\label{tab:force} Values of the $d_2$ moment and the average color force (\ref{def:force}), computed for different flavor combinations and scales $\mu$. In the definition of the force (\ref{def:force}) the proton is polarized in the $y$ direction.}
\end{table}

\subsection{The $\overline{g}_2$ structure function}
\label{sec:g2-result}

\begin{figure}
\centering
\includegraphics[width=0.98\linewidth]{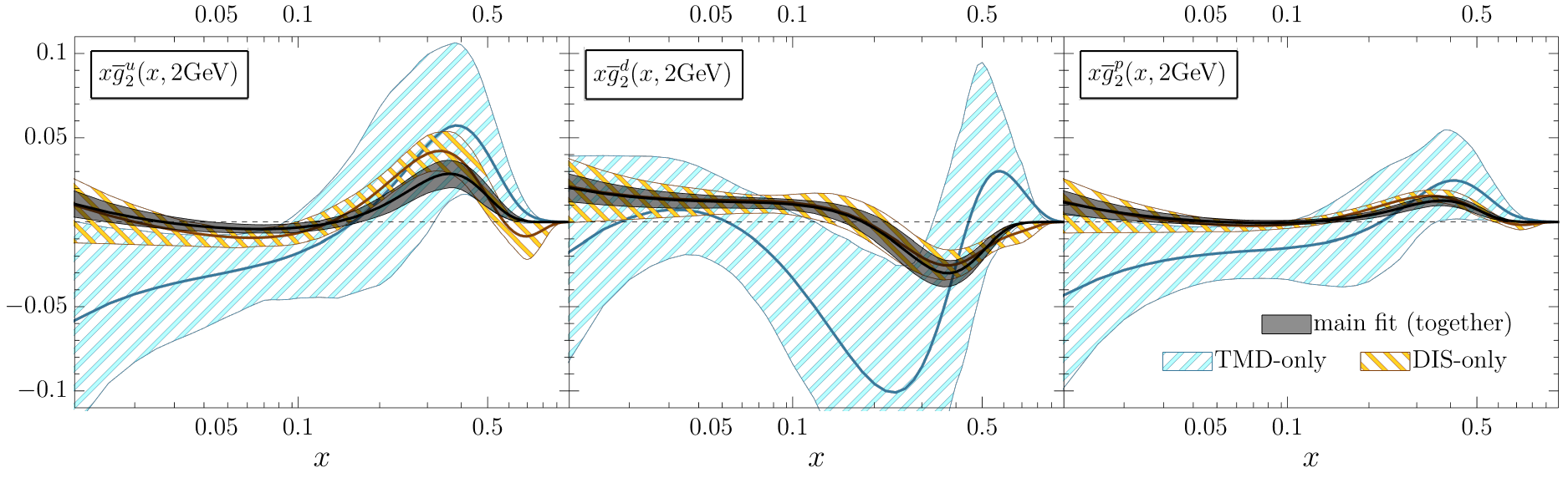}
\includegraphics[width=0.98\linewidth]{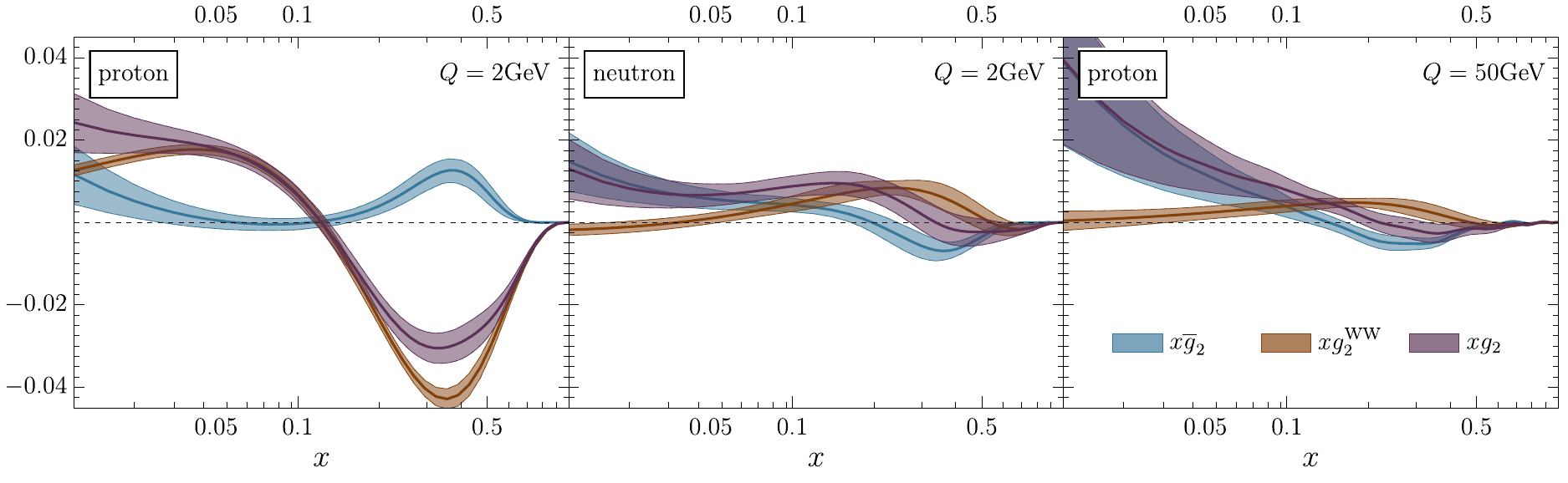}
\caption{\label{fig:g2_collection}
Plots related to the discussion concerning the $\overline{g}_2$ function. Upper panel: comparison of the shape and precision of $\overline{g}_2$ for various fits (see also sec.~\ref{sec:DISvsTMD}). Lower plot: comparison of $\overline{g}_2$ with its WW contribution, and with the complete $g_2$ structure function.
}
\end{figure}

In the upper panel of fig.\ref{fig:g2_collection}, we demonstrate the comparison of the $\overline{g}_2$ function obtained for different fit configurations. Clearly, the TMD-only and DIS-only fits produce compatible results, with the DIS fit being about 5 times more precise. The TMD-only fit is quite indefinite, and is compatible with zero in the majority of the range. Still, it catches the peaking structure of $\overline{g}_2$ at $x\sim 0.3-0.4$.

In the lower panel of fig.\ref{fig:g2_collection}, we show the comparison of the WW contribution and the genuine contributions to the $g_2(x)$ structure function. Clearly, at low $Q$, which are typical for the existing measurements, the WW-term is generally larger. Here, it is about three times larger than $\overline{g}_2$ for the proton, and around the same size for the neutron, which is in general agreement with common estimations, see refs.~\cite{Accardi:2009au, Cocuzza:2025qvf}. However, for smaller $x$, the twist-three contribution starts to dominate over the WW term, which decreases at small-$x$. However, it is important to mention that, there exists very little data with $x$<0.05 for $g_2$, and in this regime our function is merely an extrapolation somewhat impacted by the TMD data. 

The situation entirely changes at large energies, where the twist-three contribution is dominating over the WW term (see the utmost right panel, in the lower row of fig.~\ref{fig:g2_collection}). According to this prediction, future polarized experiments, such as EIC \cite{AbdulKhalek:2021gbh}, could observe genuine twist-three effects more directly. Moreover, the expected signal of the $g_2$ structure function will be much larger than an estimation by WW approximation, which is a typical method to estimate $g_2(x,Q)$ nowadays.

Another striking observation is the evident violation of the Burkhardt-Cottingham sum rule. Indeed, according to the discussion carried in sec.~\ref{sec:burkardt}, and particularly due to the expression (\ref{BC-sumrule-cut}), the Burkhardt-Cottingham sum rule is violated if $x\overline{g}_2(x)$ is not vanishing in the limit $x\to0$. Observing the plots fig.~\ref{fig:g2_collection} and \ref{fig:g2_QS_evolution}(upper), one finds that $x\overline{g}_2(x)$ grows as $x\to 0$, and therefore, the sum rule is violated. As discussed in sec.~\ref{sec:burkardt}, this is a consequence of the integral definition of $\overline{g}_2$: the extracted $\mathfrak{S}^+$ is singular at $\|x\|\to 0$, while the Burkhardt-Cottingham sum rule requires a zero of first order at least.

\subsection{Sivers function}
\label{sec:result-Sivers}

\begin{figure}
\centering
\includegraphics[width=0.98\linewidth]{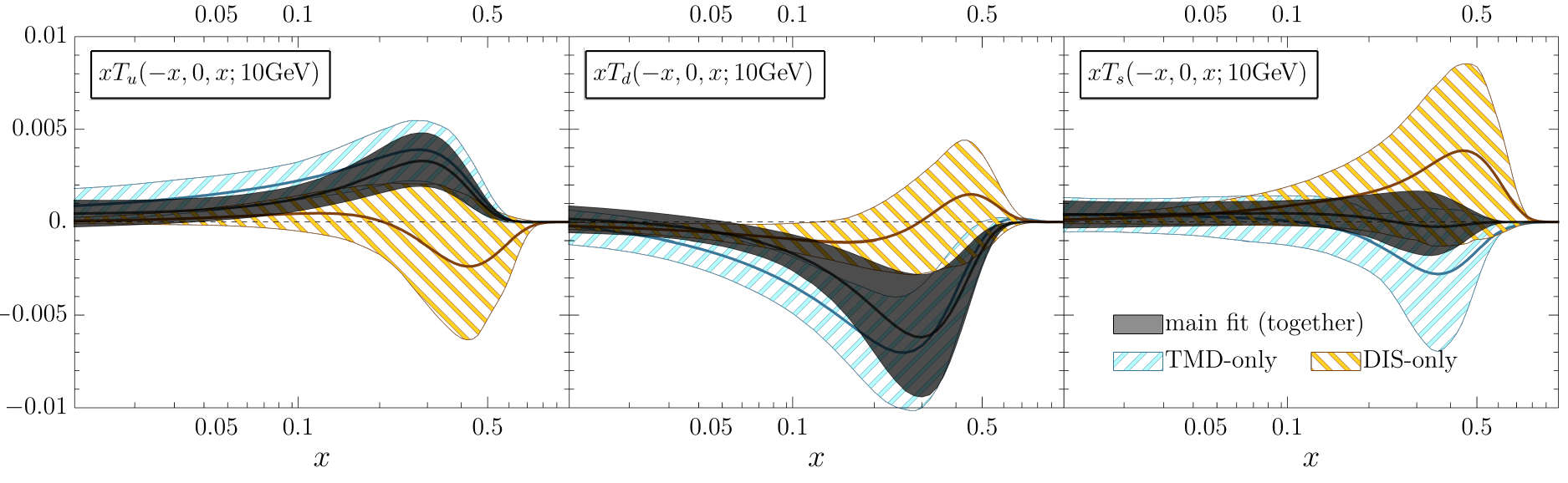}
\\
\includegraphics[width=0.98\linewidth]{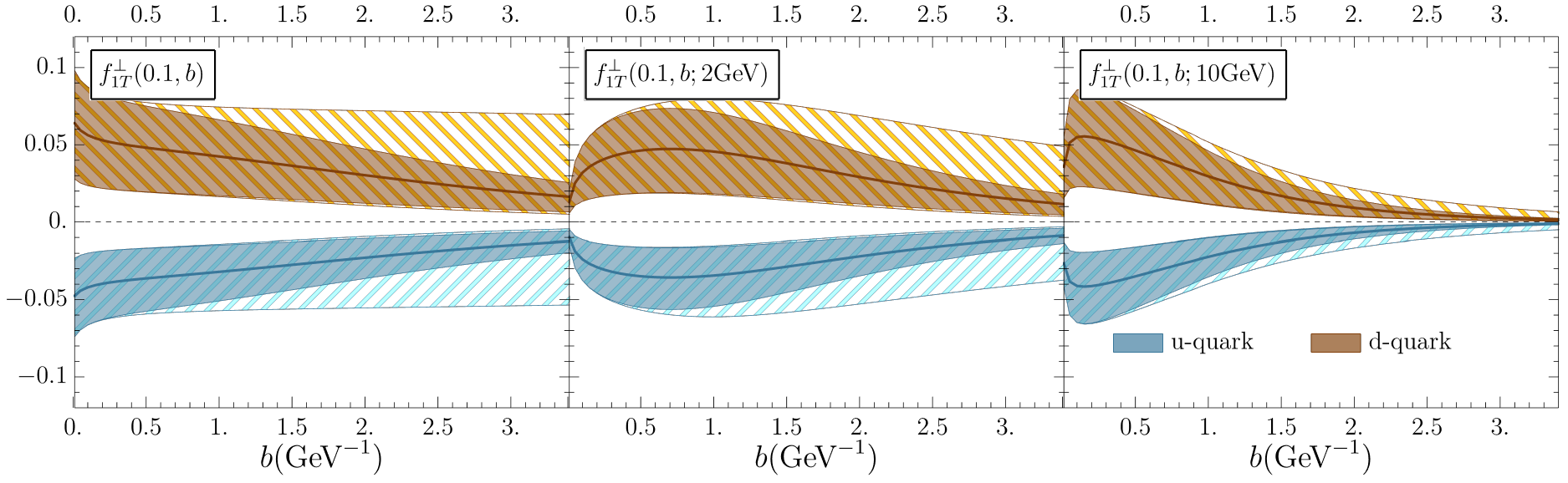}
\caption{\label{fig:sivers}
Plots related to the discussion concerning the Sivers function. Upper panel:
Comparison of the Qiu-Sterman function determined in various fits (see also sec.~\ref{sec:DISvsTMD}). Lower panel: Dependence of the Sivers function with $b$ at $x=0.1$ for different values of the evolution scale $(\mu,\mu^2)$. The left panel demonstrates the optimal distribution. The diagonal shading represents the additional uncertainty due to the uncertainty in the $b$-profile (\ref{siv:lambda_uncert}).
}
\end{figure}

The Sivers and Qiu-Sterman functions are iconic parton distributions. This fit provides a novel input to their extraction, since it incorporates other types of observables. In fig.\ref{fig:sivers} (upper panel) we compare the results of the main fit with TMD-only and DIS-only data. As one can see, the DIS-only fit results in a rather uncertain distribution, which in general disagrees with the TMD-fit. There are several reasons to explain this behavior: first of all, DIS measurements do not have access to the Qiu-Sterman function, but rather to its derivative (\ref{def:g2-tw3}); second, DIS data lack detailed flavor decomposition, whereas SIDIS has access to various hadrons.

The $b$-dependence of the Sivers function is shown in fig.~\ref{fig:sivers} (lower panel). As usual, it strongly depends on the scale of the distribution. Note that the fit of the Sivers function has been performed with a fixed $b$-dependence (\ref{ansatz-fNP}), with $\lambda=0.5$GeV. It is done in this way, since the data has a very small sensitivity to the parameter $\lambda$, and the joint fit with PDF parameters is unstable. This is also clear from the plot in fig.~\ref{fig:sivers}, where the uncertainty band is already quite large. The uncertainty for $\lambda$ can be estimated a posteriori, as described in sec.~\ref{sec:lambdas}. The corresponding bands are shown in fig.~\ref{fig:sivers} (lower panel) by a diagonal shading.

\subsection{Sivers asymmetry in Drell-Yan reaction and test of the sign change}
\label{sec:result-DY+sign}

The Sivers function is T-odd, which results into the celebrated sign-change rule (\ref{siv:sign-change}). The sign-change of the Sivers function is one of the brightest predictions of the TMD factorization approach, and thus attracts a lot of attention, see for instance refs.~\cite{Anselmino:2016uie, Echevarria:2014xaa, Bury:2021sue}.  The central problem is that the TMD fits do not have sufficient restrictive power, and allow one to utilize the same sign for SIDIS and Drell-Yan (Note that the fit should be redone under this new assumption), and still produce a relatively good description \cite{Bury:2021sue}.

\begin{figure}
\centering
\includegraphics[width=0.98\linewidth]{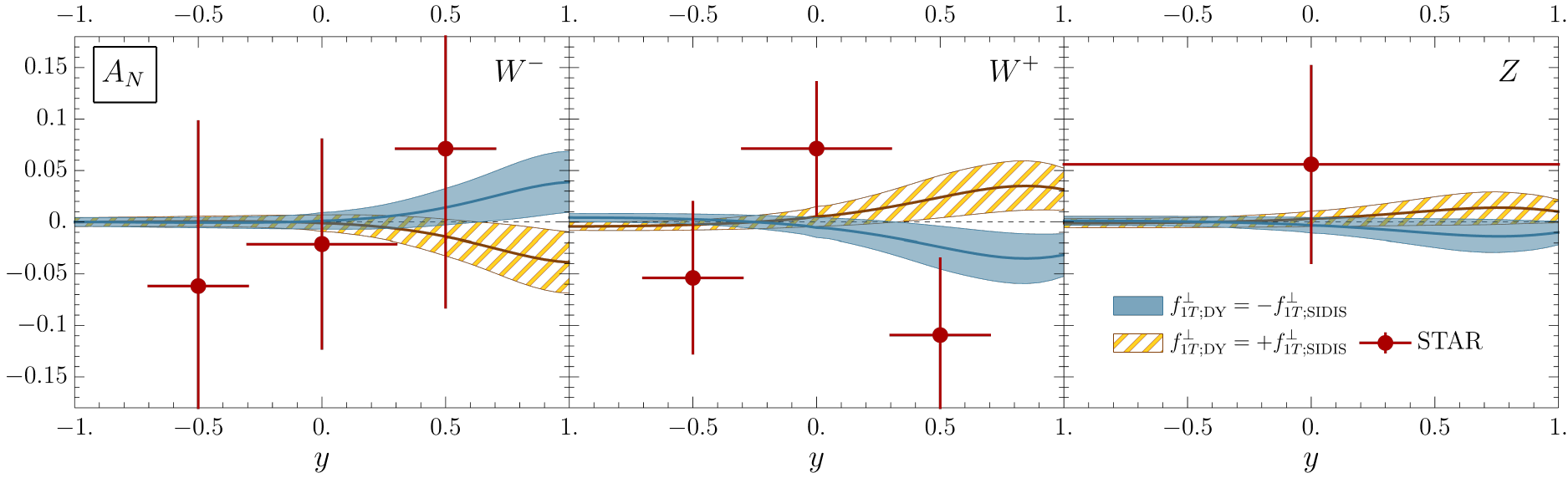}
\caption{\label{fig:WZ}
The $A_N$ asymmetry in Z/W production for the kinematics of the STAR measurement \cite{STAR:2023jwh, STAR:inprep}. The diagonally shaded bands correspond to the prediction without the sign-change relation (\ref{siv:sign-change}).
}
\end{figure}

In the present fit, we do not include Drell-Yan data into consideration, but our prediction agrees with the data for Z/W boson production measured at STAR \cite{STAR:2023jwh, STAR:inprep}, resulting in $\chi^2/N_{\text{DY}} =0.61$ (with $N_{\text{DY}}=7$), which is composed of $\chi^2/N=\{0.12,1.16,0.47\}$ for the $\{W^-, W^+, Z\}$ channels. The comparison of our prediction with the data is shown in fig.~\ref{fig:WZ}.  Changing the sign of the Sivers function (see diagonally shaded bands in fig.~\ref{fig:WZ}) we obtain $\chi^2/N_{\text{DY}}=0.77$ with $\chi^2/N=\{0.16,1.56,0.24\}$ for the $\{W^-, W^+, Z\}$ channels. Thus, given the small number of points, and imprecise measurements, one cannot derive a strict conclusion concerning the sign-change mechanism of the Sivers function.

\subsection{Worm-gear-T function}
\label{sec:result-wgt}

\begin{figure}
\centering
\includegraphics[width=0.98\linewidth]{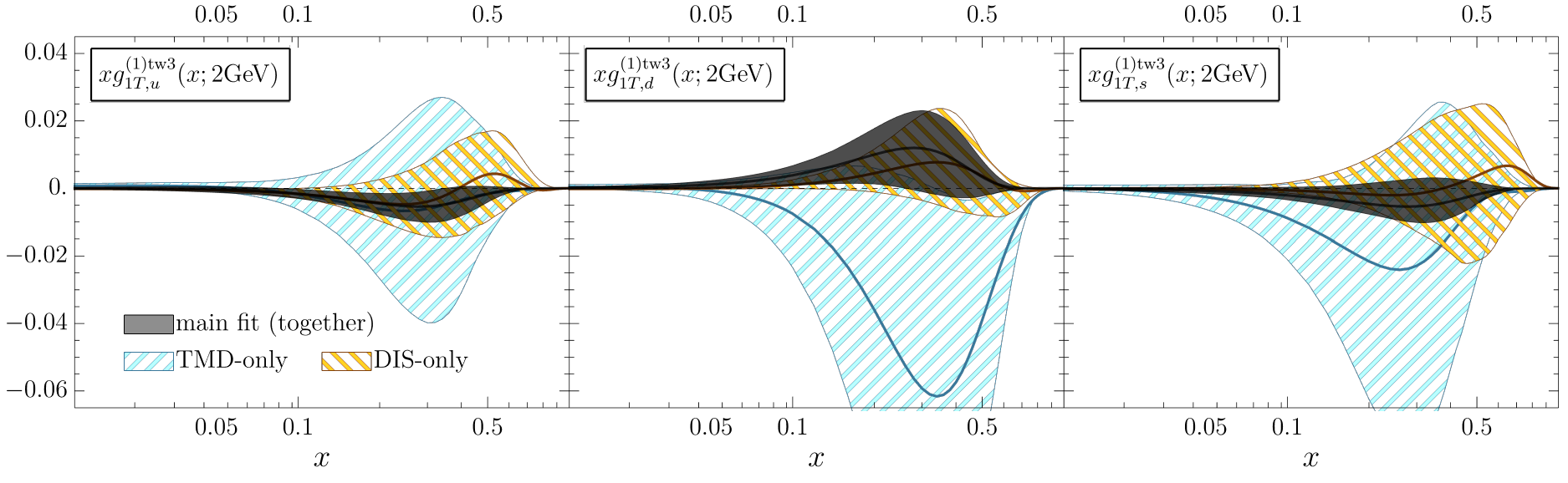}
\\
\includegraphics[width=0.98\linewidth]{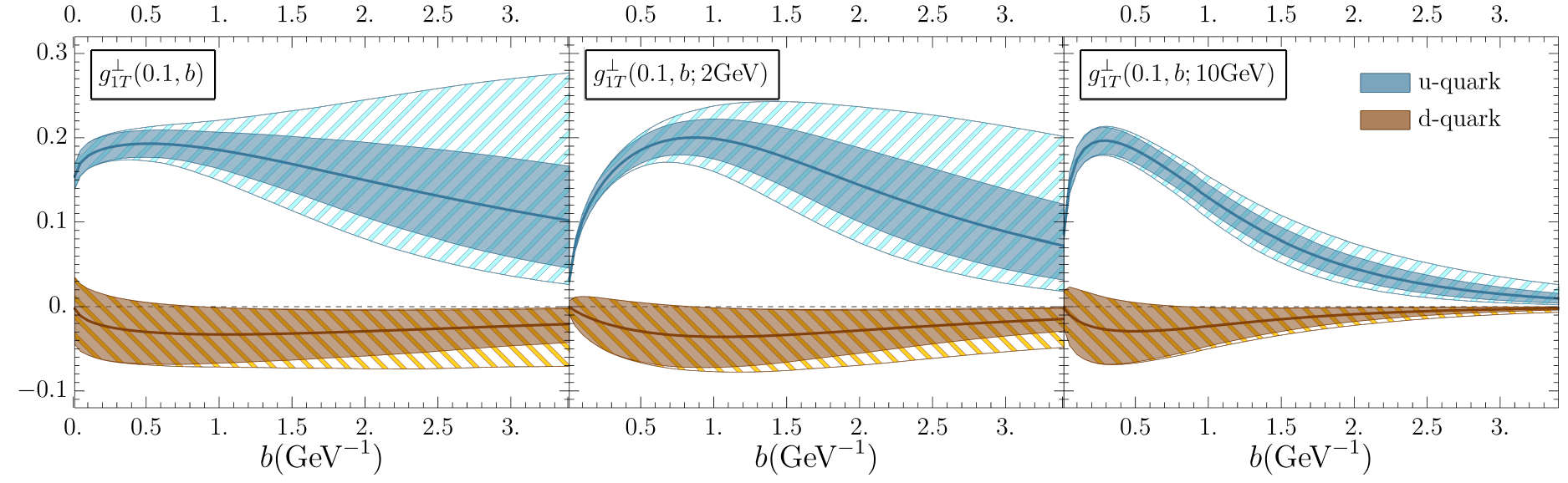}
\\
\includegraphics[width=0.4\linewidth]{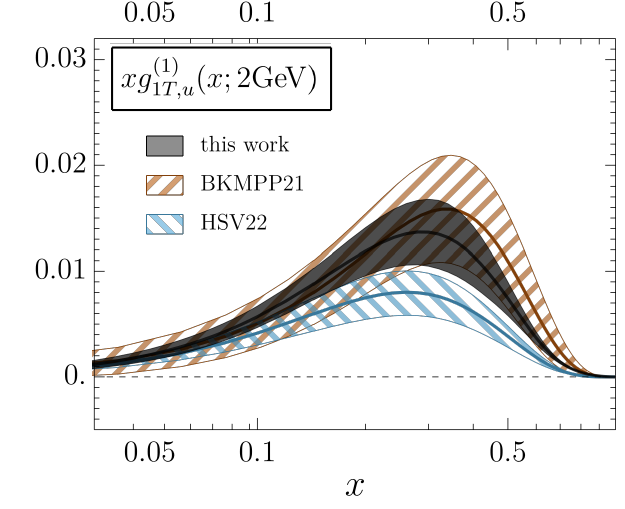}
\includegraphics[width=0.4\linewidth]{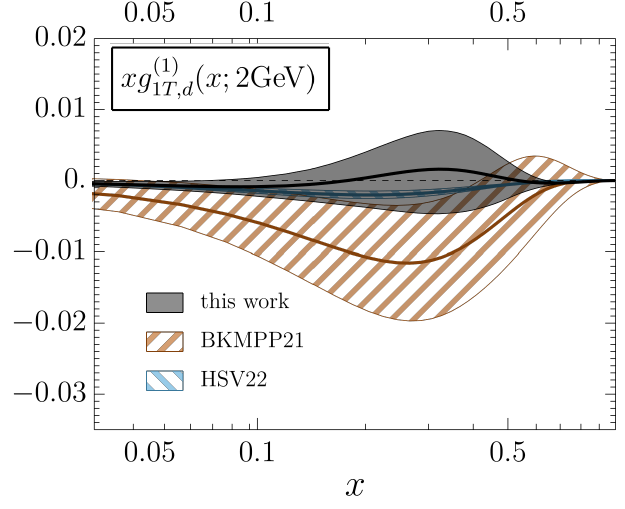}
\caption{\label{fig:wgt_comparison}
Plots related to the discussion concerning the worm-gear-T function. 
Upper panel: Comparison of different flavors of the function $g_{1T}^{(1)\text{tw3}}$ (\ref{wgt:g1T(1)+tw3}) determined in various fits (see also sec.~\ref{sec:DISvsTMD}). Middle panel: Dependence of the worm-gear-T function with $b$ at $x=0.1$ for different values of the evolution scale $(\mu,\mu^2)$. The left panel shows the optimal distribution. The diagonal shading represents the additional uncertainty due to the uncertainty in the $b$-profile (\ref{siv:lambda_uncert}). 
Lower panel: Comparison of the functions $g_{1T}^{(1)}$ (\ref{wgt:g1T(1)}) determined in this work, ref.~\cite{Bhattacharya:2021twu} (BKMPP21) and ref.~\cite{Horstmann:2022xkk} (HSV22) for $u$ and $d$ flavors.}
\end{figure}

The worm-gear-T function is much less studied than the Sivers function. Therefore, we found it very enlightening that we can determine it with a minimum data input related directly to $g_{1T}^\perp$. Most of this distribution is fixed by other data. The only specific parameter is the $b$-dependence dictated by $\lambda$. As described, in sec.~\ref{sec:lambdas} we performed the test for $\lambda_{LT}$, which led to the estimation (\ref{siv:lambda_uncert}).

There are no collinear matrix elements associated with the worm-gear-T function (like the Qiu-Sterman function is to the Sivers distribution). As such, we consider its first TMM \cite{delRio:2024vvq}, which is defined as
\begin{eqnarray}\label{wgt:g1T(1)}
g_{1T}^{(1)}(x,\mu)&=&\int^\mu d^2k_T \frac{k_T^2}{2M^2}g_{1T}(x,k_T).
\end{eqnarray}
It has two contributions, that correspond to the twist-two and twist-three parts of the small-$b$ OPE (\ref{wgt:tw2+tw3}),
\begin{eqnarray}\label{wgt:g1T(1)+tw3}
g_{1T}^{(1)}(x,\mu)&=&g_{1T}^{(1),\text{tw2}}(x,\mu)+g_{1T}^{(1),\text{tw3}}(x,\mu).
\end{eqnarray}
At LO, the twist-two term is (\ref{wgt:tw2})
\begin{eqnarray}
g_{1T}^{(1),\text{tw2}}(x,\mu)=\frac{x}{2}\int_x^1 \frac{dy}{y}g_1(y;\mu),
\end{eqnarray}
while the twist-three part is given by (\ref{wgt:tw3}) evaluated at a scale $\mu$ and multiplied by a factor $1/2$.

In the upper panel of fig.~\ref{fig:wgt_comparison}, we compare the obtained values of the twist-three contribution to $g_{1T}^{(1)}$ obtained in the main fit with the DIS-only, and TMD-only extractions. All these extractions are in good agreement with each other, and are compatible with zero. It is interesting to observe that the TMD-only extraction has a much larger uncertainty and, in the case of the $d$-quark, tends to deviate from zero. Therefore, we conclude that the worm-gear-T function is mainly constrained by DIS data, contrary to common intuition.

In the lower panel of fig.~\ref{fig:wgt_comparison}, we compare the obtained values of $g_{1T}^{(1)}$ with other extractions \cite{Bhattacharya:2021twu, Horstmann:2022xkk}. The agreement is very good, given that both extractions  \cite{Bhattacharya:2021twu, Horstmann:2022xkk} were performed using only data for $A_{LT}^{\cos(\phi_h-\phi_S)}$ and both found $\chi^2/N<1$. Interestingly, the $d$-quark part in both earlier extractions is negative, while it is mainly positive (although consistent with zero) in the present case. Note that it is in agreement with the TMD-only determination.

\subsection{Nucleon tomography}
\label{sec:nucleon-tomography}

Together with the unpolarized TMD distribution $f_1$, the Sivers function forms the density of unpolarized quarks in the proton. It is defined as
\begin{eqnarray}\label{def:density}
\rho(x,\vec k_T,\mu)=f_1(x,\vec k_T,\mu)-\frac{\epsilon^{\mu \nu}k_{T \mu}S_{T\nu}}{M}f_{1T}^\perp(x,\vec k_T,\mu),
\end{eqnarray}
where the TMD distributions in momentum representation are 
\begin{eqnarray}\label{momentum:unpol}
f_1(x,\vec k_T,\mu)&=&\int_0^\infty \frac{bdb}{2\pi} J_0(b|\vec k_T|)f_1(x,b;\mu,\mu^2),
\\\label{momentum:siv}
f_{1T}^\perp(x,\vec k_T,\mu)&=&M^2\int_0^\infty \frac{bdb}{2\pi} \frac{b}{|\vec k_T|} J_1(b|\vec k_T|)f_{1T}^\perp(x,b;\mu,\mu^2),
\end{eqnarray}
with $M$ being the mass of the hadron. The density $\rho$ is asymmetric due to the contribution of the Sivers function. This asymmetry reflects the fact that quarks of different flavors move in a slightly asymmetric fashion within the hadron. 

\begin{figure}
\centering
\includegraphics[width=0.48\linewidth]{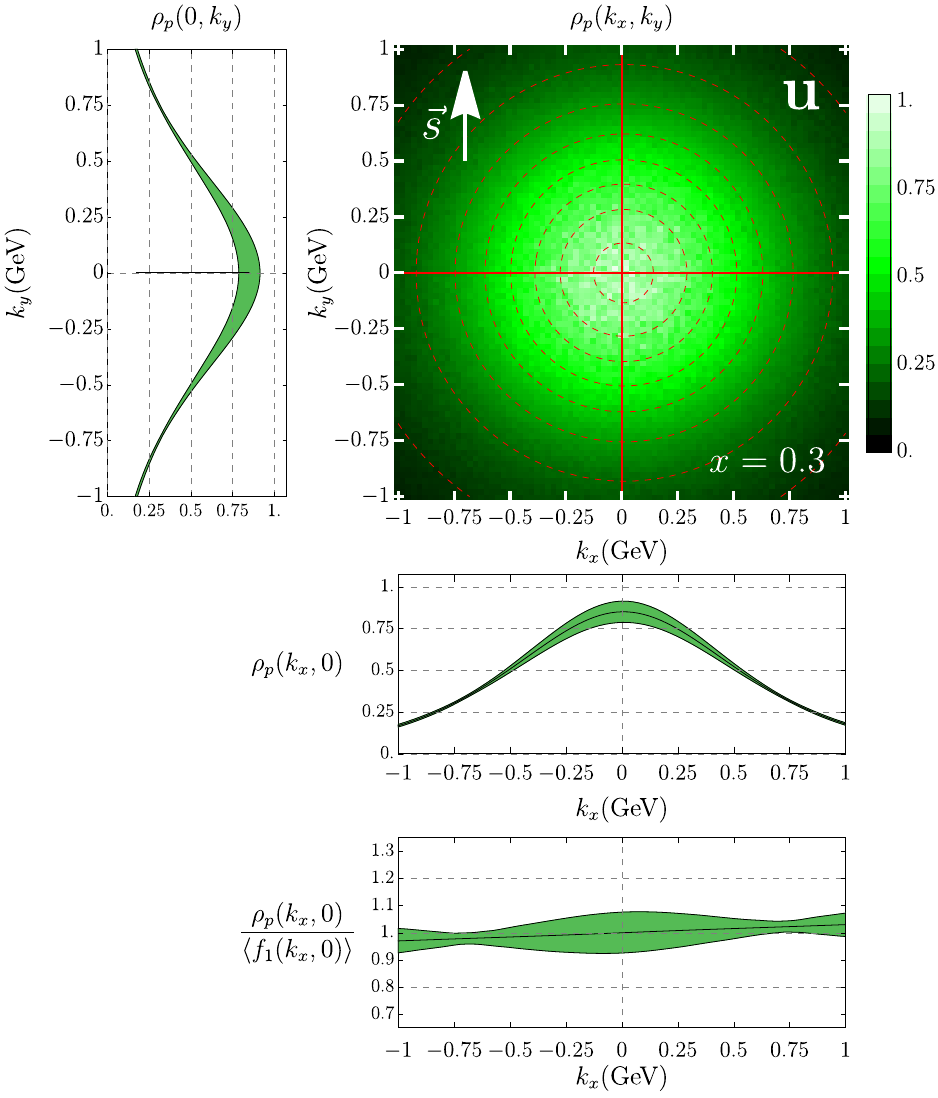}
\includegraphics[width=0.48\linewidth]{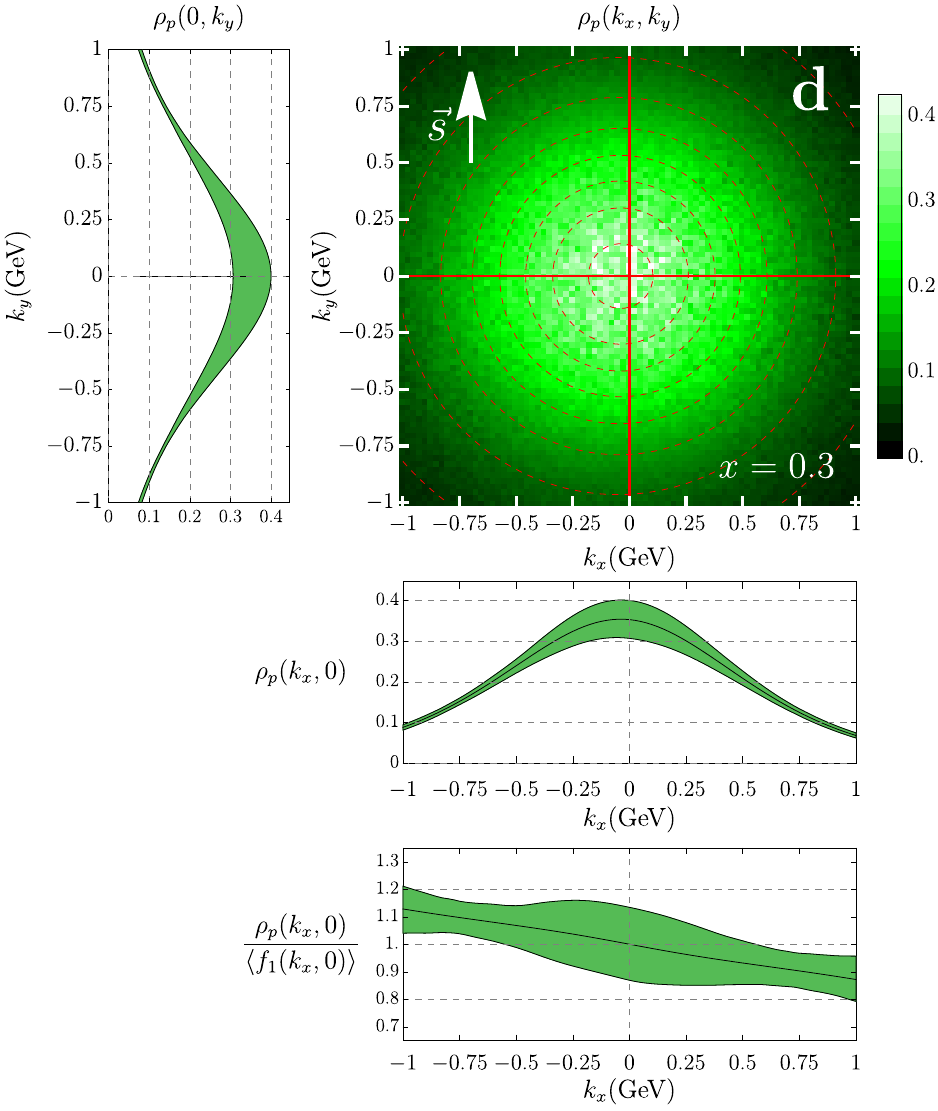}
\caption{\label{fig:tomography} The tomographic picture of the proton in the transverse momentum plane. The slice is made for $x=0.3$, $\mu=10$GeV, and the polarization of the proton in the positive $y$-direction. The Inhomogeneity in the color represents the uncertainty of the distribution. The panels illustrate the sections of the distributions along the red lines. The lowest panel represents the section weighted by the average unpolarized distribution.}
\end{figure}

\begin{figure}
\centering
\includegraphics[width=0.4\linewidth]{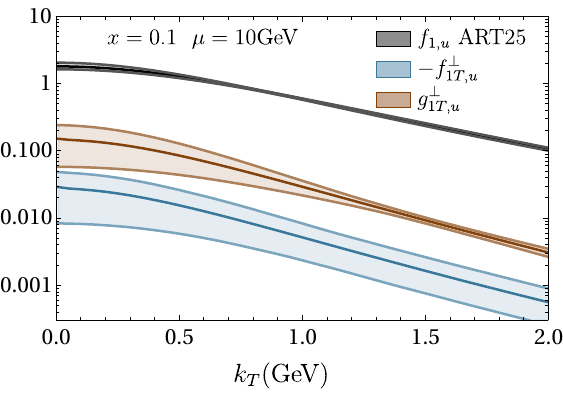}
\includegraphics[width=0.4\linewidth]{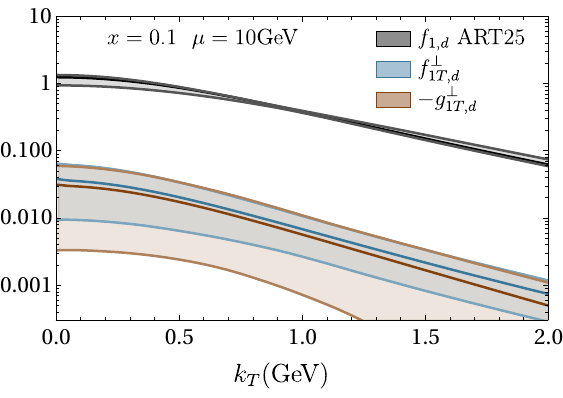}
\caption{\label{fig:siv_vskT}
Dependence of the Sivers and worm-gear-T functions on $k_T$ at $x=0.1$ and $\mu=10$GeV for the $u$ and $d$ flavors, in comparison to their corresponding unpolarized distribution.}
\end{figure}

In fig.~\ref{fig:tomography}, we show the slice of the density function at $x=0.3$ and $\mu=10$GeV. The value $x=0.3$ is selected because it roughly corresponds to the maximum of the Sivers function, and so the effect is maximized. Even so, the asymmetry is very small, because the unpolarized distribution is about 100 times larger than the Sivers distribution. This is also evident from fig.~\ref{fig:siv_vskT}, where we compare the shapes of TMD distributions in momentum space. For a clearer illustration of the asymmetry in the TMD density, in the lower panels of fig.~\ref{fig:tomography} we present the slice of density weighted by the mean value of $f_1$.

It should be pointed out that the presentation of TMD distributions in momentum space, and thus the tomographic picture, should be considered with a grain of salt. The reason is that the extraction of TMD distributions is done in position space, and is mainly controlled at small-$b$, while the large-$b$ tail remains unknown. It is known that in fits of unpolarized distributions, which have much more accurate data vs. $p_\perp$, the region $b>1.5-2$GeV$^{-1}$ is practically unrestricted. The values and uncertainty bands in this region are obtained by extrapolation. Meanwhile, the Fourier transform over region $\vec k_T\sim \vec 0$ is dictated by the large-$b$ asymptotic behavior of TMD distributions. Therefore, the uncertainty bands presented here (and in other extractions) in momentum space, are essentially underestimated in the vicinity of $k_T<0.5-1$GeV.

Note that using the worm-gear-T function, one can prepare the tomographic picture of the nucleon in terms of the longitudinally polarized partons. For that, one needs the $g_{1T}^\perp$ function in momentum space
\begin{eqnarray}
g_{1T}^\perp(x,\vec k_T,\mu)&=&M^2\int_0^\infty \frac{bdb}{2\pi} \frac{b}{|\vec k_T|} J_1(b|\vec k_T|)g_{1T}^\perp(x,b;\mu,\mu^2).
\end{eqnarray}
We do not present such plots for tomography here, because it is not very instructive at the present stage. Instead, we compare the values of $g_{1T}^\perp(x,\vec k_T)$ with the values of the Sivers and unpolarized distributions in fig.~\ref{fig:siv_vskT}. As one can see, the worm-gear-T function for the $u$-quark  is about 10 times smaller than the unpolarized TMDPDFs. For the $d$-quark it almost coincides with the Sivers function, but both have large uncertainty bands.

\subsection{Positivity constraint}
\label{sec:positivity}

In ref.~\cite{Bacchetta:1999kz}, the positivity constraints for TMD distributions were derived assuming the positive-definiteness of the polarization matrix due to its probabilistic interpretation in the parton model. In the present case, the relevant positivity constraint reads
\begin{eqnarray}\label{def:positivity}
\frac{\vec k_T^2}{M^2}\[\(g^{\phantom{\perp}}_{1T}(x,k_T)\)^2+\(f^{\perp}_{1T}(x,k_T)\)^2\]\leqslant \(f_{1}(x,k_T)\)^2.
\end{eqnarray}
This relation is derived for bare matrix elements, and thus should not hold in general. Nonetheless, it is interesting to inspect how the rules of the bare parton model are respected in our extraction. As this work is the first to simultaneously extract the Sivers and worm-gear-T functions, it is the first in which we can test the bound constraint (\ref{def:positivity}) without any approximations.

\begin{figure}
\centering
\includegraphics[width=0.8\linewidth]{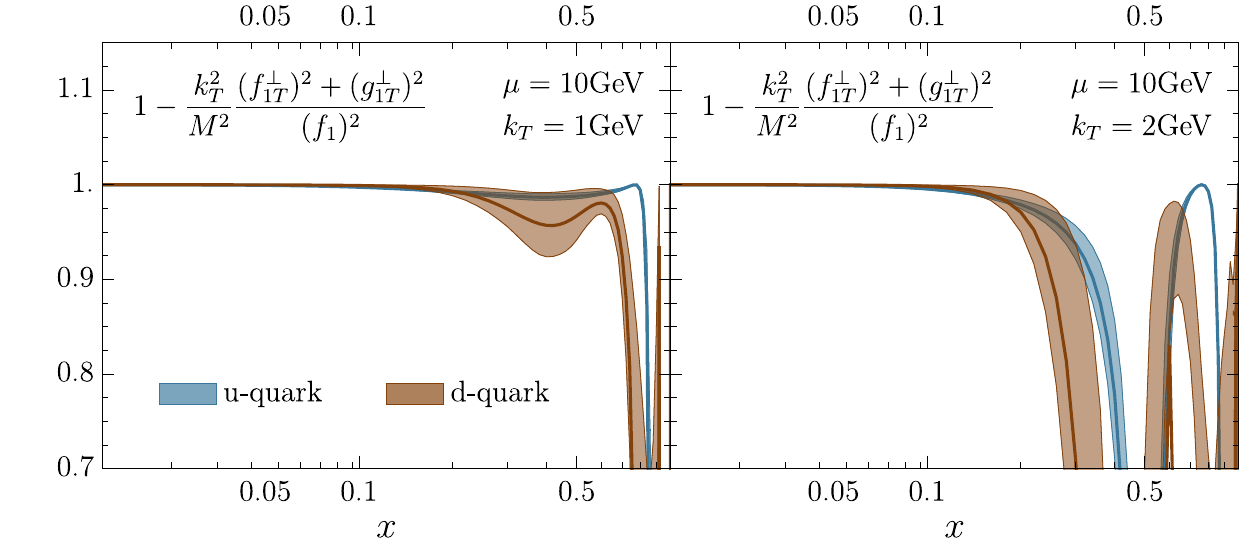}
\caption{\label{fig:positivity}
The composition of TMD distributions that form the positivity bound relation for $u$ and $d$ quarks at $\mu=10$GeV.}
\end{figure}

It is important to realize that the relation (\ref{def:positivity}) does not account for the evolution effects. However, they are crucially important because they can change the sign of a TMD distribution for large $k_T$, and usually do for $\mu \sim k_T$. In fig.~\ref{fig:positivity}, we demonstrate that the relation (\ref{def:positivity}) holds for most of the functions, and is violated at large-$x$ and large-$k_T$. The violation happens because of the non-positive definiteness of the unpolarized TMDPDF in this regime.

\subsection{Average transverse momentum and the Burkardt sum rule} 
\label{sec:burkardt}

The interpretation of the function $\rho$ (\ref{def:density}) as the density of a quark's momentum leads to the subsequent interpretation of the following quantity 
\begin{eqnarray}\label{BS:integral}
\langle \vec k_{i,f}\rangle(x,\mu)
=-\epsilon_{ij}s^j_T \int^\mu d^2 \vec k_T \frac{\vec k_T^2}{2M}f_{1T,f}^\perp(x,\vec k_T),
\end{eqnarray}
as the average transverse momentum of a parton \cite{Burkardt:2004ur, Goeke:2006ef, delRio:2024vvq}. Here, $f_{1T,f}^\perp(x,\vec k_T)$ is the Fourier transform (\ref{momentum:siv}) of the Sivers function for the flavor $f$ evaluated at the optimal scales. This quantity tells us about the asymmetry of the TMD density at a given $x$, i.e. about the position of the maximum of the transverse momentum distribution. It should not be confused with the average value of the transverse momentum $\langle \vec k^2_T\rangle(x,\mu)$, which characterizes the spread of the TMD distribution.

Note that alternatively, one can compute $\langle \vec k_{i,T}\rangle$ as the integral of the evolved TMD distribution,
\begin{eqnarray}\label{BS:integral2}
\langle \vec k_{i,T}\rangle(x,\mu)
=-\epsilon_{ij}s^j_T \int^\mu d^2 \vec k_T \frac{\vec k_T^2}{2M}f_{1T}^\perp(x,\vec k_T;\mu,\mu^2).
\end{eqnarray}
The expression (\ref{BS:integral}) and (\ref{BS:integral2}) are equivalent up to a finite-renormalization constant $1+\mathcal{O}(\alpha_2^2)$. Therefore, at our level of accuracy, they lead to the same results.

The integrals (\ref{BS:integral}, \ref{BS:integral2}) are straightforward to evaluate using eqns.~(\ref{TMD:sivers-smallb-1}, \ref{TMD:sivers-smallb-2}). One obtains the well-known result for the quark and anti-quark average transverse momenta
\begin{eqnarray}
\langle \vec k_{i,q}\rangle(x,\mu)&
=&\epsilon_{ij}s^j_T \frac{\pi M}{2}T_q(-x,0,x;\mu),
\\
\langle \vec k_{T,\bar q}\rangle(x,\mu)&
=&\epsilon_{ij}s^j_T \frac{\pi M}{2}T_q(x,0,-x;\mu).
\end{eqnarray}
Analogously, one determines the gluon transverse momentum \cite{Dai:2014ala, Alvaro:2026nip}, which reads
\begin{eqnarray}
\langle \vec k_{i,g}\rangle(x,\mu)&
=&\epsilon_{ij}s^j_T \frac{\pi M}{2x}T_{3F}^+(-x,0,x;\mu).
\end{eqnarray}
The dependence on $\mu$ of $\langle \vec k_{g,i,T}\rangle$ is complicated and involves the integral over the complete hexagon, together with mixing with the functions $\Delta T$ and $T_{3F}^-$. The dependence on $x$ and the flavors can be deduced directly from the plot of the Qiu-Sterman function, such as in fig.~\ref{fig:g2_QS_flavors} and fig.~\ref{fig:g2_QS_evolution}.

It is also practical to introduce the integrated average momentum shift
\begin{eqnarray}\label{def:av-kt}
\langle \vec k_{i,f}\rangle(\mu)=\int_0^1 dx \langle \vec k_{i,f}\rangle(x,\mu).
\end{eqnarray}
Note that the integral converges in general, although, in the gluon case, some replicas are divergent due to a $1/x$ factor. The results of the computation are presented in table \ref{tab:kt}. To evaluate this integral, we cut it at $x_{\text{min}}=10^{-4}$, and checked that the mean value is stable while decreasing the cut value (the difference with $x_{\text{min}}=10^{-5}$ is less than 1\%). 

The values presented in table \ref{tab:kt} show that the typical average transverse momentum is of the order of a few MeV. Here, the contribution of sea-quarks rises with the increase in the scale, while the combined quark momentum decreases.

\begin{table}[h]
\begin{center}    
\renewcommand{\arraystretch}{1.5}
\begin{tabular}{||c|c||c||c||}
\hline
\multicolumn{4}{||c||}{Average transverse momentum $\langle \vec k_{x,f}\rangle(\mu)$ (MeV)} 
\\\Xhline{5\arrayrulewidth}
parton & $\mu=2$GeV    & $\mu=10$GeV & $\mu=50$GeV 
\\\hline
$u$     & $7.9_{-3.5}^{+4.0}$ & $8.7_{-5.3}^{+5.8}$ & $9.3_{-7.4}^{+7.5}$
\\\hline
$d$     & $-12.6_{-7.5}^{+7.4}$ & $-10.2_{-7.2}^{+7.4}$ & $-8.5_{-8.8}^{+8.3}$
\\\hline
$s$     & $0.7_{-3.8}^{+4.2}$ & $2.7_{-5.5}^{+5.7}$ & $4.0_{-7.5}^{+7.4}$
\\\hline
$\bar u$     & $0.4_{-2.9}^{+2.9}$ & $-0.9_{-4.8}^{+4.7}$ & $-1.8_{-6.6}^{+6.9}$
\\\hline
$\bar d$     & $2.5_{-5.3}^{+5.8}$& $-1.0_{-6.7}^{+6.3}$ & $-3.0_{-7.9}^{+8.3}$
\\\hline
$\bar s$     & $10.1_{-5.5}^{+6.0}$& $7.4_{-6.6}^{+6.5}$ & $5.6_{-8.5}^{+8.8}$
\\\hline
$c+b+\bar c+\bar b$     & $-0.9_{-0.5}^{+0.5}$ & $-2.9_{-1.2}^{+1.2}$ & $-3.8_{-1.5}^{+1.5}$
\\\Xhline{5\arrayrulewidth}
Quarks total     & $8.2_{-4.4}^{+5.2}$ &  $3.8_{-3.9}^{+4.6}$ & $1.8_{-3.9}^{+4.2}$
\\\Xhline{5\arrayrulewidth}
Gluon     & $-34.0_{-13.7}^{+13.7}$ & $23.6_{-9.1}^{+8.7}$ & $-18.0_{-6.9}^{+6.5}$
\\\Xhline{5\arrayrulewidth}
Total    & $-25.7_{-12.0}^{+12.7}$ & $-19.7_{-8.8}^{+9.1}$  & $-16.2_{-7.3}^{+7.4}$
\\\hline
\end{tabular}
\end{center}
\caption{\label{tab:kt} Values of the average transverse momentum (\ref{def:av-kt}) in the $x$-direction, for the proton spin oriented along the $y$-direction, computed for different flavor combinations and scales $\mu$. }
\end{table}

The Burkardt sum rule \cite{Burkardt:2004ur} states that, within a polarized proton, the total average transverse momentum of partons is zero. I.e.
\begin{equation}
\langle \vec k_{\text{tot}}\rangle=\sum_{f=q,\bar{q},g}\langle \vec k_{f}\rangle=0,
\end{equation}
where the summation runs though all active flavors. In terms of genuine distributions, the sum rule can be rewritten as 
\begin{equation}\label{def:Burk_SR}
\langle \vec k_{\text{tot}}\rangle=\int_{0}^{1}\frac{dx}{x}T^{+}_{3F}(-x,0,x)+\sum_{f}\int_{-1}^{1}dx\:T_{f}(-x,0,x)=0.
\end{equation}
This sum rule is based on the parton approximation, and generally could not hold in QCD, in particular it could be violated by QCD evolution. Indeed, in table \ref{tab:kt}, the value of $\langle \vec k\rangle_{\text{tot}}$ is non-zero, but presents large uncertainties. We remind the reader that, in the present fit, the gluon distributions are extracted indirectly, and are parameterized only by one additional parameter, while the other parameters are inherited from the quark distributions. Therefore, the uncertainty is undoubtedly underestimated, which is also evident from the discussion below.

However, it is known \cite{Zhou:2015lxa} that the total average transverse momentum $\langle \vec k\rangle_{\text{tot}}$ is an eigenfunction of the evolution equation\footnote{We have checked that the numerical evolution by \texttt{Snowflake} confirms this equation with an accuracy superior to $10^{-3}$, which is smaller than the $x\to0$ extrapolation uncertainty.} (\ref{def:ev-ns-singlet}, \ref{def:ev-singlet}), at least at LO
\begin{eqnarray}
\mu^2 \frac{d}{d\mu^2}\langle \vec k_{\text{tot}}\rangle(\mu)=-a_s(\mu)\frac{C_A}{2}\langle \vec k_{\text{tot}}\rangle(\mu).
\end{eqnarray}
Therefore, if the Burkardt sum rule is satisfied at some scale, it is satisfied at all scales, at least at the LO approximation.

The preservation of the Burkardt sum rule (\ref{def:Burk_SR}) by the evolution implies that it can be implemented at the level of the initial ansatz, and can be used as an additional constraint to the distributions. Evaluating the integrals for our ansatz (\ref{ansatz1}-\ref{ansatz6}) we find the value of $\beta_1$ that exactly satisfies the Burkardt sum rule:
\begin{eqnarray}
\beta_1=
\sum_f \frac{1}{2}\(\alpha_0^f+(1-2a_0)\frac{\alpha_{11}^f-\alpha_{13}^f+\alpha_{33}^f}{3-2a_0+2a_1+2a_3}\).
\end{eqnarray}
Setting this value, we obtain the values of $\chi^2$ (compare with (\ref{chi2:main}))
\begin{eqnarray}\label{chi2:BS}
\frac{\chi^2_{d2}}{N_{d2}}=1.01^{+0.12}_{-0.13},\quad
\frac{\chi^2_{g2}}{N_{g2}}=1.06^{+0.07}_{-0.07},\quad
\frac{\chi^2_{UT}}{N_{UT}}=1.07^{+0.06}_{-0.06},\quad
\frac{\chi^2_{LT}}{N_{LT}}=0.94^{+0.03}_{-0.03},
\end{eqnarray}
with
\begin{eqnarray}\label{chi2:tot+BC}
\frac{\chi^2_{\text{tot}}}{N_{\text{tot}}}=1.04_{-0.04}^{+0.04}.
\end{eqnarray}
This is only slightly worse than the main fit. From this, we draw the conclusion that the uncertainty of the $T_{3F}^\pm$ is strongly underestimated, most probably due to the oversimplified parametrization, along with the lack of evidence of the violation of the Burkardt sum rule.


\section{Conclusions} 
\label{sec:conclusions}

In this work, we have performed the joint analysis of the four observables $d_2$, $g_2$, $A_{UT}^{\sin(\phi_h-\phi_S)}$ and $A_{LT}^{\cos(\phi_h-\phi_S)}$, and extracted genuine twist-three distributions. This is the first analysis of its kind, and it undoubtedly demonstrates high potential for this direction of investigation. Our analysis links various branches of theoretical QCD and unifies the description of multiple experiments that were previously considered independent. The determined values of the genuine twist-three PDFs provide access to many dynamical characteristics of the proton, some of which are also discussed in this paper.

One of the main conclusions of this work is the phenomenological confirmation of the universality of factorization theorems and twist-three distributions. We have demonstrated that different kinds of data—specifically polarized DIS, alongside single- and double-spin asymmetries in SIDIS—are described by a single set of functions. Moreover, we have shown that one can utilize a subset of observables (e.g., only DIS data) to successfully predict other observables (e.g., SIDIS). Consequently, the unification of these data sets leads to a significant increase in the precision of the extracted functions. Using this setup as a baseline, one can include additional twist-three observables into consideration, adding extra constraints and correlations between different parts of the 2D domain of twist-three distributions.

As the present analysis represents the first such study (apart from the proof-of-concept in ref.~\cite{Vladimirov:2025qrh}), we have attempted to use conservative practices as much as possible while maintaining a minimum amount of restrictions. Additionally, we have presented various validation tests and conclude that the signal of non-zero genuine twist-three distributions is statistically significant.

The numerical effects of twist-three dynamics are generally small. Their typical size constitutes about 1\% of the typical size of unpolarized valence twist-two effects. However, given the precision reached in the investigation of leading terms, the contribution of twist-three effects could be included in high-precision phenomenology alongside other corrections, since it could impact various studies within QCD and the Standard Model.

The inclusion of the complete twist-three evolution into phenomenology is one of the principal achievements of this work. The effects of evolution are highly important, especially in the description of TMD data, since TMD distributions are sensitive to collinear distributions up to very high scales. It is also evident that the mathematical properties of the twist-three evolution equations should be further investigated. In particular, we found that the evolution introduces a pole at $\|x\|\to0$, which was also observed in earlier works. This pole leads to important physical consequences, such as the growth of the $g_2$ structure function at large $Q^2$ and the violation of the Burkhardt-Cottingham sum rule. Thus, the properties of the twist-three evolution equations should be studied deeper in future works.

The determination of the genuine twist-three PDFs provides a practical foundation for multiple investigations involving hadron structure. It gives access to the values of the average transverse momentum of partons and the forces applied to them, in addition to revealing the magnitude of quantum interference between parton states. Clearly, the exploration of these effects is essential for understanding the dynamics of QCD, and the full impact of these observations will continue to unfold.

\acknowledgments
The authors would like to thank Simone Rodini for enlightening comments and aid in numerical computations. We are also grateful to S.Bhattacharya, V.Braun, C.Cocuzza, B.Parsamyan, and A.Prokudin, for discussions and data. Furthermore, we thank the STAR collaboration, and particularly Oleg Eyser, for providing us with preliminary results of measurements. This project is supported by grants No. PID2022-136510NB-C31 funded by MCIN/AEI/10.13039/501100011033 and the Spanish Ministerio de Ciencia e Innovaci\'on. G.P.M acknowledges funding from the European Social Fund Plus, the Castilla y León Operational Program, and the Regional Government of Castilla y León, through the Regional Ministry of Education. A.V. is funded by the Atracci\'on de Talento Investigador program of the Comunidad de Madrid (Spain) No. 2020-T1/TIC-20204.  

\appendix
\section{Additional plots}
\begin{figure}
\centering
\includegraphics[width=0.99\linewidth]{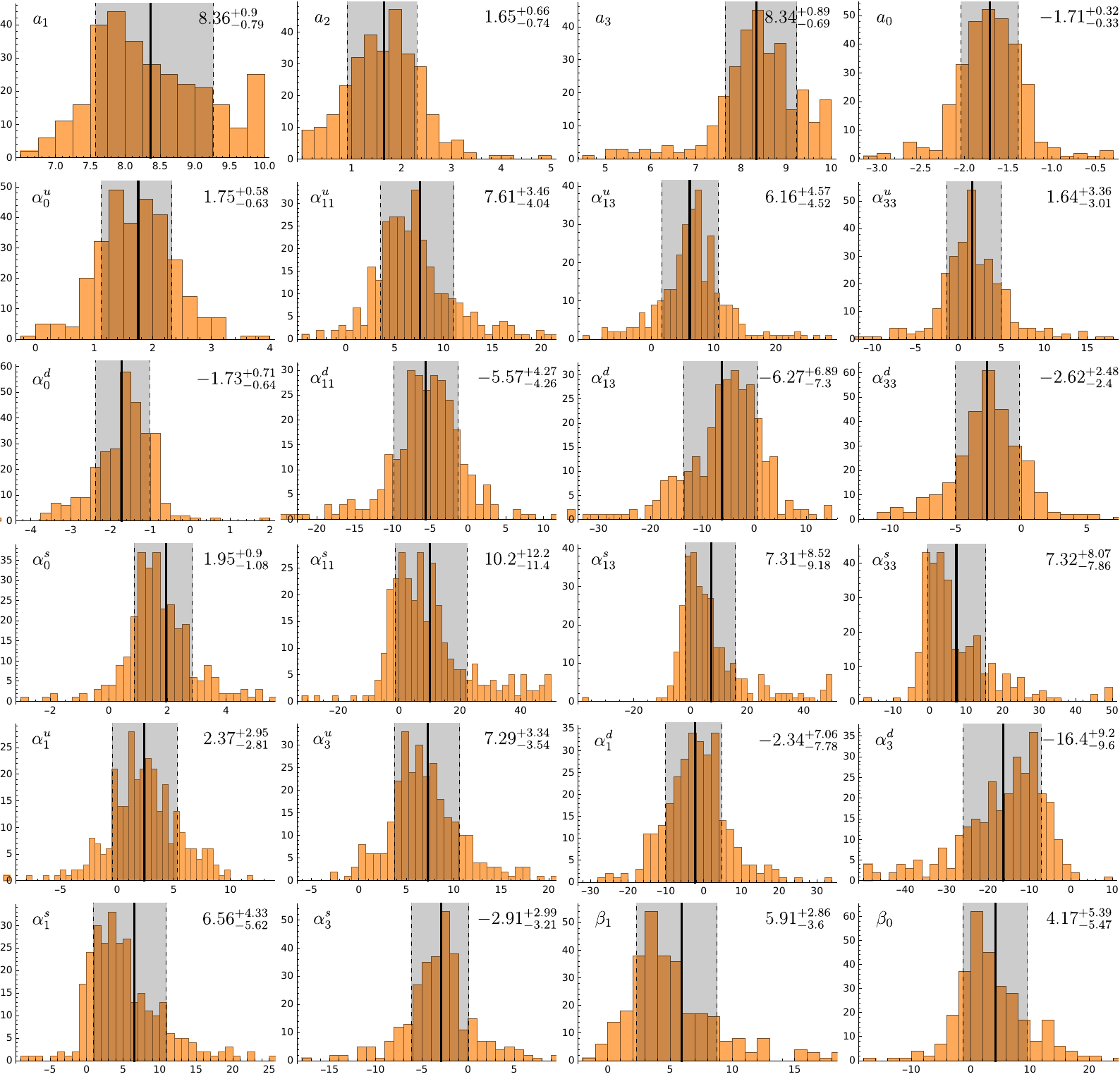}
\caption{\label{fig:parameter_distr} Distribution of parameters in the main fit.}
\end{figure}

\begin{figure}
\centering
\includegraphics[width=0.85\linewidth]{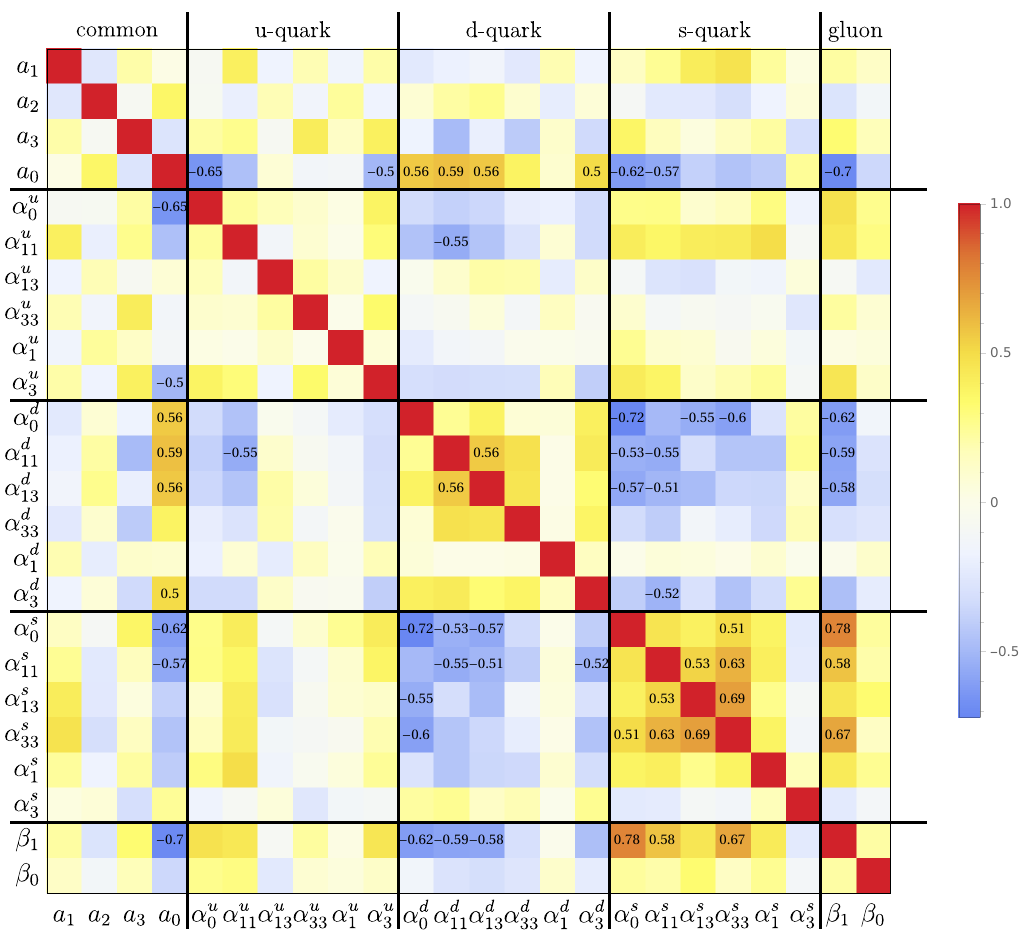}
\caption{\label{fig:parameter_correlation} The correlation matrix between parameters of the model in the main fit. The elements with a correlation bigger than $|0.5|$ are expressed by numbers.}
\end{figure}

\begin{figure}
\centering
\includegraphics[width=0.98\linewidth]{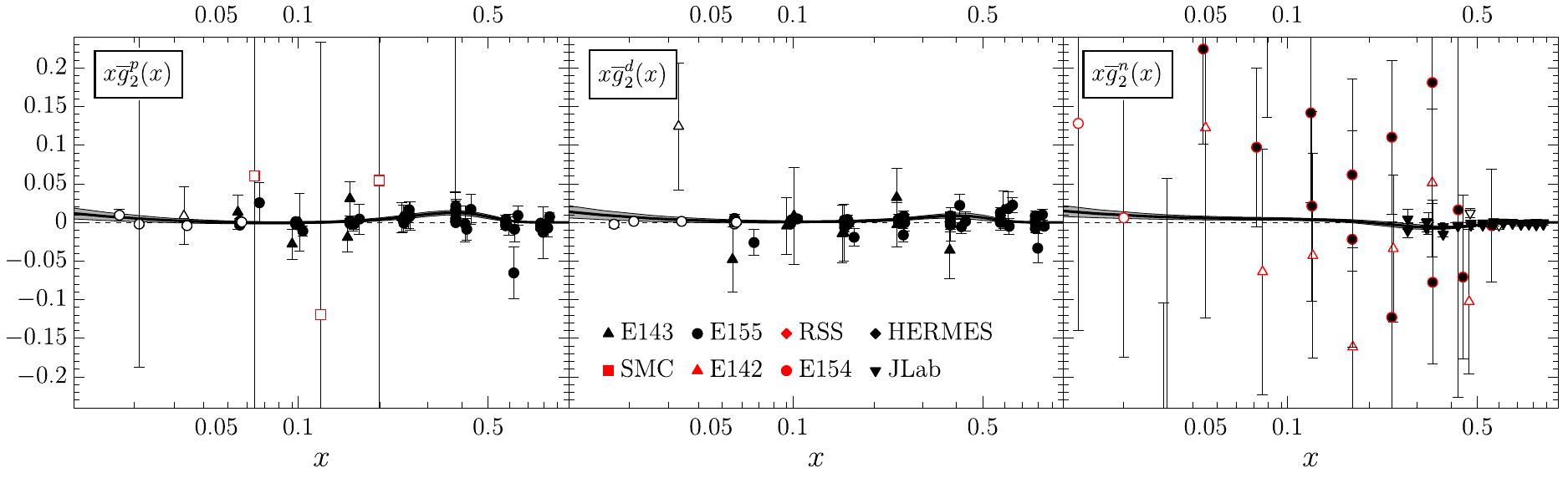}
\caption{\label{fig:g2_all} Comparison of the value for $\bar g_2$ with experimental data. Filled (empty) points were included (excluded) in the fit. The majority of these data were excluded due to excessively large uncertainties. In fig.~\ref{fig:d2_exp}, we show the agreement of our curves with the most precise measurements. }
\end{figure}

\begin{figure}
\begin{minipage}[b]{.5\textwidth}
\centering
\includegraphics[width=0.8\linewidth]{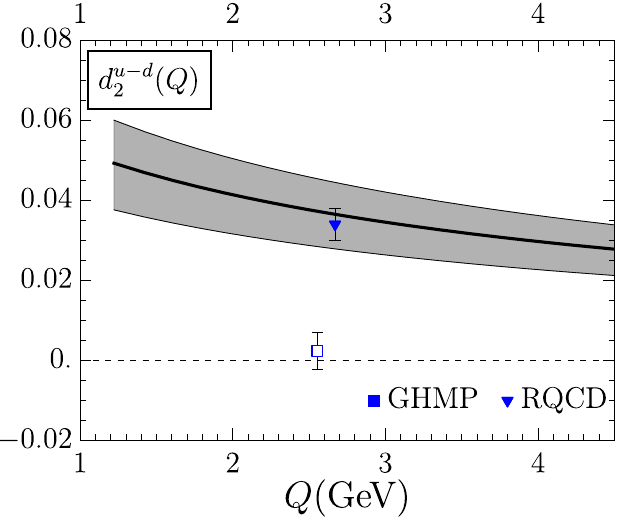}
\caption{\label{fig:d2_lattice} Comparison of the value of $d_2$ for the $u-d$ combination with lattice measurements \cite{Burger:2021knd, Gao:2026wlz}. Filled (empty) points were included (excluded) from the fit.}  
\end{minipage}
\hfill
\end{figure}

\begin{figure}
\centering
\includegraphics[width=0.98\linewidth]{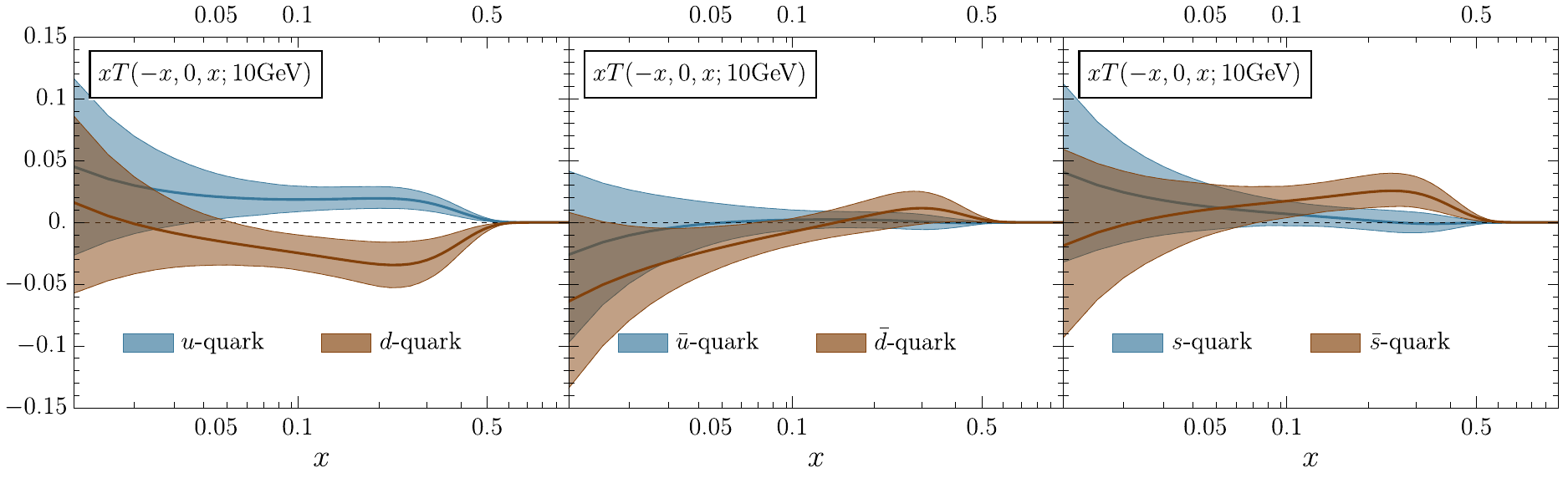}
\includegraphics[width=0.98\linewidth]{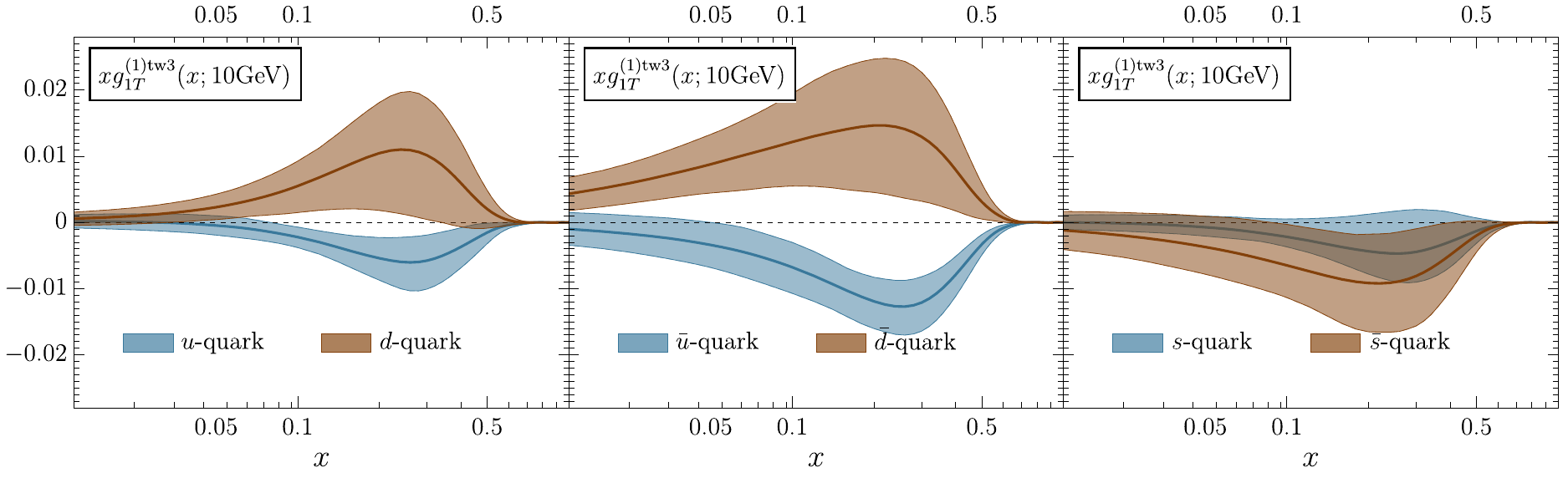}
\caption{\label{fig:g2_QS_flavors}
Comparison of various twist-three matrix elements for different flavors at $\mu=$10GeV. Upper (lower) plot shows the Qiu-Sterman ($g_{1T}^{(1)\text{tw3}}$) function.
}
\end{figure}

\begin{figure}
\centering
\includegraphics[width=0.98\linewidth]{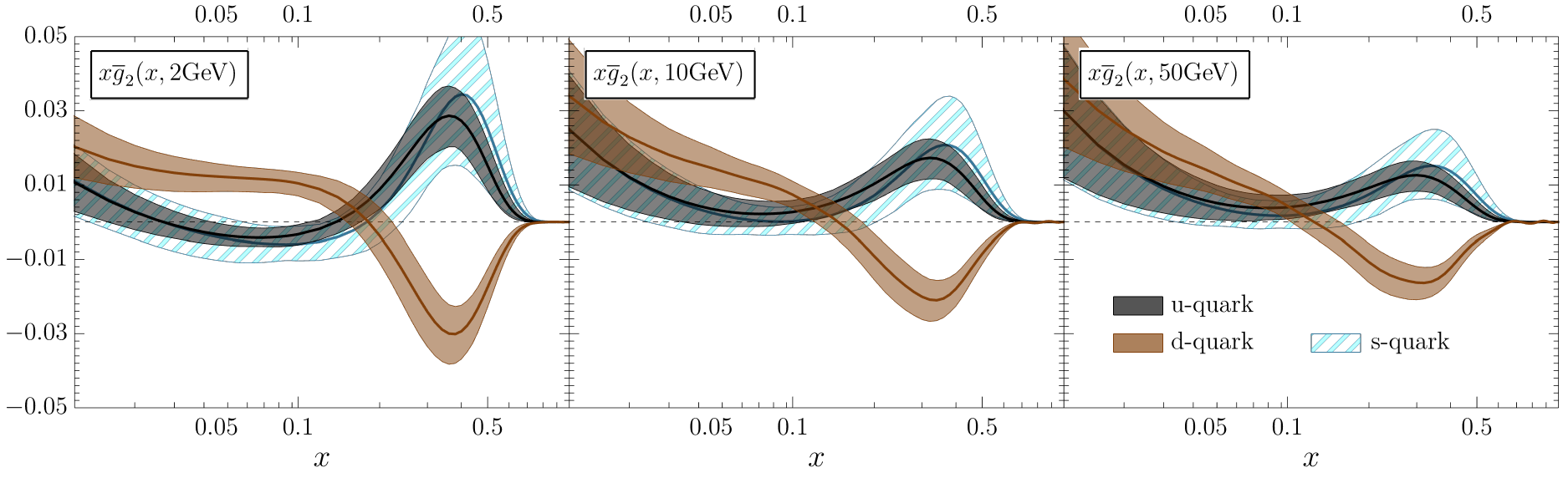}
\includegraphics[width=0.98\linewidth]{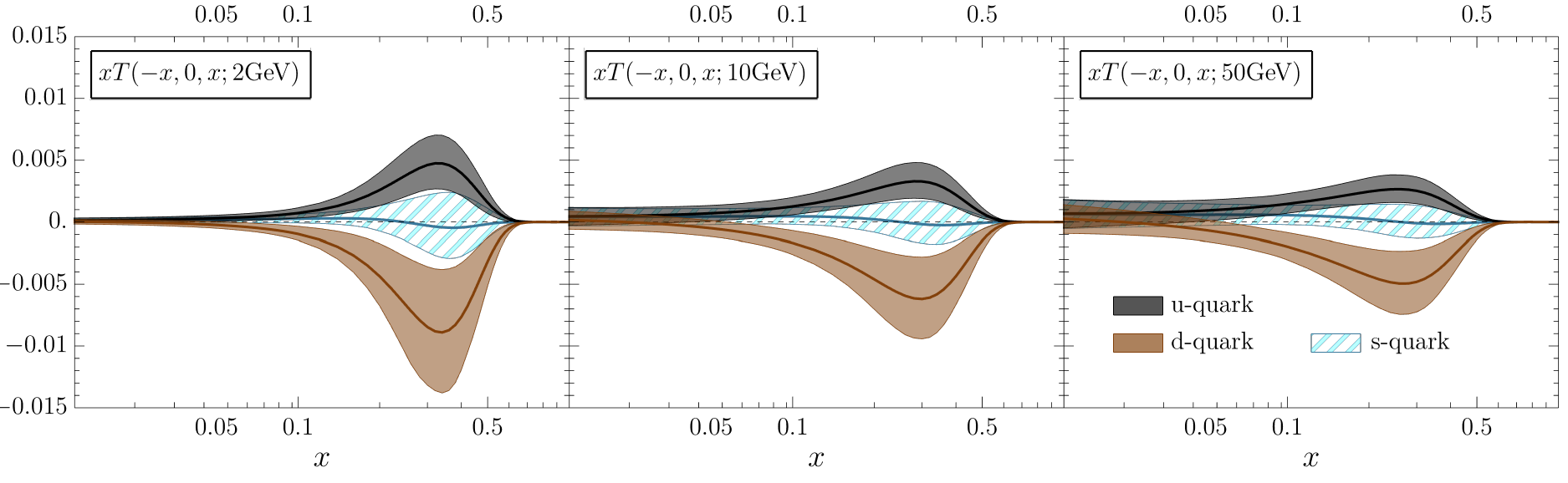}
\includegraphics[width=0.98\linewidth]{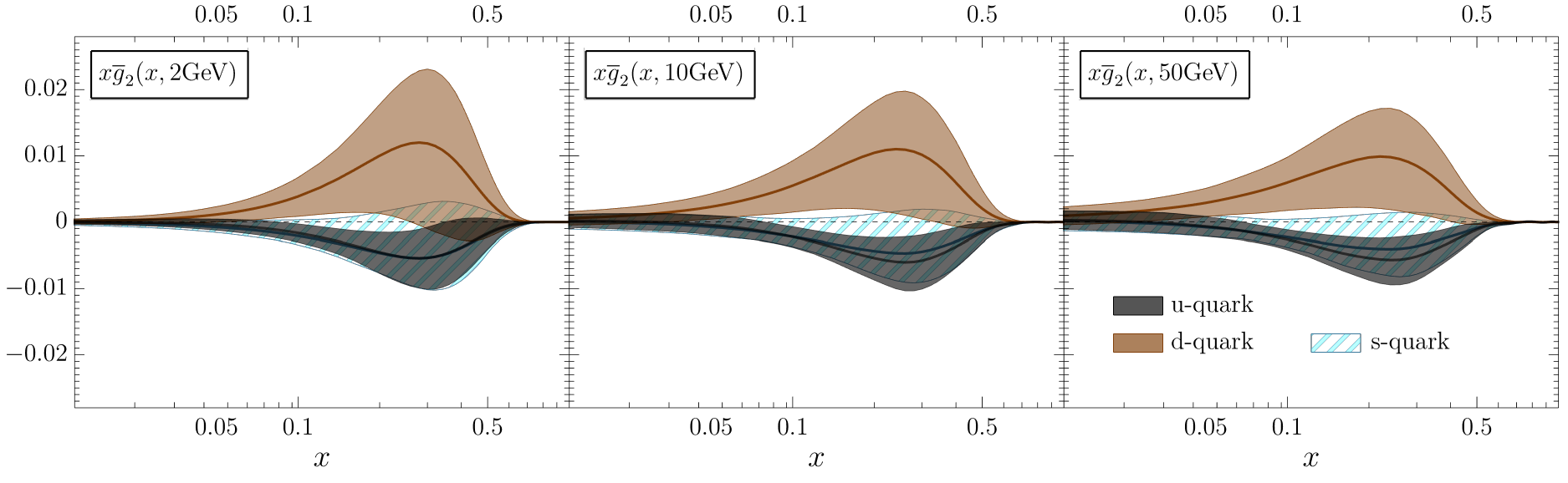}
\caption{\label{fig:g2_QS_evolution}
Evolution of various twist-three matrix elements for $u$ and $d$-quarks. Upper (middle/lower) plot shows the $\bar g_2$ (Qiu-Sterman/$g_{1T}^{(1)\text{tw3}}$) function.
}
\end{figure}

\begin{figure}[t]
\centering
\includegraphics[width=0.99\linewidth]{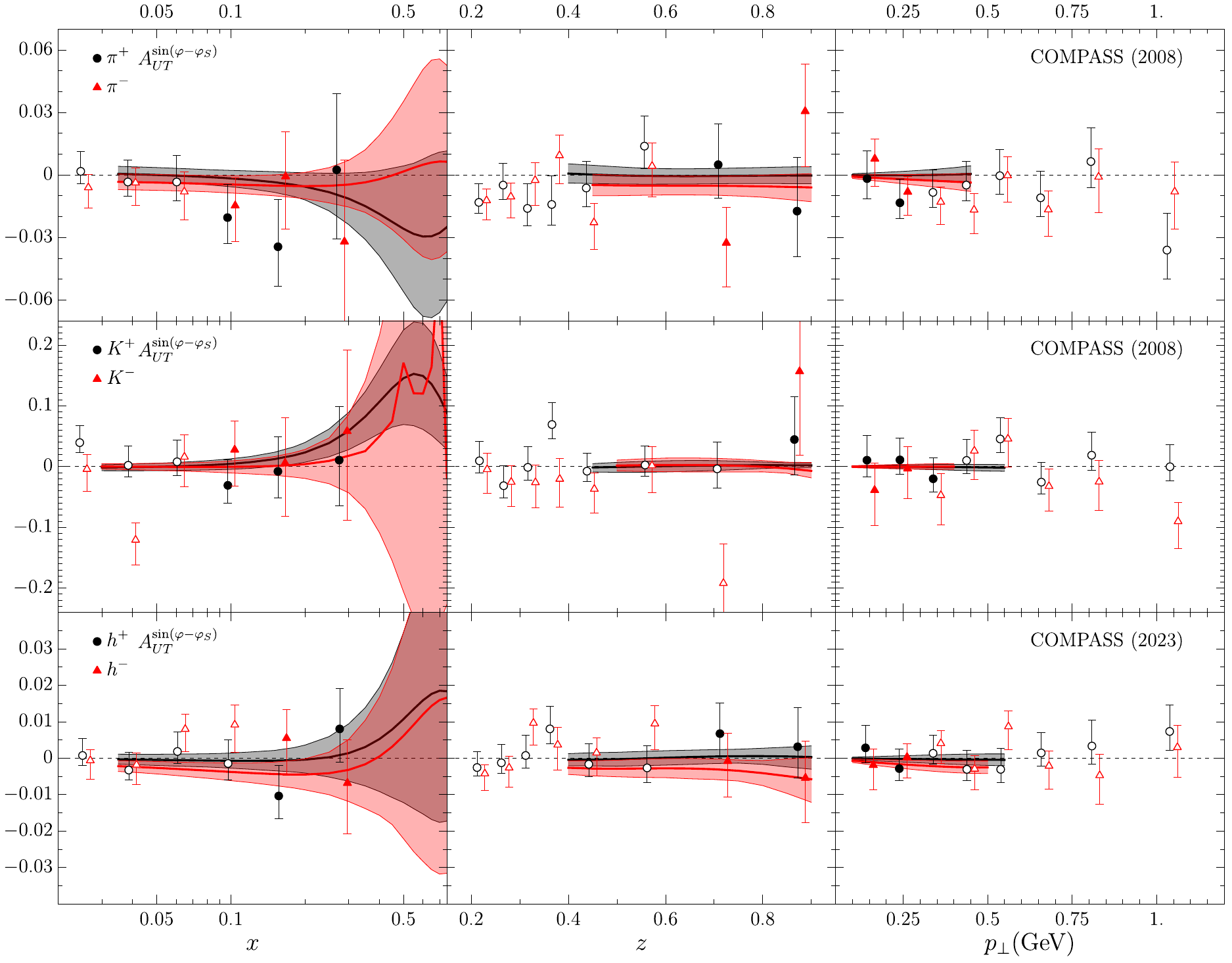}
\caption{\label{fig:COMPASS1} Comparison of the theory prediction with COMPASS data \cite{COMPASS:2008isr, COMPASS:2023vhr} for $A_{UT}^{\sin(\varphi-\varphi_S)}$. The theory prediction is computed for the average values of the integrated variables, and goes beyond the factorization limits. The filled points were included in the fit, since they are within the applicability region of the factorization theorem \ref{data:Q2>2+TMD}}
\end{figure}

\begin{figure}[t]
\centering
\includegraphics[width=0.99\linewidth]{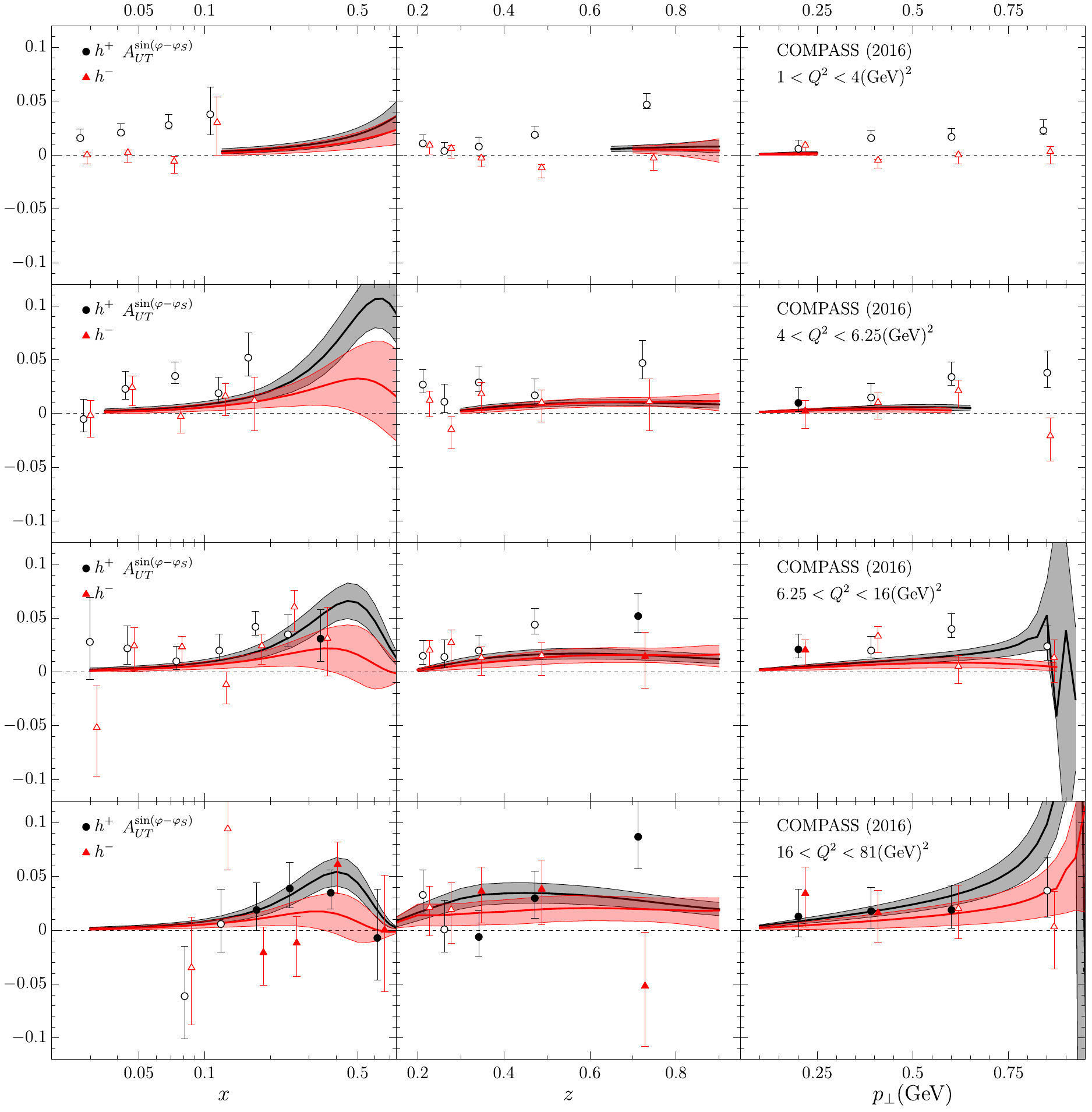}
\caption{\label{fig:COMPASS2} Comparison of the theory prediction with COMPASS data \cite{COMPASS:2016led} for $A_{UT}^{\sin(\varphi-\varphi_S)}$. The theory prediction is computed for the average values of the integrated variables, and goes beyond the factorization limits. The filled points were included in the fit, since they are within the applicability region of the factorization theorem \ref{data:Q2>2+TMD}}
\end{figure}

\begin{figure}[t]
\centering
\includegraphics[width=0.99\linewidth]{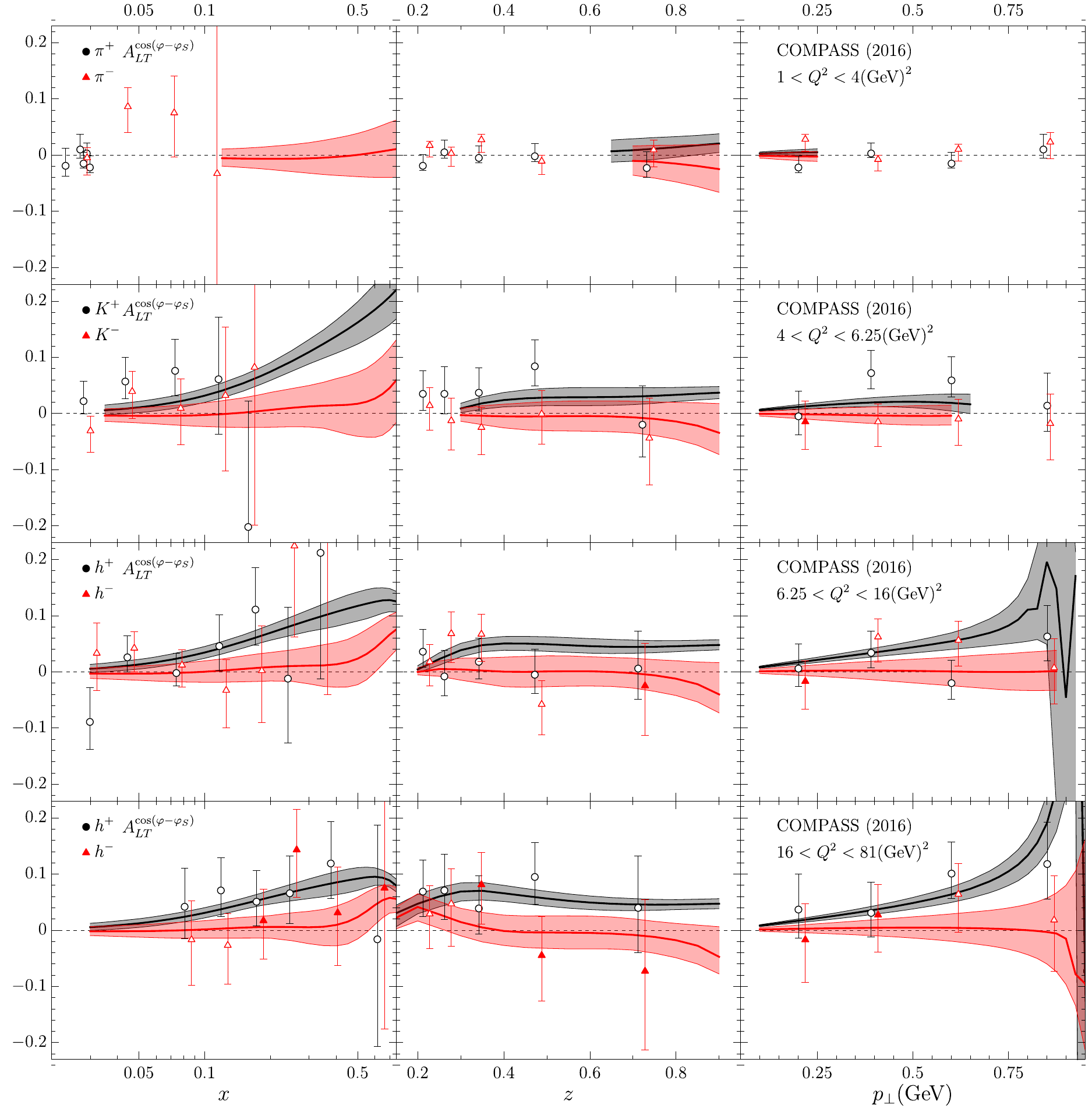}
\caption{\label{fig:COMPASS3} Comparison of the theory prediction with COMPASS data \cite{COMPASS:2016led} for $A_{LT}^{\cos(\varphi-\varphi_S)}$. The theory prediction is computed for the average values of the integrated variables, and goes beyond the factorization limits. The filled points were included in the fit, since they are within the applicability region of the factorization theorem \ref{data:Q2>2+TMD}}
\end{figure}

\begin{figure}[t]
\centering
\includegraphics[width=0.99\linewidth]{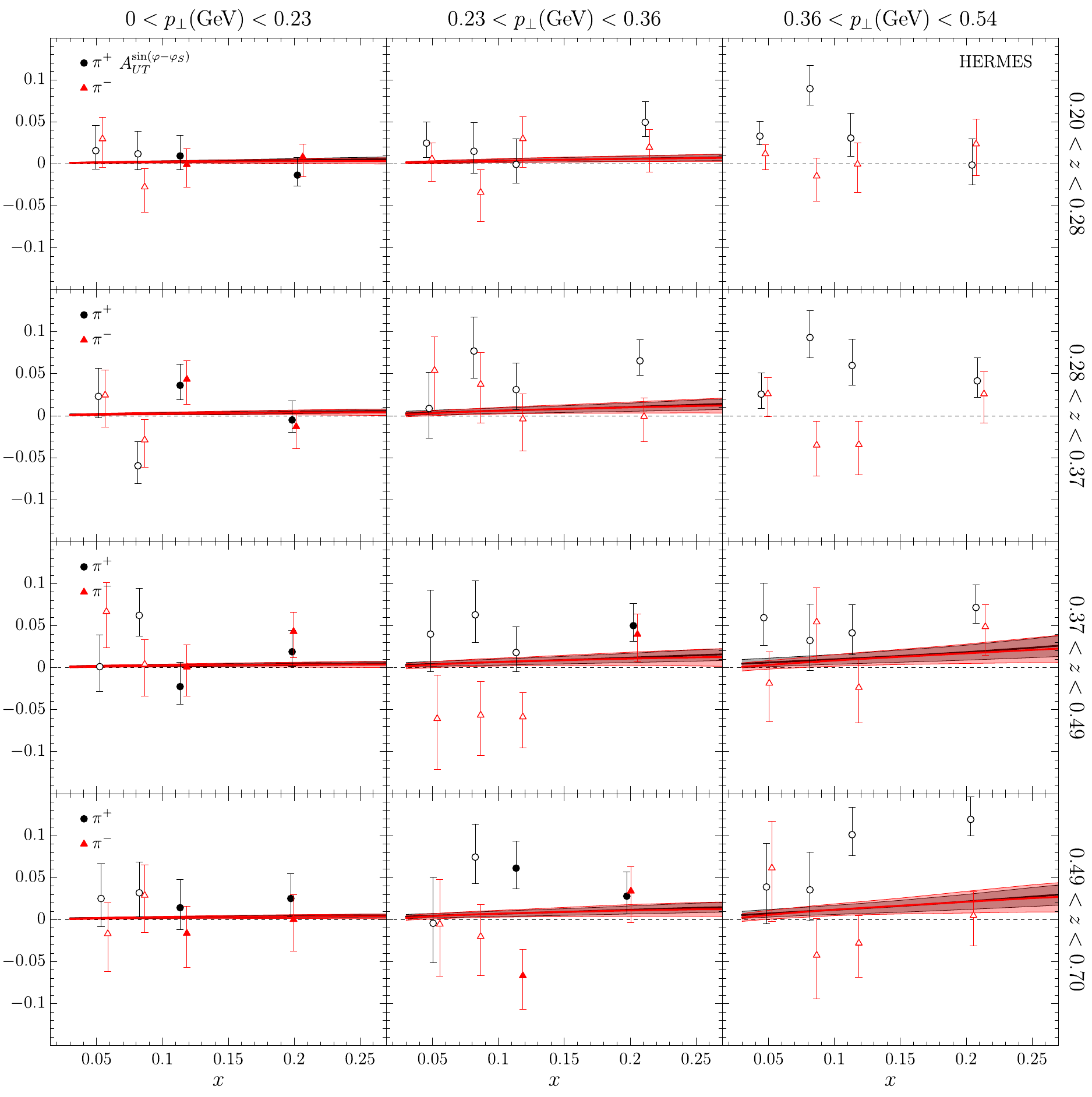}
\caption{\label{fig:HERMES1} Comparison of the theory prediction with HERMES data \cite{HERMES:2020ifk} for $A_{UT}^{\sin(\varphi-\varphi_S)}$, $\pi^\pm$-production. The theory prediction is computed for the average values of the integrated variables, and goes beyond the factorization limits. The filled points were included in the fit, since they are within the applicability region of the factorization theorem \ref{data:Q2>2+TMD}}
\end{figure}

\begin{figure}[t]
\centering
\includegraphics[width=0.99\linewidth]{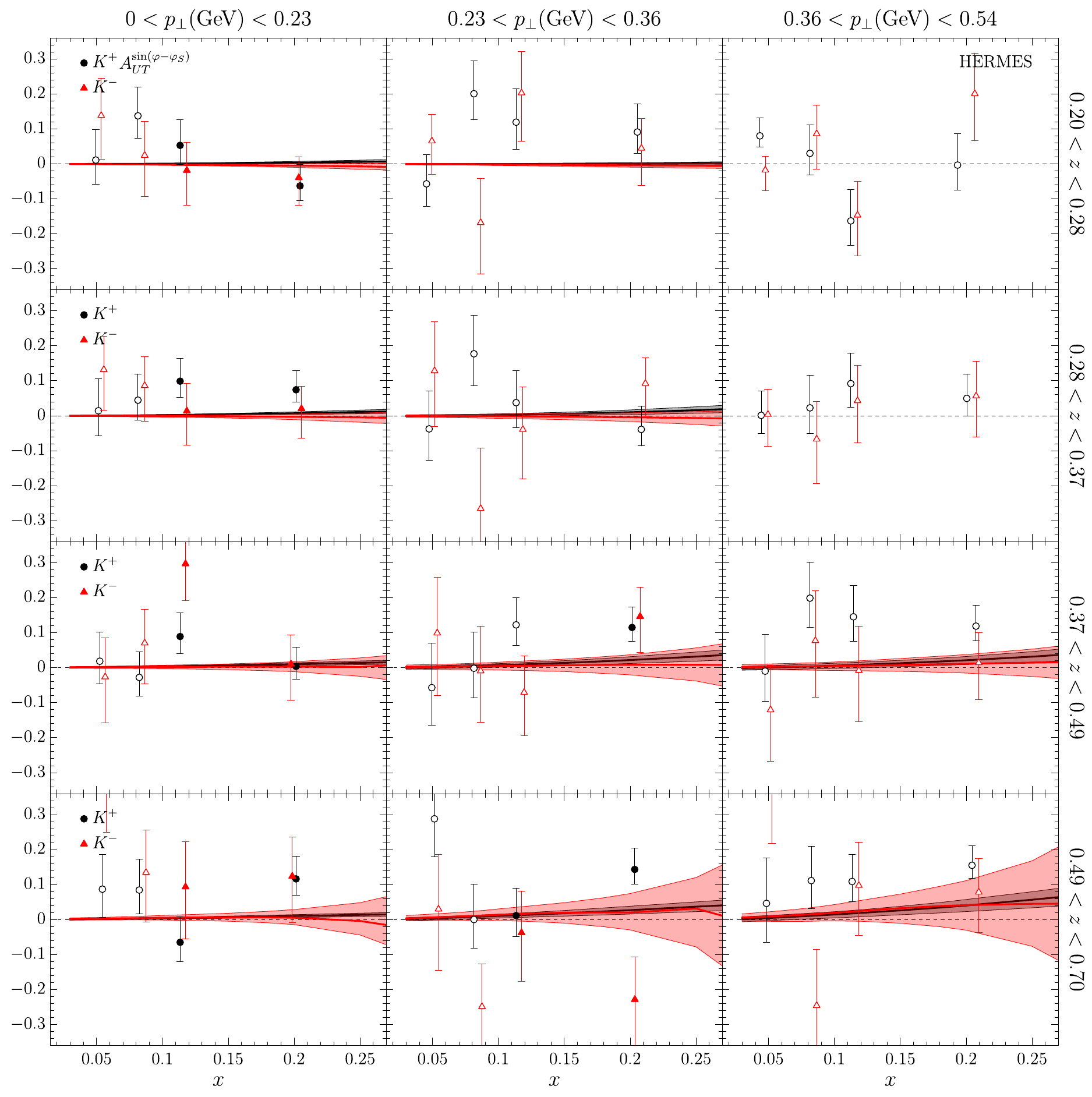}
\caption{\label{fig:HERMES2} Comparison of the theory prediction with HERMES data \cite{HERMES:2020ifk} for $A_{UT}^{\sin(\varphi-\varphi_S)}$, $K^\pm$-production. The theory prediction is computed for the average values of the integrated variables, and goes beyond the factorization limits. The filled points were included in the fit, since they are within the applicability region of the factorization theorem \ref{data:Q2>2+TMD}}
\end{figure}

\begin{figure}[t]
\centering
\includegraphics[width=0.99\linewidth]{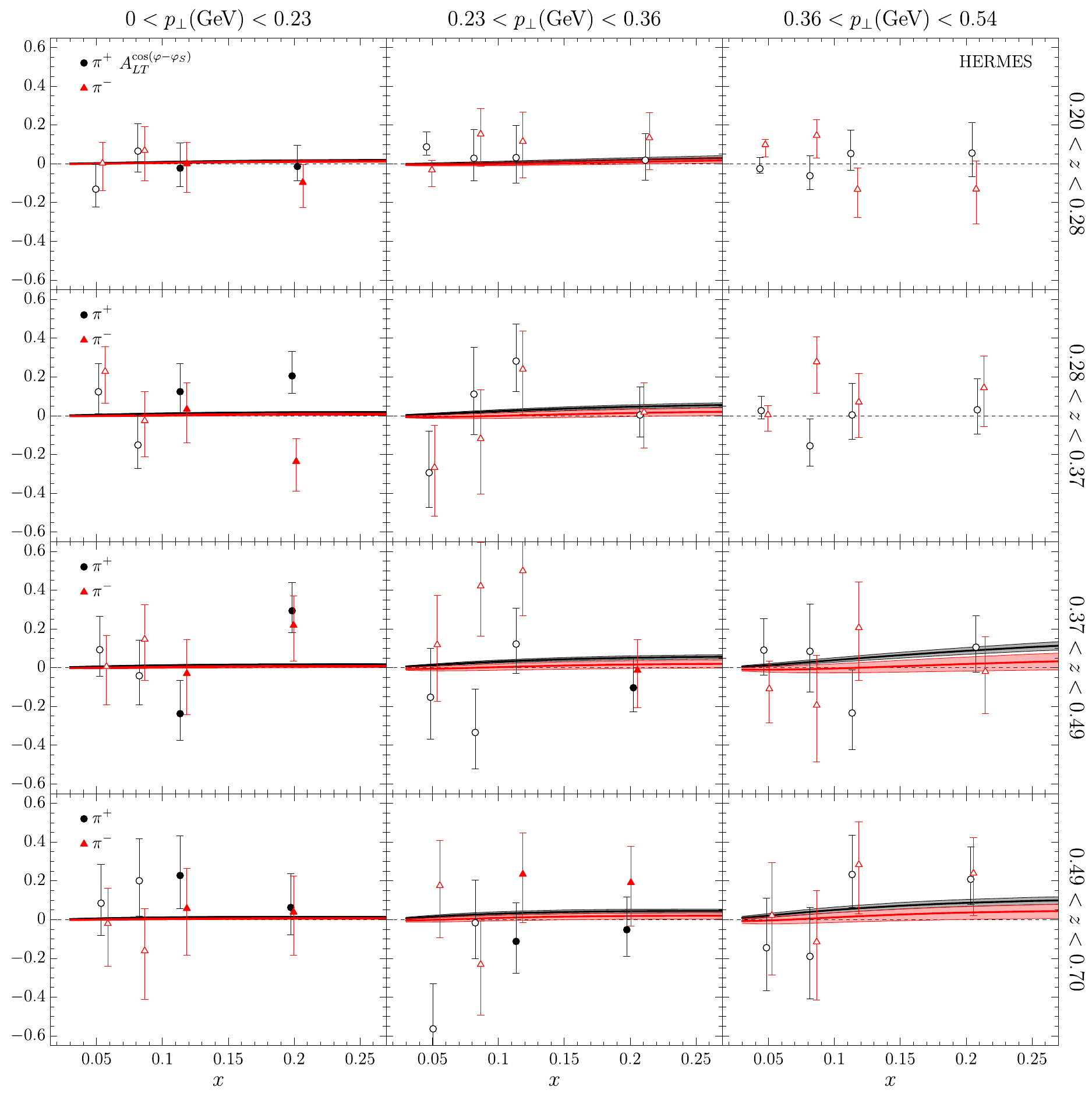}
\caption{\label{fig:HERMES3} Comparison of the theory prediction with HERMES data \cite{HERMES:2020ifk} for $A_{LT}^{\cos(\varphi-\varphi_S)}$, $\pi^\pm$-production. The theory prediction is computed for the average values of the integrated variables, and goes beyond the factorization limits. The filled points were included in the fit, since they are within the applicability region of the factorization theorem \ref{data:Q2>2+TMD}}
\end{figure}

\begin{figure}[t]
\centering
\includegraphics[width=0.99\linewidth]{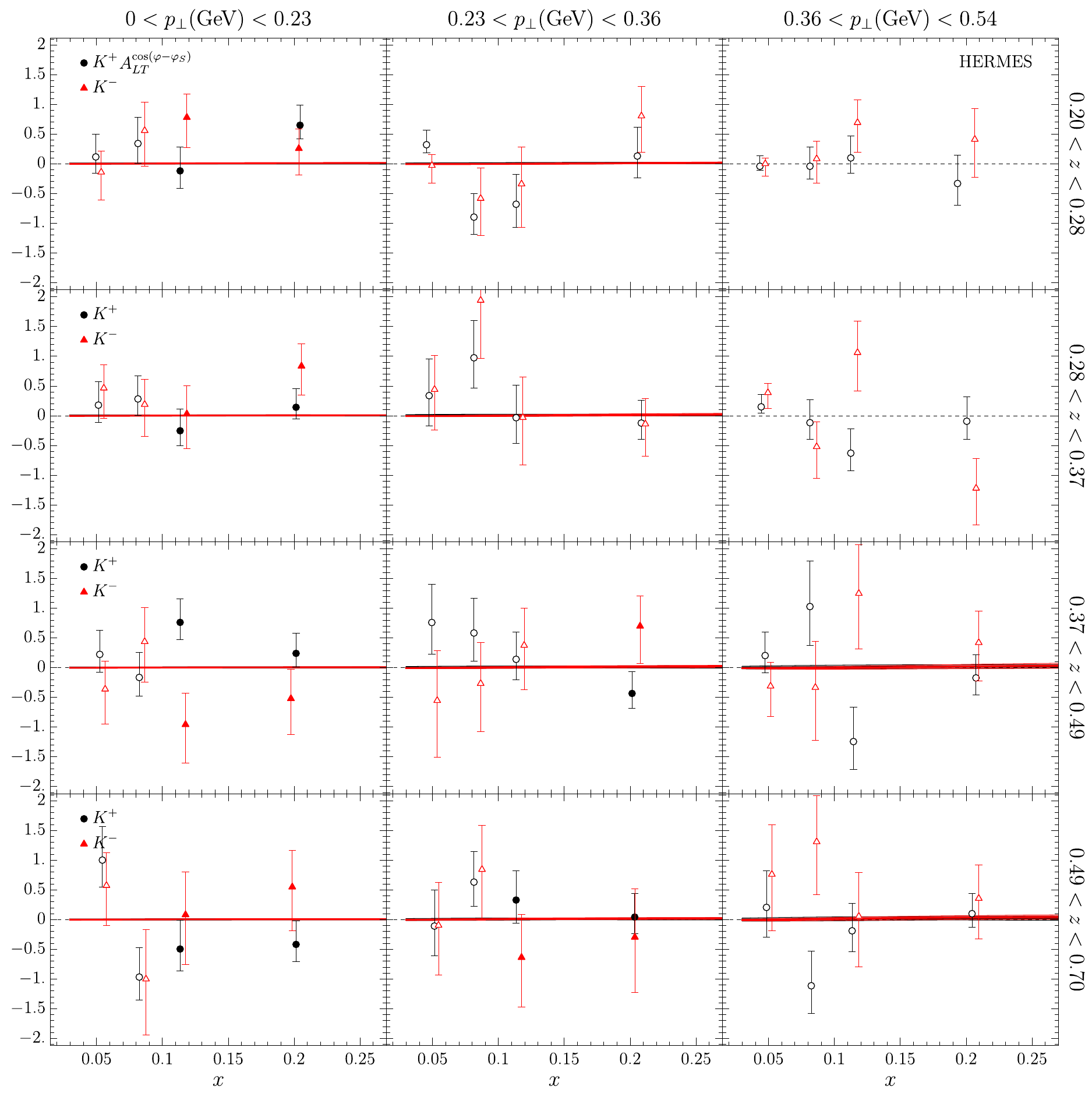}
\caption{\label{fig:HERMES4} Comparison of the theory prediction with HERMES data \cite{HERMES:2020ifk} for $A_{LT}^{\cos(\varphi-\varphi_S)}$, $K^\pm$-production. The theory prediction is computed for the average values of the integrated variables, and goes beyond the factorization limits. The filled points were included in the fit, since they are within the applicability region of the factorization theorem \ref{data:Q2>2+TMD}}
\end{figure}

\begin{sidewaysfigure}
\includegraphics[width=\textwidth]{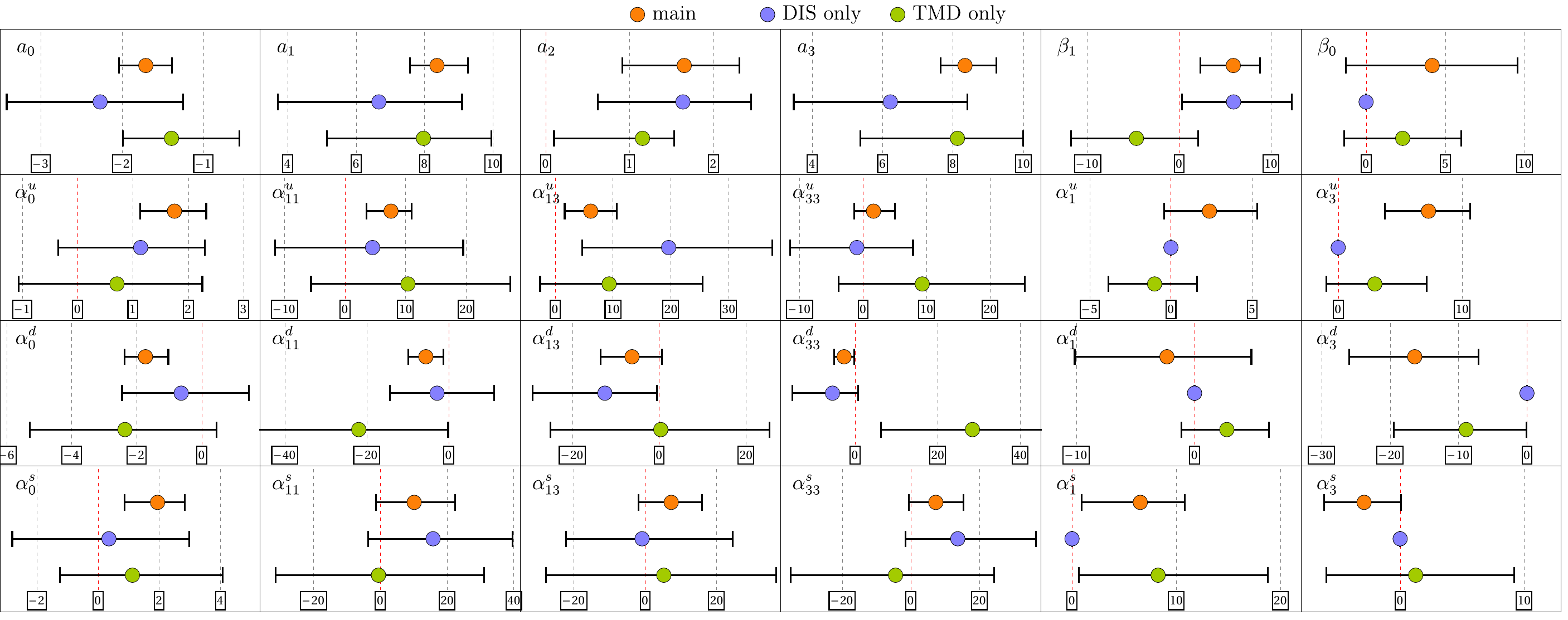}
\caption{\label{fig:parameter_comparison} Comparison of the values of the parameters extracted in various fits.}
\end{sidewaysfigure}

\begin{sidewaysfigure}
\includegraphics[width=\textwidth]{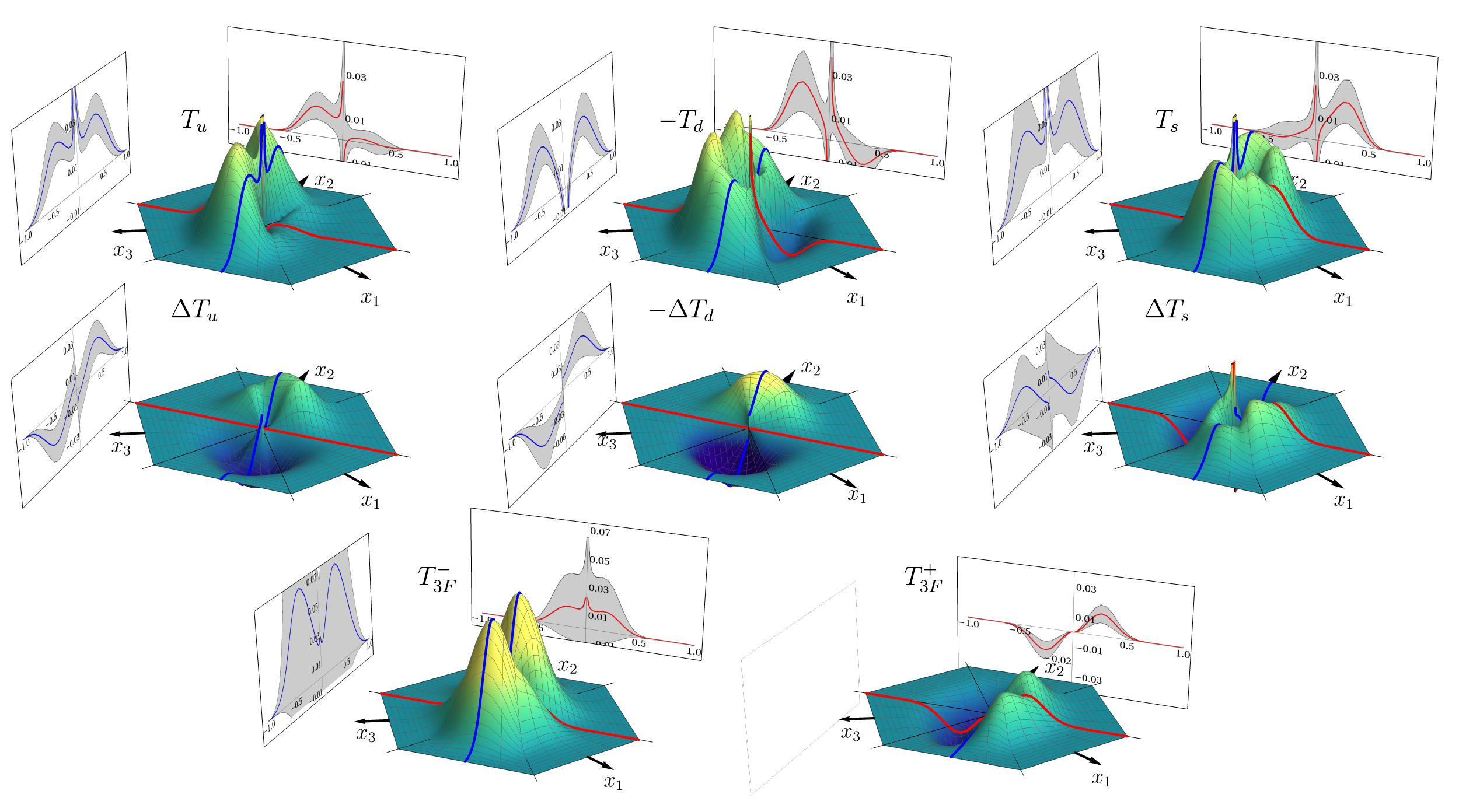}
\caption{\label{fig:AllDistributions} Extracted genuine twist-three distributions $T$, $\Delta T$ and $T_{3F}^\pm$ at $\mu=4$GeV. Note that the distributions $T_d$ and $\Delta T_d$ are multiplied by the factors $(-1)$ for better visibility.}
\end{sidewaysfigure}

\begin{sidewaysfigure}
\includegraphics[width=\textwidth]{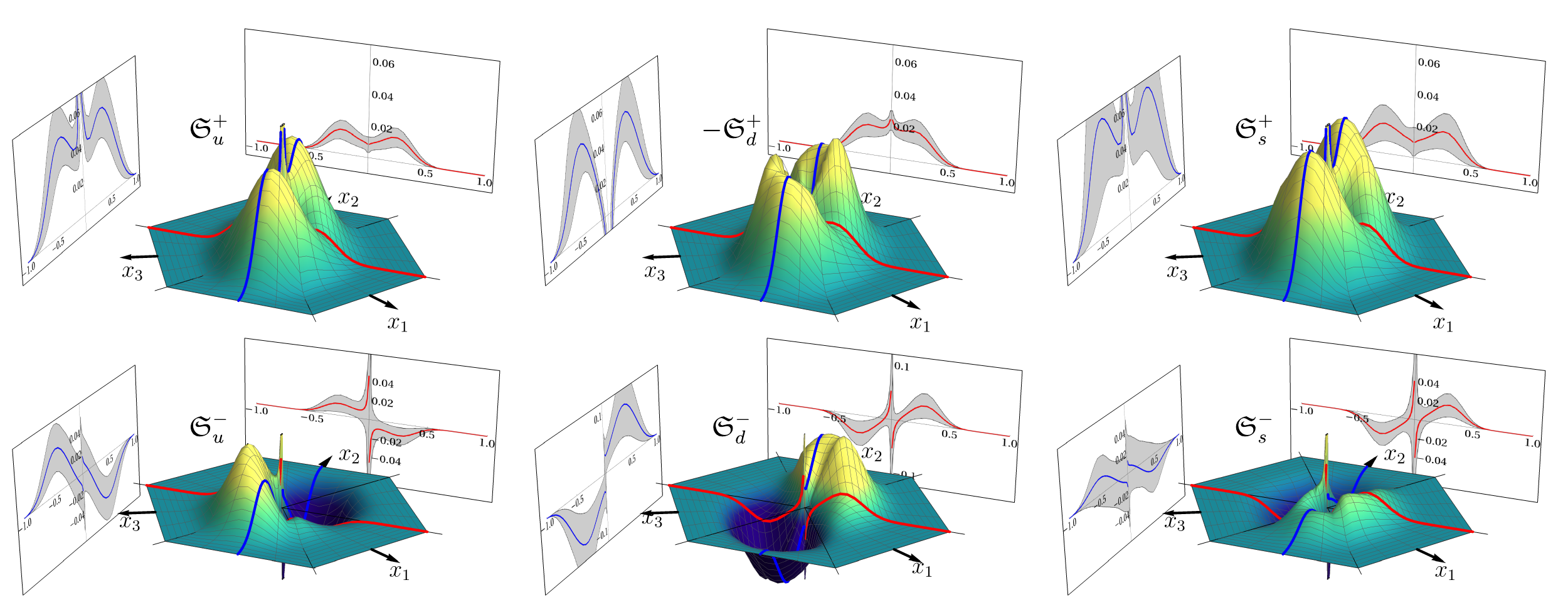}
\caption{\label{fig:AllDistributions_C} Extracted genuine twist-three distributions $\mathfrak{S}^\pm$ at $\mu=4$GeV.}
\end{sidewaysfigure}

\begin{sidewaysfigure}
\includegraphics[width=\textwidth]{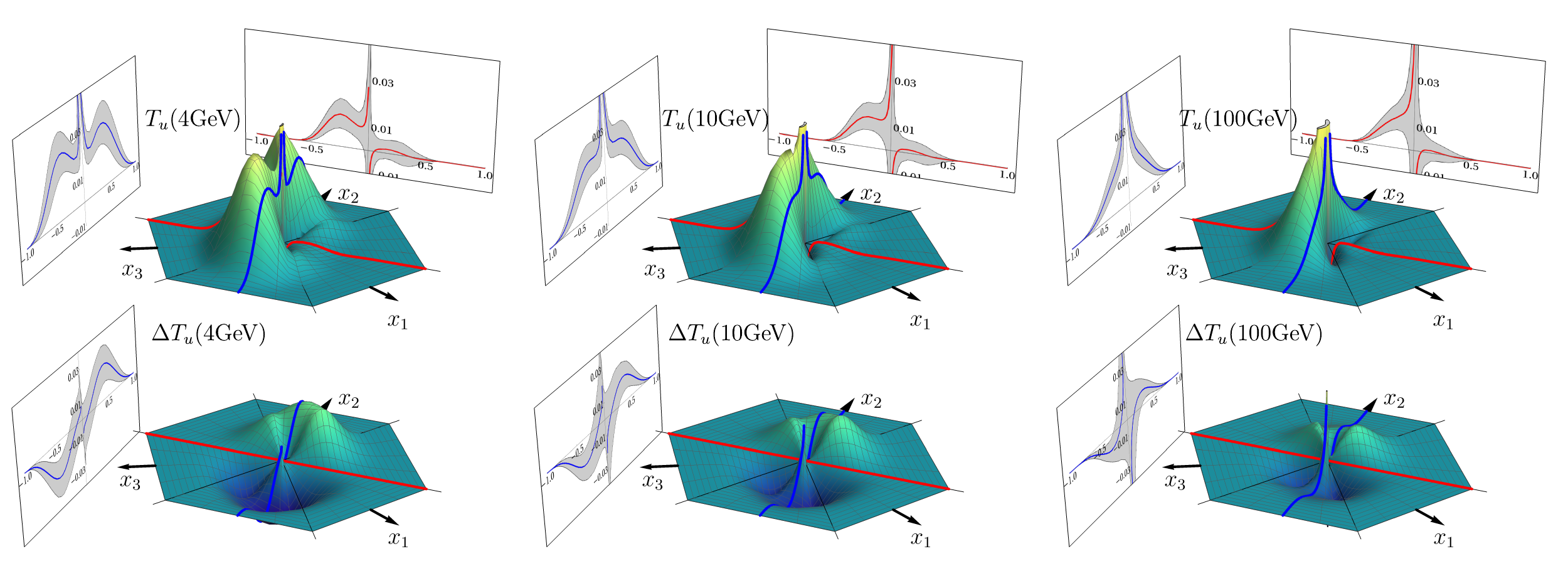}
\caption{\label{fig:T_evolution} The distributions $T_u$ and $\Delta T_u$ at different values of $\mu$.}
\end{sidewaysfigure}

\clearpage

\bibliographystyle{JHEP}
\bibliography{bibFILE}

\end{document}